\documentclass[aps,pre,preprint,groupedaddress,longbibliography]{revtex4-1}
\pdfoutput=1
\usepackage{amsmath}
\usepackage[utf8]{inputenc}
\usepackage[english]{babel}
\usepackage{amssymb}
\usepackage{graphicx}
\usepackage{amsthm}
\usepackage{makecell}
\usepackage{calc}
\usepackage{natbib}
\usepackage{lipsum}
\usepackage{amsfonts}

\usepackage[usenames,dvipsnames]{color} 
\usepackage{ifthen}
\usepackage{soul}
\newboolean{showcomments}
\setboolean{showcomments}{true}

\newcommand{\dennice}[1]{\ifthenelse{\boolean{showcomments}}
{\textcolor{purple}{Dennice says: #1}}{}}
\newcommand{\addcites}[0]{\ifthenelse{\boolean{showcomments}}
{\textcolor{blue}{(add cite(s))}}{}}

\newcommand{\cliu}[1]{\ifthenelse{\boolean{showcomments}}
{\textcolor{blue}{Chang says: #1}}{}}

\newcommand{\rev}[1]{
{\textcolor{black}{#1}}{}}

\usepackage{pifont}

\usepackage{tikz}

\AtBeginDocument{%
    \newwrite\bibnotes
    \def\bibnotesext{Notes.bib}
    \immediate\openout\bibnotes=\jobname\bibnotesext
    \immediate\write\bibnotes{@CONTROL{REVTEX41Control}}
    \immediate\write\bibnotes{@CONTROL{%
    apsrev41Control,author="08",editor="1",pages="1",title="0",year="1"}}
     \if@filesw
     \immediate\write\@auxout{\string\citation{apsrev41Control}}%
    \fi
}%

\newtheorem{thm}{Theorem} 

\newtheorem{pro}{Proposition} 

\newtheorem{remark}{Remark}
\newcommand\norm[1]{\left\lVert#1\right\rVert}

\newtheorem{lemma}{Lemma}

\newenvironment{myproof}{{\textbf{Proof:}}}{\hfill$\square$}

\def\mline{\vrule width4pt height2.5pt depth -2pt}

\def\bdot{\raise.2em\hbox to .15em{.}}
\def\dashed{\mline\hskip3.5pt\mline\thinspace}
\def\dashdot{\mline\ \bdot\,\  \mline\thinspace}

\begin{document}


\title{An input--output inspired method for permissible perturbation amplitude of transitional wall-bounded shear flows}


\author{Chang Liu}\email[]{changliu@jhu.edu}
\author{Dennice F. Gayme}\email[]{dennice@jhu.edu}
\affiliation{Department of Mechanical Engineering, Johns Hopkins University, Baltimore, MD 21218, USA}


\date{\today}

\begin{abstract}
The precise set of parameters governing the transition to turbulence in wall-bounded shear flows remains an open question; many theoretical bounds have been obtained, but there is not yet a consensus between these bounds and experimental/simulation results. In this work, we focus on a method to provide a provable Reynolds number dependent bound on the amplitude of perturbations a flow can sustain while maintaining the laminar state. Our analysis relies on an input--output approach that partitions the dynamics into a feedback interconnection of the linear and nonlinear dynamics (i.e., a Lur\'e system that represents the nonlinearity as static feedback). We then construct quadratic constraints of the nonlinear term that is restricted by system physics to be energy-conserving (lossless) and to have bounded input--output energy. Computing the region of attraction of the laminar state (set of safe perturbations) and permissible perturbation amplitude are then reformulated as Linear Matrix Inequalities (LMI), which provides a more computationally efficient solution than prevailing nonlinear approaches based on the sum of squares programming. The proposed framework can also be used for energy method computations and linear stability analysis. We apply our approach to low dimensional nonlinear shear flow models for a range of Reynolds numbers. The results from our analytically derived bounds are consistent with the bounds identified through exhaustive simulations. However, they have the added benefit of being achieved at a much lower computational cost and providing a provable guarantee that a certain level of perturbation is permissible.
\end{abstract}

\maketitle


\section{Introduction}
\label{sec:introduction}

Linear  analysis has been widely used to study transition in a range of flows \citep{drazin2004hydrodynamic,schmid2012stability}. However, it has been known to fail in predicting the Reynolds number at which transition occurs in wall-bounded shear flows, which are important in a wide range of applications. For example, linear stability analysis indicates that the laminar state of the plane Couette flow is stable against infinitesimal perturbation for any Reynolds number; i.e., $Re_L=\infty$ \citep{romanov1973stability}, while experimental observations indicate that transition occurs at a critical Reynolds number of $Re_C=360\pm 10$ \citep{tillmark1992experiments}. This mismatch has been  attributed to the fact that the infinitesimal perturbation inherent in linear stability analysis does not capture the true growth of the perturbation either due to nonlinear effects \citep{waleffe1995transition} as well as to the known algebraic growth \citep{reddy1993energy,schmid2012stability} resulting from the non-normality of the linearized Navier-Stokes (NS) operator \citep{trefethen1993hydrodynamic,henningson1994role,trefethen2005spectra}.

Energy methods employ Lyapunov based analysis of the nonlinear flow field and therefore overcome the limitations to infinitesimal perturbations and linear behavior~\citep{joseph2013stability,straughan2013energy}. Classical energy methods employ the perturbation kinetic energy as a radially unbounded Lyapunov function, which produces a certificate (rigorous proof) of globally asymptotic stability of the base flow at a given Reynolds number. Defining transition to turbulence in terms of loss of this globally asymptotic stability using a quadratic Lyapunov function provides a conservative bound on the transition Reynolds number predicted by the energy method (here denoted $Re_E$). Thus, $Re_E$ is typically much lower than the critical Reynolds number observed in experiments; e.g., $Re_E\approx 20.7$ for plane Couette flow (See e.g., Figure 5.11(b) in Ref. \citep{schmid2012stability}). Energy methods have recently been expanded to a broader class of polynomial Lyapunov functions, which has led to less conservative bounds for a range of flow configurations~\citep{goulart2012global,chernyshenko2014polynomial,huang2015sum,fuentes2019global}. For example, \citet{fuentes2019global} employed quartic polynomials as a Lyapunov function to verify the global stability of 2D plane Couette flow at Reynolds numbers below $Re=252.4$, which is substantially higher than the $Re_E=177.2$ bound attained through classical energy stability methods. Much of that work has been enabled through the sum of squares (SOS) techniques that provide a computational approach for computing polynomial Lyapunov functions~\citep{prajna2002introducing,papachristodoulou2013sostools}. However, both the energy stability method and its generalization provide no information about the flow regime $Re_E<Re<Re_L$, where the base flow is stable against infinitesimal perturbations, but some finite perturbations can lead to transition,  for example at the $Re_C$ values observed in experiments.

In general, at a given $Re$ in the flow regime $Re_E<Re<Re_L$, there exists a critical perturbation amplitude above which transition to turbulence is observed for particular forcing shapes and another permissible perturbation amplitude, $\delta_{\text{p}}$, below which all perturbations will decay \citep{baggett1997low}. These perturbation amplitudes are of particular importance in understanding the transition to turbulence and in the design of flow control approaches. However, they are difficult to determine in practice. The most common approach involves extensive numerical simulations  \citep{kreiss1994bounds,reddy1998stability,Schneider2007,eckhardt2007turbulence,Schneider2010,Chantry2014} or experiments \citep{grossmann2000onset,hof2003scaling,peixinho2007finite,mullin2011experimental}. However, an inherently finite set of experiments or numerical simulations cannot provide a provable bound on either the permissible level of perturbation to maintain a laminar flow state or the critical perturbation that leads to transition. A more rigorous (but likely conservative) bound  on the permissible perturbation amplitude can be obtained through computing a region of attraction based on Lyapunov methods; see, e.g., Chapter 8.2 of Ref. \citep{khalil2002nonlinear}.  Lyapunov based methods have been applied in a wide range of stability based analyses for different flow regimes including global stability analysis \citep{goulart2012global,chernyshenko2014polynomial,huang2015sum,fuentes2019global}, bounding long time averages \citep{chernyshenko2014polynomial,fantuzzi2016bounds}, controller synthesis for laminar wakes \citep{lasagna2016sum,Huang2017},  and finding dynamically important periodic orbits \citep{lakshmi2020finding}. However, computation of the Lyapunov function and the associated analysis approaches typically rely on SOS methods, which are known to be computationally expensive when the dimension of the system is large \citep{zheng2018fast}.

Alternative approaches to determining permissible perturbations for a given flow condition have combined 
optimization methods with NS solvers to obtain initial conditions resulting in the largest nonlinear energy growth at a given final time $T$; i.e., the nonlinear optimal transient growth \citep{Kerswell2014,Kerswell2018}. This method has been effective in determining the shape of perturbation that is most efficient in triggering the transition to turbulence \citep{Pringle2010,duguet2010towards,Pringle2012,rabin2012triggering,Duguet2013}. However, this method requires an a priori specification of a large enough $T$ to ensure that it captures the full behavior as $T\rightarrow \infty$ \citep{Kerswell2014}, which leads to a trade-off between accuracy and computational time.

Low dimensional shear flow models have been used to provide insight into the critical Reynolds number and the permissible perturbation amplitude for a given flow without the full computational burden of the NS equations \citep{trefethen1993hydrodynamic,kreiss1994bounds,gebhardt1994chaos,waleffe1995transition,baggett1995mostly,baggett1997low,Moehlis2004,Moehlis2005,lebovitz2013edges,Joglekar2015}. These models are constructed to capture the transitional behavior of wall-bounded shear flows. In particular, the nine-dimensional shear flow model obtained from a Galerkin projection of NS equations \citep{Moehlis2004} was designed to reproduce the bifurcations, periodic orbits \citep{Moehlis2005}, and edge of chaos phenomena \citep{Kim2008,Joglekar2015} observed in direct numerical simulations (DNS) of wall-bounded shear flows. This nine-mode model \citep{Moehlis2004} has been widely studied as a prototype shear flow model, see e.g. \citep{Moehlis2004,Moehlis2005,Kim2008,Joglekar2015,goulart2012global,chernyshenko2014polynomial}. In particular, the question of transition in this flow has been assessed in terms of both its global stability \citep{goulart2012global}, bounds on the long-time average of the energy dissipation \citep{chernyshenko2014polynomial} as well as through exhaustive simulations to determine both permissible and critical perturbations as a function of the Reynolds number \citep{Joglekar2015}. The reduced-order and ability of these models to capture important flow characteristics have led to extensive use of such models to both gain insight into the underlying physics and test analysis tools. However, a number of challenges remain even in characterizing these reduced-order models, including the inability to attain a rigorous bound through simulation and the large computational cost of the prevailing SOS based analysis tools.

\begin{figure}
    \centering
    \includegraphics[width=3in]{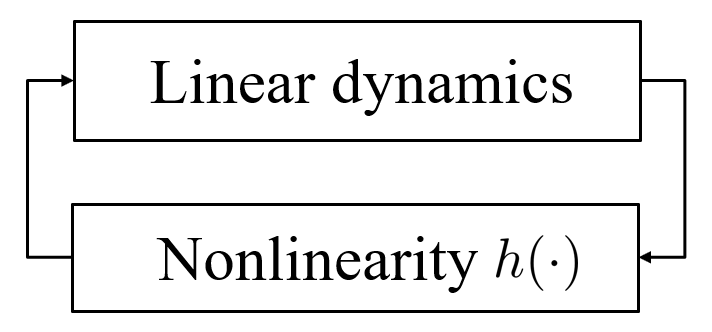}
    \caption{Illustration of partitioning the dynamics into a feedback interconnection of linear and nonlinear dynamics; i.e., a Lur\'e system.}
    \label{fig:IO_critical_amplitude}
\end{figure}

In this work, we address the problem of determining a permissible perturbation amplitude through an alternative view of the stability properties of these nonlinear systems in terms of general input--output properties of the system, see  e.g. \cite{Bamieh2001, ahmadi2019framework,Jovanovic2005,McKeon2010,jovanovic2020bypass}. A common approach to input--output based analysis involves partitioning the system into a linear system that is forced by the system nonlinearity $h(\cdot)$, as shown in Figure \ref{fig:IO_critical_amplitude}. This point of view in which the nonlinearity acts as a forcing that mixes the nonlinear modes forms the basis of a number of previous analyses of the system transfer function or resolvent, see e.g. \citep{Bamieh2001,Jovanovic2005,McKeon2010,Sharma2013,mckeon2013experimental,mckeon2017engine,liu2019vorticity,liu2019input,jovanovic2020bypass}.  This reformulation of the problem leads to a Lur\'e system \citep{kalman1963lyapunov,boyd1994linear,khalil2002nonlinear,li2007improved,li2008concise} in which a linear time-invariant system is connected to a memoryless nonlinear system. This decomposition enables the use of control theoretic tools to provide insight into the input--output stability of the interconnected system based on the properties of the constitutive linear (transfer function/resolvent) and nonlinear relations $h(\cdot)$ in the two blocks in Figure \ref{fig:IO_critical_amplitude} and their interconnection structure \cite{popov1961absolute,kalman1963lyapunov,zames1966input,khalil2002nonlinear}.

In the context of analyzing the stability and of synthesizing controllers for shear flows, the most widely used theory involves ensuring that the interconnection structure is passive. Passive systems are stable in the sense of Lyapunov (i.e., bounded inputs lead to bounded outputs) under certain conditions, see e.g., Lemma 6.5-6.7 of Ref. \citep{khalil2002nonlinear}, and, therefore, the concept of passivity is often used for stability analysis and in control design. 
This concept is useful in terms of analyzing systems of the form in Figure \ref{fig:IO_critical_amplitude} \rev{because} the passivity theorem (e.g., Theorem 6.1 in Ref. \citep{khalil2002nonlinear}) states that if two systems are passive, the feedback interconnection of these two passive systems remains passive. This property allows one to analyze and control the full nonlinear system through each subsystem; e.g., passivity-based control \citep{van2000l2,ortega2013passivity}. In shear flows, as the nonlinearity is known to be energy-conserving \citep{joseph2013stability} (lossless), which is a special case of passive, this theory is an appealing analysis tool for these systems. \citet{sharma2011relaminarisation} invoked this theory to synthesize a feedback controller to render the linear system passive in order to stabilize the full nonlinear system governing turbulent channel flow at $Re_{\tau}=100$ (i.e. to relaminarize it). Similar approaches have been applied to the Blasius boundary layer \citep{Damaren2016,Damaren2018} and for control of channels with sensing and actuation limited to the wall \citep{heins2016passivity}. The notion of passivity has also been used in recent work to study a wider class of input--output properties \cite{ahmadi2019framework}.

The dynamics of the interconnected system can also be evaluated using the concept of sector bounds (see e.g., Chapter 6 of Ref. \citep{khalil2002nonlinear}), wherein the nonlinear map of the state $h(\boldsymbol{x})$ mapping the zero state to the origin can be contained within a sector in the $(\boldsymbol{x}, h(\boldsymbol{x}))$ plane. This sector bound on nonlinearity combined with the sector occupying the nonlinear system provides important information about the input--output stability of the interconnected system \cite{zames1966input} and forms the basis of a number of stability analysis tools for nonlinear systems, e.g., Popov and circle criteria \citep{popov1961absolute,zames1966input,khalil2002nonlinear}.  Passive systems provide a special case of sector bounded systems; see e.g., Definitions 6.1 and 6.2 of Ref. \citep{khalil2002nonlinear}.

Sector bound requirements have proven conservative in problems in which the form of the nonlinearity is known or there are slope restrictions on the sector bound \citep{park1997revisited,park2019less}. Less conservative results can be obtained through relaxing the sector bounds requirement and instead imposing local bounds that enable an analysis of the system over a local region rather than by global analysis \citep{weissenberger1968application,hindi1998analysis,valmorbida2018regional}. This approach was used to compute the region of attraction for a dynamical system with logarithmic and fractional nonlinearity by \citet{valmorbida2018regional}. \citet{kalur2020stability,kalur2020nonlinear} similarly employed a local bound on quadratic nonlinearity to perform local stability and energy growth analyses of four-dimensional Waleffe-Kim-Hamilton (WKH) shear flow model \citep{waleffe1995transition}.

In this work, we \rev{employ} the notions of \rev{passivity and }relaxed sector bound constraints to develop a linear matrix inequalities (LMI) based approach to computing a provable bound on the permissible perturbation amplitude $\delta_{\text{p}}$ for a wide class of shear flow models \rev{in which the nonlinearity is passive (in this case energy-conserving) and can be locally sector bounded}.\rev{ We first express known properties of the nonlinearity, e.g. that is  energy-conserving (lossless) and has bounded input--output energy in a local region as LMI system constraints. We then formulate the computation of a region of attraction as an LMI, which allows us to analyze systems with quadratic constraints using linear techniques by expressing conditions related to the positive semi-definiteness of symmetric matrices. This approach has been widely applied in analyzing systems using concepts of passivity and sector bounds; see e.g., examples for fluids problem \citep{ahmadi2019framework,kalur2020nonlinear,kalur2020stability} and a general formulation \citep{boyd1994linear}. The LMI is a natural generalization of linear inequalities where LMI is defined based on the positive semi-definiteness of symmetric matrices.} While our approach is similar to the approach taken in analyzing the WKH model in Refs. \citep{kalur2020stability,kalur2020nonlinear}, we provide a tighter bound, which is expected to lead to a less conservative estimation of the region of attraction. We also take the further step of computing the permissible perturbation amplitude, i.e. the $\delta_{\text{p}}$ below which any perturbation is guaranteed to decay for a full range of shear flow models including the more comprehensive nine-dimensional model \cite{Moehlis2004}. In particular, we compute the Reynolds number dependent permissible perturbation amplitude $\delta_{\text{p}}$ for seven low dimensional shear flow models \citep{trefethen1993hydrodynamic,baggett1995mostly,waleffe1995transition,baggett1997low,Moehlis2004} and compare it with results obtained from extensive numerical simulation using the same models \citep{baggett1997low,Joglekar2015}. The proposed method results in permissible perturbation amplitudes as a function of the Reynolds number for shear flow models \citep{trefethen1993hydrodynamic,baggett1995mostly,waleffe1995transition,baggett1997low,Moehlis2004} that are conservative, yet consistent with those estimated from simulations with randomly chosen initial conditions \citep{baggett1997low,Joglekar2015}. \rev{The analysis provides a generalization of both linear analysis and classical energy methods. In addition, this approach overcomes the lack of rigor associated with simulation based
approaches in that our results provide a provable guarantee that the system will converge to the laminar state for any perturbation amplitude below $\delta_{\text{p}}$.  The LMI based method is more computationally efficient than SOS programming because we restrict the characteristics of the nonlinearity in order to reduce the search space for candidate Lyapunov functions. }  We illustrate the computational efficiency of the method through comparisons with the SOS based approaches for the nine-dimensional shear flow model \citep{Moehlis2004}, which has the largest dimension of the models tested.

The remainder of the paper is organized as follows. Section \ref{sec:nonlinear_IO} describes the problem set-up and derivation of the Linear Matrix Inequalities (LMI) based constraints on the nonlinearity, which are then employed to determine permissible perturbation amplitude. In Section \ref{sec:application_shear}, we apply this framework to shear flow models \citep{trefethen1993hydrodynamic,baggett1995mostly,waleffe1995transition,baggett1997low,Moehlis2004} and compare the obtained permissible perturbation amplitudes with these obtained from extensive simulations \citep{baggett1997low,Joglekar2015} and SOS programming. Section \ref{sec:conslusion} concludes this paper and  discusses future work directions.

\section{Input--Output Based Analysis Framework}
\label{sec:nonlinear_IO}

\begin{figure}
    \centering
    \includegraphics[width=3in]{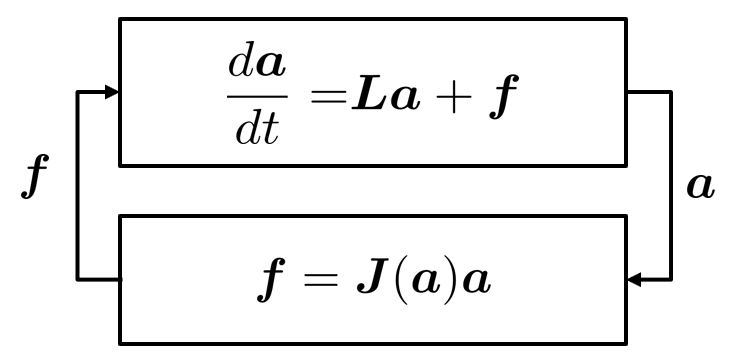}
    \caption{Lur\'e partition of dynamics described in equation (\ref{eq:nonlinear_af}).}
    \label{fig:feedback_forcing}
\end{figure}

The dynamics of a general shear flow can be written in the form,
\begin{align}
    \frac{d\boldsymbol{a}}{dt}=&\boldsymbol{L}\boldsymbol{a}+\boldsymbol{f},
    \label{eq:nonlinear_af}
\end{align}
where $\boldsymbol{a}\in\mathbb{R}^n$ is the state variable,   $\boldsymbol{L}\in\mathbb{R}^{n\times n}$ represents the linear operator arising from a linearization about a flow state, and $\boldsymbol{f}\in\mathbb{R}^n$ are the remaining nonlinear terms. This Lur\'e partition of the equations, illustrated in Figure \ref{fig:feedback_forcing}, views the nonlinearity as feedback forcing to the linear system in the spirit of several previous works using input--output and resolvent analysis, see e.g. \cite{Bamieh2001,Jovanovic2005,McKeon2010,Sharma2013,mckeon2013experimental,mckeon2017engine,liu2019vorticity,liu2019input,jovanovic2020bypass}.

The nonlinear interactions for the class of shear flows of interest here have certain properties that can be exploited in analyzing the block diagram of Figure \ref{fig:feedback_forcing}. Here we focus our analysis on the spatial discretization of the governing equations, which results in a set of ordinary differential equations that approximate the dynamics in equation \eqref{eq:nonlinear_af}. The nonlinearity  is quadratic in the state variable for shear flows and the reduced-order models of interest here. In this setting, such a nonlinearity can be written as $ \boldsymbol{f}=\boldsymbol{J}(\boldsymbol{a})\boldsymbol{a}$, where $\boldsymbol{J}(\boldsymbol{a})\in\mathbb{R}^{n\times n}$ is a state-dependent matrix such that $\boldsymbol{J}(\boldsymbol{0})=\boldsymbol{0}$, and $n$ denotes the number of points used in the discretization of the state variable.

In subsection \ref{subsec:charac_nonlinear}, we use both this quadratic form of the nonlinear interactions and the fact that the nonlinearity is known to be energy-conserving (lossless) \citep{joseph2013stability,sharma2011relaminarisation,sharma2009model,heins2016passivity,Damaren2016,Damaren2018,ahmadi2019framework,Constantin1995} in order to derive constraints that we will later use in our LMI based algorithm in subsection \ref{subsec:permissible_LMI} to evaluate system stability. We take the approach of characterizing the nonlinearity using local rather than (global) sector bounds on two of its properties in order to define an LMI based condition on local stability of the interconnection structure. Our focus on the local rather than global constraints provides relaxation of the strict conditions in classical energy methods in order to understand the behavior of systems whose solutions (laminar states) are stable for finite perturbations but not globally asymptotically stable. In particular, in Lemma \ref{thm:bound} we provide quadratic bounds on the input--output amplification of the nonlinear term $\boldsymbol{f}$ within a neighborhood. Then in Theorem \ref{thm:lmi}, we use these bounds along with a corresponding Lyapunov function to define a region of attraction for the trajectories under the nonlinear mapping. Finally, determining the associated permissible perturbation amplitude to maintain the laminar state is formulated as an LMI constrained optimization problem. Our main theoretical result demonstrates that a feasible solution of this optimization problem provides a permissible perturbation amplitude for the given model.

\subsection{Characterizing the nonlinear interactions}
\label{subsec:charac_nonlinear}

Prior to presenting the main result, we provide a closed-form expression describing the energy-conserving property using the properties of the operator $\boldsymbol{J}(\boldsymbol{a})$ and a related set of quadratic constraints that capture the properties of the nonlinearity. We then derive an upper bound on the quadratic nonlinearity in a local region, which is presented in Lemma \ref{thm:bound}. These results are used in the proof of Theorem \ref{thm:lmi} that provides an LMI based approach to computing the permissible perturbation amplitude for dynamical systems of the form in equation \eqref{eq:nonlinear_af}.

The nonlinear terms in wall-bounded shear flows (see e.g., employed in Refs. \citep{joseph2013stability,sharma2011relaminarisation,sharma2009model,heins2016passivity,Damaren2016,Damaren2018,ahmadi2019framework,Constantin1995}) and all of the shear flow models discussed herein \citep{baggett1997low,goulart2012global} are known to be lossless, \rev{which is a special case of passivity. We can therefore analyze the dynamics in terms of the partition of the dynamics into feedback interconnection between its constitutive linear and nonlinear parts, as shown in Figure \ref{fig:feedback_forcing}. In particular, passivity theory allows us to connect the behavior of the nonlinear and linear parts of the system to overall stability within a local region.} For the system described in equation \eqref{eq:nonlinear_af} and Figure \ref{fig:feedback_forcing}, this lossless property can be expressed as:
\begin{align}
    \boldsymbol{a}^T\boldsymbol{f}=0,
    \label{eq:energy_IO}
\end{align}
i.e., $\boldsymbol{a}^T\boldsymbol{J}(\boldsymbol{a}) \boldsymbol{a}=0$, which implies that  $\boldsymbol{J}(\boldsymbol{a})$ is a skew-symmetric matrix. A skew-symmetric matrix $\boldsymbol{J}(\boldsymbol{a})$ of odd dimension is known to have a zero eigenvalue and a corresponding non-trivial nullspace; see e.g., Theorem 5.4.1 in \citet{eves1980elementary}. The non-trivial element in the left null space of $\boldsymbol{J}(\boldsymbol{a})$  is the orthogonal complement of the nonlinear term $\boldsymbol{f}$; i.e. $\boldsymbol{n}$ such that:
\begin{align}
    \boldsymbol{n}^T\boldsymbol{f}=\boldsymbol{n}^T\boldsymbol{J}(\boldsymbol{a})\boldsymbol{a}=0.
    \label{eq:orthogonal}
\end{align}

\rev{The energy-conserving property in equation \eqref{eq:energy_IO} and the orthogonal complement in equation \eqref{eq:orthogonal} are associated with two constants of motion $E:=\frac{1}{2}\boldsymbol{a}^T\boldsymbol{a}$ and $C:=\boldsymbol{n}^T\boldsymbol{a}$ for the dynamical system associated with the nonlinearity: $\frac{d\boldsymbol{a}}{dt}=\boldsymbol{f}$. Such constants of motion are commonly exploited in stability analysis of passive systems, e.g. this notion is employed in the energy-Casimir method that has been widely employed in nonlinear stability analysis of ideal fluids; see e.g., \citet{holm1985nonlinear}; \citet[Section 7]{salmon1988hamiltonian}; \citet[Section VI]{morrison1998hamiltonian}; \citet{mu2001arnol}. The feedback interconnection decomposition of the linear and nonlinear dynamics (i.e., a Lur\'e system) allows us to incorporate constraints associated with these constants of motion in the analysis of full nonlinear dynamical system $\frac{d\boldsymbol{a}}{dt}=\boldsymbol{L}\boldsymbol{a}+\boldsymbol{f}$.}

\rev{We next rewrite the constraints described by equation \eqref{eq:orthogonal} as the following LMI:}
\begin{align}
    \boldsymbol{a}^T\boldsymbol{M}_i\boldsymbol{f}=0,\;i=1,2,...,n,
    \label{eq:input_output_af}\\
    \boldsymbol{f}^T\boldsymbol{T}_j\boldsymbol{f}=0,\;j=1,2,...,n,
    \label{eq:input_output_ff}
\end{align}
where $\boldsymbol{M}_i:=\boldsymbol{e}_i\boldsymbol{n}^T$, $\boldsymbol{T}_j:=\boldsymbol{e}_j\boldsymbol{n}^T+\boldsymbol{n}\boldsymbol{e}_j^T$ and $\boldsymbol{e}_i$ denotes the standard basis vector, i.e. a column vector with the $i^{\text{th}}$ element equal to one, and all other elements equal to zero. We can rewrite equation \eqref{eq:energy_IO} in the form of equation \eqref{eq:input_output_af} by defining $\boldsymbol{M}_0:=\boldsymbol{I}$, which leads to $\boldsymbol{a}^T\boldsymbol{M}_0\boldsymbol{f}=0$.

\begin{figure}
    \centering
    \includegraphics[width=3in]{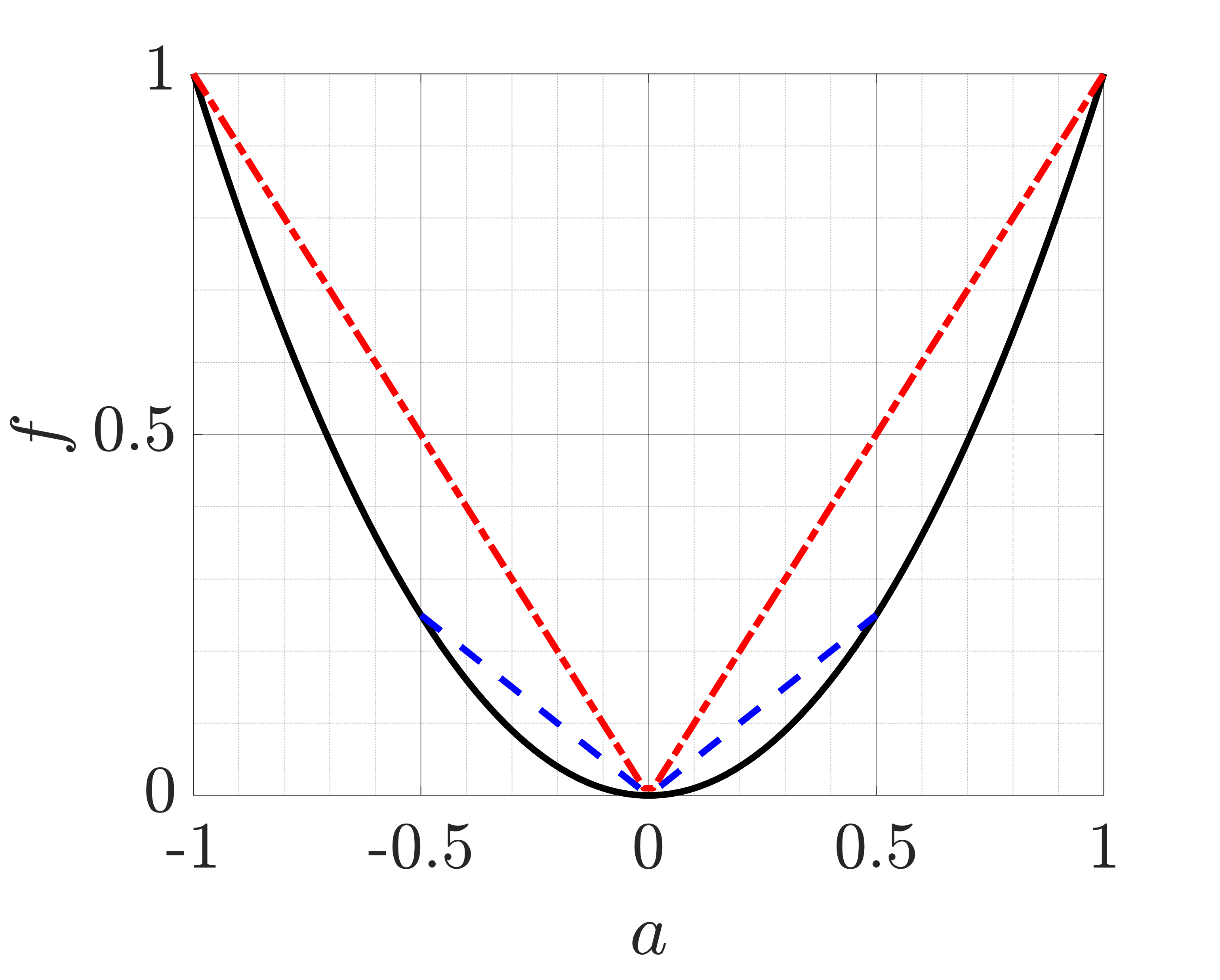}
    \caption{Illustration of local sector bounds for a quadratic nonlinear function $f=a^2$ (\parbox{0.115in}{\color{black}$\mline\mline$ }) which is bounded by a sector region  $f^2\leq 0.5^2a^2$ (\parbox{0.115in}{\color{blue}$\dashed$} ) when $a^2\leq 0.5^2$ and bounded by another sector region $f^2\leq a^2$ (\parbox{0.2in}{\color{red} $\dashdot$}  ) when $a^2\leq 1$.}
    \label{fig:local_bound}
\end{figure}

We next provide two sets of local bounds on the nonlinearity\rev{ that form the relaxed sector bounds that enable us to study the local stability  associated with a finite  amplitude perturbation, which is of interest in this work.  Figure \ref{fig:local_bound} illustrates the concept of local sector bounds for a quadratic nonlinear function $f=a^2$ that is bounded by a sector region $f^2\leq 0.5^2a^2$ when $a^2\leq 0.5^2$ and bounded by another sector region $f^2\leq a^2$ when $a^2\leq 1$.} The first\rev{set of local bounds}, provided in  Lemma \ref{thm:bound}(a), is in terms of a decomposition of the nonlinear term $\boldsymbol{f}$ into components $f_m:=\boldsymbol{e}_m^T\boldsymbol{f}$, which enables additional degrees of freedom in characterizing the system properties.  Lemma \ref{thm:bound}(b) instead provides an upper bound on the norm of $\boldsymbol{f}$. Both bounds are provided in terms of quadratic forms that are valid in a local region  $\|\boldsymbol{a}\|_2\leq\delta$, where $\|\boldsymbol{a}\|_2:=\sqrt{\sum_{i=1}^n a_i^2}=\sqrt{\boldsymbol{a}^T\boldsymbol{a}}$ denotes the $l_2$ norm of the state vector $\boldsymbol{a}$.  The associated symmetric matrices are independent of the state variable. The bound that is provided in Lemma \ref{thm:bound}(a) is similar to equation (16) of \citet{kalur2020stability} and equation (15) of \citet{kalur2020nonlinear}, but is shown to be tighter than that proposed in either of these works (see Remark \ref{remark:bound_kalur}).

\begin{lemma}
(a) Given a vector $\boldsymbol{f}\in\mathbb{R}^n$ that can be decomposed into  $f_m:=\boldsymbol{e}_m^T\boldsymbol{f}$ associated with a quadratic form $f_m=\boldsymbol{a}^T\boldsymbol{R}_m\boldsymbol{a}$ with a symmetric matrix $\boldsymbol{R}_m\in\mathbb{R}^{n\times n}$. In a local region  $\|\boldsymbol{a}\|^2_2\leq \delta^2$, each $f_m^2$ is bounded as:
\begin{align}
    f_m^2\leq \delta^2 \boldsymbol{a}^T\boldsymbol{R}_m\boldsymbol{R}_m\boldsymbol{a},\;m=1,2,...,n. 
    \label{eq:bound_fm}
\end{align}
(b) Given $\boldsymbol{f}=\boldsymbol{J}(\boldsymbol{a})\boldsymbol{a}$ with $\boldsymbol{J}(\boldsymbol{a})\in\mathbb{R}^{n\times n}$ and a local region $\|\boldsymbol{a}\|^2_2\leq \delta^2$,  $\|\boldsymbol{f}\|_2^2$ is bounded as:
\begin{align}
     \|\boldsymbol{f}\|_2^2\leq \delta^2 \boldsymbol{a}^T\boldsymbol{J}_F\boldsymbol{a},
     \label{eq:input_output_ff_bound}
 \end{align}
 where $\boldsymbol{J}_F\in\mathbb{R}^{n\times n}$ is a symmetric matrix such that $\boldsymbol{a}^T\boldsymbol{J}_F\boldsymbol{a}=\|\boldsymbol{J}(\boldsymbol{a})\|_F^2$ and $\|\boldsymbol{J}(\boldsymbol{a})\|_F:=\sqrt{\sum_{i=1}^n\sum_{j=1}^n|[\boldsymbol{J}(\boldsymbol{a})]_{i,j}|^2}$ denotes the Frobenius norm.
 \label{thm:bound}
\end{lemma}

\begin{myproof}

Part (a): In a local region $\|\boldsymbol{a}\|_2^2\leq \delta^2$, we have:
\begin{subequations}    \label{eq:bound_fm_cos}
\begin{align}
    f_m^2=&(\boldsymbol{a}^T\boldsymbol{R}_m\boldsymbol{a})(\boldsymbol{a}^T\boldsymbol{R}_m\boldsymbol{a})\\
    =&\|\boldsymbol{a}\|_2^2\;\|\boldsymbol{R}_m\boldsymbol{a}\|_2^2\;\frac{\boldsymbol{a}^T\boldsymbol{R}_m\boldsymbol{a}}{\|\boldsymbol{a}\|_2\;\|\boldsymbol{R}_m\boldsymbol{a}\|_2}\;\frac{\boldsymbol{a}^T\boldsymbol{R}_m\boldsymbol{a}}{\|\boldsymbol{a}\|_2\;\|\boldsymbol{R}_m\boldsymbol{a}\|_2}\\
    =&\|\boldsymbol{a}\|_2^2\;\|\boldsymbol{R}_m\boldsymbol{a}\|_2^2\;\text{cos}^2\theta_m\\
    \leq& \|\boldsymbol{a}\|_2^2\;\|\boldsymbol{R}_m\boldsymbol{a}\|_2^2\label{eq:bound_fm_cos_ineq}\\
    \leq&\delta^2\boldsymbol{a}^T\boldsymbol{R}_m\boldsymbol{R}_m\boldsymbol{a},\;m=1,2,...,n.
\end{align}
\end{subequations}
Here we used $\frac{\boldsymbol{a}^T\boldsymbol{R}_m\boldsymbol{a}}{\|\boldsymbol{a}\|_2\;\|\boldsymbol{R}_m\boldsymbol{a}\|_2}=:\text{cos}\theta_m$ and $\text{cos}^2\theta_m\leq 1$ with $\theta_m$ representing the angle between vectors $\boldsymbol{a}$ and $\boldsymbol{R}_m\boldsymbol{a}$.  The last step uses the bound on the local region  $\|\boldsymbol{a}\|_2^2\leq \delta^2$ to attain the upper bound on $f_m^2$ in equation (\ref{eq:bound_fm}). 

\noindent Part (b): Using the definition of $\boldsymbol{f}$,
\begin{subequations}
\begin{align}
   \|\boldsymbol{f}\|_2^2=&\|\boldsymbol{J}(\boldsymbol{a})\boldsymbol{a}\|_2^2\\
 \leq&\|\boldsymbol{a}\|^2_2 \;\|\boldsymbol{J}(\boldsymbol{a})\|^2_{2,2} \label{eq:induced_norm}\\
 \leq &\|\boldsymbol{a}\|^2_2\; \|\boldsymbol{J}(\boldsymbol{a})\|^2_F \label{eq:frobenius}\\
 \leq &\delta^2 \boldsymbol{a}^T\boldsymbol{J}_F\boldsymbol{a},\label{eq:J_F}
 \end{align}
 \end{subequations}
 where $\|\boldsymbol{J}(\boldsymbol{a})\|_{2,2}:=\underset{\boldsymbol{a}\neq \boldsymbol{0}}{\text{max}}\frac{\|\boldsymbol{J}(\boldsymbol{a})\boldsymbol{a}\|_2}{\|\boldsymbol{a}\|_2}$ represents the matrix norm induced by the $l_2$ vector norm and the inequality in equation (\ref{eq:induced_norm}) is directly obtained using the definition of the induced norm. The inequality in equation (\ref{eq:frobenius}) invokes the matrix norm property $\|\boldsymbol{J}(\boldsymbol{a})\|_{2,2}\leq \|\boldsymbol{J}(\boldsymbol{a})\|_F$; see, e.g., Problem 5.6.P23 in Ref. \citep{horn2012matrix}. As each element of $\boldsymbol{J}(\boldsymbol{a})$ is a linear function of $\boldsymbol{a}$, the square of the Frobenius norm $\|\boldsymbol{J}(\boldsymbol{a})\|_F^2$ can be written as a quadratic form $\|\boldsymbol{J}(\boldsymbol{a})\|_F^2=\boldsymbol{a}^T\boldsymbol{J}_F\boldsymbol{a}$ where $\boldsymbol{J}_F$ is independent of $\boldsymbol{a}$. Rewriting the expression in this manner and imposing the bound on the local region  $\|\boldsymbol{a}\|_2^2\leq \delta^2$ lead to the upper bound in equation (\ref{eq:input_output_ff_bound}).
\end{myproof}

\begin{remark}
We can obtain the bound in equation (16) of \citet{kalur2020stability} and equation (15) of \citet{kalur2020nonlinear} from the result (\ref{eq:bound_fm}) in Lemma \ref{thm:bound}(a) in the following manner. Starting from (\ref{eq:bound_fm}) in Lemma \ref{thm:bound}(a), we further apply the inequalities,
\begin{subequations}
\begin{align}
    f_m^2&\leq\delta^2\boldsymbol{a}^T\boldsymbol{R}_m\boldsymbol{R}_m\boldsymbol{a}\nonumber\\
    &\leq \delta^2\boldsymbol{a}^T\rho(\boldsymbol{R}_m\boldsymbol{R}_m)\boldsymbol{a}    \label{eq:remark_bound_kalur_a}\\
    &\leq \delta^2\rho(\boldsymbol{R}_m)^2 \boldsymbol{a}^T \boldsymbol{a}
    \label{eq:remark_bound_kalur_b}
\end{align}
\end{subequations}
with $\rho(\cdot)$ representing the spectral radius and the resulting (\ref{eq:remark_bound_kalur_b}) is the upper bound in \cite{kalur2020nonlinear,kalur2020stability}. The inequality in equation (\ref{eq:remark_bound_kalur_a}) results from the Rayleigh quotient theorem (See e.g., Theorem 4.2.2 in Ref. \citep{horn2012matrix}) and the definition of the spectral radius, and this inequality achieves equality if and only if all eigenvalues of $\boldsymbol{R}_m\boldsymbol{R}_m$ are equal to $\rho(\boldsymbol{R}_m\boldsymbol{R}_m)$.
The inequality in equation (\ref{eq:remark_bound_kalur_b}) results from the Gelfand formula (Corollary 5.6.14 of Ref. \citep{horn2012matrix}) and submultiplicativity of the matrix norm (Chapter 5.6 of Ref. \citep{horn2012matrix}). Whenever the condition to achieve equality in equation (\ref{eq:remark_bound_kalur_a}) or (\ref{eq:remark_bound_kalur_b}) are violated, our bounds in equation \eqref{eq:bound_fm} of Lemma \ref{thm:bound}(a) is tighter than \citep{kalur2020nonlinear,kalur2020stability}.
\label{remark:bound_kalur}
\end{remark}

\subsection{LMI based permissible perturbation amplitude computations}
\label{subsec:permissible_LMI}

We now present the main theoretical result of the paper, in which we pose the problem of determining a permissible perturbation amplitude $\delta_{\text{p}}$ through testing the feasibility of an LMI constrained optimization problem.  The result is presented in the following theorem, which first provides the neighborhood over which perturbations decay. A maximization over said regions is used to determine an estimate of the permissible perturbation amplitude.

\begin{thm}
Given the nonlinear dynamical system described in equation (\ref{eq:nonlinear_af}) satisfying the conditions in (\ref{eq:energy_IO}) and Lemma \ref{thm:bound} along with $\|\boldsymbol{a}\|_2\leq
\delta$, $\delta>0$.  

 If there exists a symmetric matrix $\boldsymbol{P}\in \mathbb{R}^{n\times n}$ satisfying
\begin{subequations} \label{eq:LMI}
\begin{align}
    \boldsymbol{P}-\epsilon \boldsymbol{I}\succeq & 0,\label{eq:P_positive_definite} \\ \epsilon>& 0,\;\; \label{eq:LMIb}\\
    \boldsymbol{G}\preceq & 0,\;\;\label{eq:LMIc}\\
    s_m\geq & 0,
    \;\;m=0,1,...,n,
\end{align}
\end{subequations}
where $(\cdot)\succeq 0$ and $(\cdot)\preceq 0$, respectively, represent positive and negative semi-definiteness of the associated operator and $\boldsymbol{G}$ is defined as:
\begin{widetext}
\begin{align*}
    \boldsymbol{G}:=\begin{bmatrix}
    \boldsymbol{L}^T\boldsymbol{P}+\boldsymbol{P}\boldsymbol{L}+\epsilon \boldsymbol{I}+s_0\delta^2\boldsymbol{J}_F+\displaystyle\sum_{m=1}^n s_m\delta^2 \boldsymbol{R}_{m}\boldsymbol{R}_m  & \boldsymbol{P}+\displaystyle\sum_{i=0}^{n}\lambda_i\boldsymbol{M}_i \\
    \boldsymbol{P}+\displaystyle\sum_{i=0}^{n}\lambda_i\boldsymbol{M}_i^T& -s_0\boldsymbol{I}-\displaystyle\sum_{m=1}^n s_m \boldsymbol{e}_m\boldsymbol{e}_m^T+\displaystyle\sum_{j=1}^{n} \kappa_j \boldsymbol{T}_j
    \end{bmatrix},
    \label{eq:G_definition}
\end{align*}
\end{widetext}
then $\|\boldsymbol{a}(t=0)\|_2\leq \delta_{f}\Rightarrow \underset{t\rightarrow \infty}{\text{lim}}\boldsymbol{a}(t)=0$, where $\delta_{f}:=\delta\sqrt{\frac{\mu_{\text{min}}(\boldsymbol{P})}{\mu_{\text{max}}(\boldsymbol{P})}}$ with $\mu_{\text{min}}(\cdot)$ and $\mu_{\text{max}}(\cdot)$ denoting the minimal and maximal eigenvalues.

\label{thm:lmi}
\end{thm}

\begin{myproof}

When inequalities in equation (\ref{eq:LMI}) are feasible, $\boldsymbol{P}$ can be used to define  $V:=\boldsymbol{a}^T\boldsymbol{P}\boldsymbol{a}\geq \epsilon\boldsymbol{a}^T\boldsymbol{a}>0,\;\;\forall\boldsymbol{a}\neq \boldsymbol{0}$. We now demonstrate that $V$ is a Lyapunov function for the system described in equation (\ref{eq:nonlinear_af}) in the region $\|\boldsymbol{a}\|_2\leq \delta$. According to Lemma \ref{thm:bound}, we have $\delta^2\boldsymbol{a}^T\boldsymbol{R}_m\boldsymbol{R}_m\boldsymbol{a}-f_m^2\geq 0,\;m=1,2,...,n$ and $ \delta^2\boldsymbol{a}^T\boldsymbol{J}_F\boldsymbol{a}-\boldsymbol{f}^T\boldsymbol{f}\geq 0$, and, therefore, we can further obtain $\;\forall \boldsymbol{a}\neq \boldsymbol{0}$ in the region $\|\boldsymbol{a}\|_2\leq \delta$:
\begin{subequations}    \label{eq:dV_ROA}
\begin{align}
    \frac{dV}{dt}\leq&\frac{dV}{dt}+s_0(\delta^2\boldsymbol{a}^T\boldsymbol{J}_F\boldsymbol{a}-\boldsymbol{f}^T\boldsymbol{f})\nonumber\\
    &\;\;\;\;\;\,+\sum_{m=1}^ns_m(\delta^2\boldsymbol{a}^T\boldsymbol{R}_m\boldsymbol{R}_m\boldsymbol{a}-f_m^2)  \\
    =&\begin{bmatrix}\boldsymbol{a}\\
    \boldsymbol{f}\end{bmatrix}^T
    \boldsymbol{G}
    \begin{bmatrix}\boldsymbol{a}\\
    \boldsymbol{f}\end{bmatrix}-\epsilon\boldsymbol{a}^T\boldsymbol{a}\\
    \leq&-\epsilon\boldsymbol{a}^T\boldsymbol{a}<0.
\end{align}
\end{subequations}  
Thus, by Lyapunov's stability theorem (see e.g., Theorem 4.1 in Ref. \citep{khalil2002nonlinear}) the origin $\boldsymbol{a}=\boldsymbol{0}$ is asymptotically stable. In addition, a region of attraction of the origin is given by $D_c:=\{\boldsymbol{a}| V=\boldsymbol{a}^T\boldsymbol{P}\boldsymbol{a}\leq c\}\subseteq B_{\delta}:=\{\boldsymbol{a}|\;\|\boldsymbol{a}\|_2\leq \delta\}$, where we select $c>0$ to define the maximum level set of $V$ contained in $ B_\delta$.  

Given $\delta_{f}:=\delta\sqrt{\frac{\mu_{\text{min}}(\boldsymbol{P})}{\mu_{\text{max}}(\boldsymbol{P})}}$, the Rayleigh quotient theorem implies that $\mu_{\text{min}}(\boldsymbol{P})\boldsymbol{a}^T\boldsymbol{a}\leq \boldsymbol{a}^T\boldsymbol{P}\boldsymbol{a}\leq \mu_{\text{max}}(\boldsymbol{P})\boldsymbol{a}^T\boldsymbol{a}$ (see e.g., Theorem 4.2.2 in Ref. \citep{horn2012matrix}). Therefore $B_{\delta_{f}}:=\{\boldsymbol{a}|\;\|\boldsymbol{a}\|_2\leq \delta_{f}\}\subseteq D_c$ and as such, $\|\boldsymbol{a}(t=0)\|_2\leq \delta_{f}\Rightarrow \underset{t\rightarrow \infty}{\text{lim}}\boldsymbol{a}(t)=0$ as stated in the theorem.
\end{myproof}

\begin{figure}
    \centering
    \includegraphics[width=2.2in]{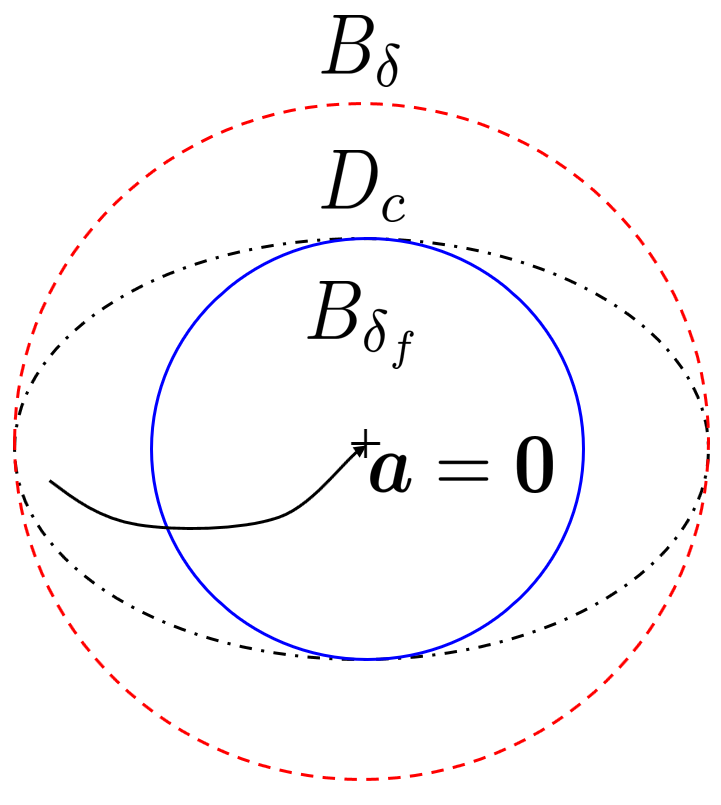}
    \caption{A two-dimensional illustration of the set relationship, $B_{\delta_{f}}\subseteq D_c\subseteq B_{\delta}$, employed in the proof of Theorem \ref{thm:lmi}. $B_{\delta}$ (\parbox{0.15in}{\color{red} $\dashed$}): a local region as a condition to bound the nonlinearity in Lemma \ref{thm:bound}; $D_c$ (\parbox{0.165in}{ \color{black}$\dashdot$} ): region of attraction of the origin $\boldsymbol{a}=\boldsymbol{0}$ illustrated with a trajectory (\parbox{0.115in}{\color{black}$\rightarrow$} ); $B_{\delta_f}$ (\parbox{0.115in}{\color{blue}$\mline\mline$ }): a circular region contained inside $D_c$.}
    \label{fig:ROA_level_proof}
\end{figure}

Figure \ref{fig:ROA_level_proof} provides a two-dimensional illustration of the set relationship $B_{\delta_{f}}\subseteq D_c\subseteq B_{\delta}$ employed in the proof of Theorem \ref{thm:lmi}. Theorem \ref{thm:lmi} is essentially trying to find a local Lyapunov function $V$ contained within the $B_\delta$ in which the nonlinearity is bounded. The permissible perturbation amplitude is defined as the radius of the largest multidimensional sphere $B_{\delta_f}$ contained within the associated region of attraction $D_c$. The permissible perturbation amplitude can therefore be computed as the solution of the optimization problem:
\begin{align}
    &\delta_{\text{p}}:=\underset{\delta}{\text{max}}\;\;\delta_{f}\label{eq:optimze_delta}\\
    &\text{subject to}\;(\ref{eq:LMI}).\nonumber
\end{align}

\begin{remark} 
As seen in the depiction of the region of attraction $D_c$ in Figure \ref{fig:ROA_level_proof}, the permissible perturbation amplitude $\delta_{\text{p}}$ given in equation \eqref{eq:optimze_delta} is conservative in the sense that certain directions can sustain perturbations larger than $\delta_f$. The form of  $\boldsymbol{P}$ can be further explored to gain further information regarding the directions that are the most sensitive to perturbations. The notion of perturbation structures that are most likely to lead to transition has been explored in other works, see e.g. \citep{Kerswell2014,Kerswell2018,Pringle2010,duguet2010towards,Pringle2012,rabin2012triggering,Duguet2013,Kim2008,Joglekar2015}. Here we focus on providing formal guarantees on the magnitude of the permissible perturbation amplitude, which has been previously studied using extensive simulations in \citep{baggett1997low,Joglekar2015}. 
\label{remark:ROA}
\end{remark}

The formulation and analysis described above provide a means to evaluate both classical energy and linear stability by restricting the form of $\boldsymbol{G}$ in equation \eqref{eq:LMIc}. In particular, neither classical energy nor linear stability analysis includes the local bounds on the nonlinear terms defined in Lemma \ref{thm:bound}, which take the form of the non-negative multipliers $s_m,\;m=0,1,...,n$ in equation \eqref{eq:LMIc}. Our formulation further imposes equality constraints in describing the orthogonal complement of the nonlinear term in equation (\ref{eq:orthogonal}), which take the form of equations \eqref{eq:input_output_af} and \eqref{eq:input_output_ff} that are associated with the multipliers $\lambda_i$, $i=1,2,...,n$ and $\kappa_j$, $j=1,2,...,n$. Classical energy methods do include the constraint associated with energy conservation in equation (\ref{eq:energy_IO}), described through the term associated with the multiplier $\lambda_0$, which leads to the following simplified form of equation \eqref{eq:LMIc} for energy stability analysis
\begin{align}
     \boldsymbol{G}_{E}:=\begin{bmatrix}
     \boldsymbol{L}^T\boldsymbol{P}+\boldsymbol{P}\boldsymbol{L}+\epsilon \boldsymbol{I} & \boldsymbol{P}+\lambda_0\boldsymbol{I} \\
     \boldsymbol{P}+\lambda_0\boldsymbol{I} & \mathbb{O}
     \end{bmatrix}\preceq 0,
     \label{eq:energy_stability}
\end{align}
where $\mathbb{O}\in \mathbb{R}^{n\times n}$ is the zero matrix. By the generalized Schur's complement (See e.g., Theorem 4.3 in Ref. \citep{gallier2010schur}), the expression in (\ref{eq:energy_stability}) is true if and only if both $\boldsymbol{P}+\lambda_0\boldsymbol{I}=\mathbb{O}$ and $\boldsymbol{L}^T\boldsymbol{P}+\boldsymbol{P}\boldsymbol{L}+\epsilon\boldsymbol{I}\preceq 0$. Combining these relations with the condition $\boldsymbol{P}-\epsilon\boldsymbol{I}\succeq 0$ in equation \eqref{eq:P_positive_definite} leads to:
\begin{align}
    \boldsymbol{L}^T+\boldsymbol{L}\prec0,
    \label{eq:energy_LL}
\end{align}
where $\prec$ represents negative definiteness. Equation (\ref{eq:energy_LL}) is equivalent to the condition for energy stability derived in Ref. \citep{goulart2012global} with a Lyapunov function of $V=\frac{1}{2}\boldsymbol{a}^T\boldsymbol{a}$. Setting $s_m=0$, $m=0,1,...,n$ in the LMI formulation removes the local region $\|\boldsymbol{a}\|_2\leq \delta$ restriction in Lemma \ref{thm:bound}. This means that the Lyapunov function, $V=\frac{1}{2}\boldsymbol{a}^T\boldsymbol{a}$, is radially unbounded and, therefore, the origin (equilibrium point) of the system in (\ref{eq:nonlinear_af}) with the nonlinearity satisfying  (\ref{eq:energy_IO}) is globally asymptotically stable ($\delta_{\text{p}}=\infty$), see e.g., Theorem 4.2 in Ref. \citep{khalil2002nonlinear}.  Equation (\ref{eq:energy_stability}) was used to perform global stability analysis for the WKH model by~\citet{kalur2020stability,kalur2020nonlinear}.

Linear stability analysis corresponds to a further restriction on $\mathbf{G}_E$ in  (\ref{eq:energy_stability}), where the off-diagonal elements are replaced by zero matrices (i.e., the nonlinear term $\boldsymbol{f}$ in the model dynamics \eqref{eq:nonlinear_af} and its energy-conserving constraint in equation \eqref{eq:energy_IO} are removed). In this case, the form of $\boldsymbol{G}$ in equation \eqref{eq:LMIc} is
\begin{align}
    \boldsymbol{G}_{L}:=
    \boldsymbol{L}^T\boldsymbol{P}+\boldsymbol{P}\boldsymbol{L}+\epsilon \boldsymbol{I}\preceq 0,
\end{align}
and Theorem \ref{thm:lmi} is equivalent to Lyapunov based linear stability analysis; see e.g., Theorems 4.6 and 4.7 of Ref. \citep{khalil2002nonlinear}.

In the next section, we will employ the proposed framework to compute the permissible perturbation amplitude as a function of the Reynolds number and compare the resulting functions to those obtained  from simulations of a range of shear flow models that have been widely used as benchmark problems in the study of transition and low Reynolds number shear flows.

\section{Numerical Results}
\label{sec:application_shear}

In this section, we first focus on comparisons of the perturbation as a function of Reynolds numbers for six of the low (2-4) dimensional models studied through extensive numerical simulations in~\citep{baggett1997low} (subsection \ref{subsec:application_shear}). We then perform a more detailed analysis of the  nine-dimensional shear flow model \citep{Moehlis2004}  including comparisons  of the computational requirements and solutions obtained through SOS based analysis (subsection \ref{subsec:com_SOS}). 

For all of the results herein, we implement the LMIs in equation (\ref{eq:LMI}) of Theorem \ref{thm:lmi} in YALMIP \citep{lofberg2004yalmip} version R20190425 in MATLAB R2018b and solve the optimization problem in equation \eqref{eq:optimze_delta} using the Semi-definite Programming (SDP) solver SeDuMi \citep{sturm1999using} version 1.3.\rev{ We solve the LMI problem and the SOS problem discussed in subsection \ref{subsec:com_SOS} by converting it to an SDP, which can be solved using off the shelf optimization methods. The feasible region of SDP is the cone of positive semi-definite (PSD) matrices; i.e., a region that is closed under linear combinations of PSD matrices with non-negative coefficients \citep[Chapter 4.6.2]{vandenberghe1996semidefinite,boyd2004convex}. The dimension of this PSD cone involved in the optimization problem provides a measurement of computational resources required for the solver; e.g., employed in \citep{zheng2018fast}. We therefore report this as a benchmark of computational efficiency in subsection \ref{subsec:com_SOS}.} We note that for comparison purposes, all computations are performed on the same computer with a 3.4 GHz Intel Core i7-3770 Central Processing Unit (CPU) and 16GB Random Access Memory (RAM). We set the value of $\epsilon$ in equation \eqref{eq:LMIb} to $0.01$; however, the specific value of $\epsilon$ does not alter the results due to  the homogeneity of the inequalities in equation (\ref{eq:LMI}). For each model, we solve the optimization problems in (\ref{eq:optimze_delta}) over $40$ logarithmically spaced Reynolds numbers $Re\in[1,2000]$. This optimization problem is solved through testing its feasibility over $400$ logarithmically spaced $\delta\in[10^{-6},1]$ and then selecting the largest $\delta_f$ that provides a feasible solution (i.e, satisfies the conditions in equation \eqref{eq:LMI}) as $\delta_{\text{p}}$, i.e. we find the solution to equation \eqref{eq:optimze_delta}. \rev{The range of $\delta\in[10^{-6},1]$ is selected to ensure that we cover the range of permissible perturbation amplitude in the transitional regime (e.g., $Re\geq100$) observed from simulation results for these shear flow models considered here \citep{baggett1997low,Joglekar2015}.} We use this approach of solving for particular values $\delta$ at each $Re$ as this renders the set of LMI constraints convex, which is more numerically tractable than the alternative bilinear optimization problem. Finally, we use the least-squares fit to find the exponents $A$ and $\sigma$ in $\delta_{\text{p}}(Re)=10^A Re^{\sigma}$, which is the same functional form used in \citep{baggett1997low,Joglekar2015}. We select the same functional form in order to directly compare the scaling exponents $\sigma$ obtained from extensive simulations with randomly chosen initial conditions computed by \citet{baggett1997low} and \citet{Joglekar2015}.

For all of the low dimensional shear flow models in section \ref{subsec:application_shear}, all of the eigenvalues of $\boldsymbol{L}$, corresponding to the linearization around the laminar state (origin), have negative real parts for all Reynolds numbers. In other words, the laminar state is linearly stable; i.e., $Re_L=\infty$. However, as is common in linear systems such as these where the linear operator (matrix) is non-normal, i.e., ($\boldsymbol{L}\boldsymbol{L}^T\neq \boldsymbol{L}^T\boldsymbol{L}$), the energy stability requirement $\boldsymbol{L}+\boldsymbol{L}^T\prec 0$ in equation (\ref{eq:energy_LL}) is violated at certain Reynolds number $Re_E< Re_L$ for all of the models considered here. The nonlinear terms $\boldsymbol{f}$ for all of these models satisfy the energy-conserving property described by equation (\ref{eq:energy_IO}).

\subsection{Application to shear flow models}
\label{subsec:application_shear}

We now introduce the set of low dimensional shear flow models and the procedure that is used in applying  Theorem \ref{thm:lmi} and equation (\ref{eq:optimze_delta}).  We employ the notation and naming convention (abbreviations based on authors' last names) used in \citet{baggett1997low} for consistency as we compare our results to the simulation results in that work. In particular, we introduce and explain the application of Theorem \ref{thm:lmi} to the two-dimensional TTRD (Trefethen, Trefethen, Reddy, and Driscoll) model proposed in \citet{trefethen1993hydrodynamic} and the two variations, TTRD' and TTRD'', introduced in \citep{baggett1997low}. We then provide the details of the three-dimensional BDT (Baggett, Driscoll, and Trefethen) model introduced in \citet{baggett1995mostly} and explain the pertinent values for the application of Theorem \ref{thm:lmi}. Finally, we describe the four-dimensional W (Waleffe) proposed by \citet{waleffe1995transition} and its three-dimensional variation W' introduced in \citep{baggett1997low}.  For all of the models described in this subsection, we use the same coefficients as  \citep{baggett1997low} for a direct comparison with their results.

The three variations of the TTRD model are two-dimensional models of the form,
\begin{align}
    \frac{d}{dt}\begin{bmatrix} u\\ v\end{bmatrix}&=\begin{bmatrix}
    -Re^{-1} & 1 \\
    0 & -Re^{-1}
    \end{bmatrix}\begin{bmatrix} u\\ v\end{bmatrix}+\boldsymbol{f_{(\cdot)}},
\end{align}
where the function $\boldsymbol{f_{(\cdot)}}$ describing the nonlinearity for the respective TTRD, TTRD', and TTRD'' variations of the model are given by:
{\renewcommand{\theequation}{TTRD}
\begin{align}
    \boldsymbol{f}_{\text{\tiny TTRD}}:=
   \norm{ \begin{bmatrix} u\\ v\end{bmatrix}}_2
    \begin{bmatrix}
    0 & -1 \\
    1 & 0
    \end{bmatrix}
    \begin{bmatrix} u\\ v\end{bmatrix},
    \label{eq:TTRD}
\end{align}
}
\addtocounter{equation}{-1}
{\renewcommand{\theequation}{TTRD'}
\begin{align}
    \boldsymbol{f}_{\text{\tiny TTRD'}}:=&\begin{bmatrix}
    0 & -u \\
    u & 0
    \end{bmatrix}
    \begin{bmatrix} u\\ v\end{bmatrix},
    \label{eq:TTRD'}
\end{align}
}
{\renewcommand{\theequation}{TTRD''}
\begin{align}
    \boldsymbol{f}_{\text{\tiny TTRD''}}:=&\begin{bmatrix}
    0 & -v \\
    v & 0
    \end{bmatrix}
    \begin{bmatrix} u\\ v\end{bmatrix}.
    \label{eq:TTRD''}
\end{align}
}\addtocounter{equation}{-2}

In order to apply the theory in Section \ref{sec:nonlinear_IO} to the TTRD model, we need to deal with the fact that the nonlinear term (\ref{eq:TTRD}) involves the $l_2$ norm of the state variable, and, therefore, Lemma \ref{thm:bound} is not directly applicable. The following Proposition \ref{pro:bound_2norm} provides corresponding upper bounds on $\boldsymbol{f}_{\text{\tiny TTRD}}$ in a form similar to those in Lemma \ref{thm:bound}. 
\begin{pro}
Given a vector $\boldsymbol{f}\in\mathbb{R}^n$ that can be decomposed into  $f_m:=\boldsymbol{e}_m^T\boldsymbol{f}$ with expression $f_m=\|\boldsymbol{a}\|_2\boldsymbol{r}_m^T\boldsymbol{a},\;m=1,2,...,n$ with $\boldsymbol{r}_m\in\mathbb{R}^n$. 

(a) In a local region  $\|\boldsymbol{a}\|^2_2\leq \delta^2$, each $f_m^2$ is bounded as
\begin{align}
    f_m^2=\|\boldsymbol{a}\|^2_2\boldsymbol{a}^T\boldsymbol{r}_m\boldsymbol{r}_m^T\boldsymbol{a}\leq \delta^2\boldsymbol{a}^T\boldsymbol{r}_m\boldsymbol{r}_m^T\boldsymbol{a}.
\end{align}

(b) In a local region  $\|\boldsymbol{a}\|^2_2\leq \delta^2$, $\|\boldsymbol{f}\|_2^2$ is bounded as 
\begin{align}
    \|\boldsymbol{f}\|_2^2\leq \delta^2 \sum_{m=1}^n\boldsymbol{a}^T\boldsymbol{r}_m\boldsymbol{r}_m^T\boldsymbol{a}.
\end{align}
\label{pro:bound_2norm}
\end{pro}
Taking the bounds in Proposition \ref{pro:bound_2norm} and employing the substitution $\boldsymbol{R}_m\boldsymbol{R}_m=\boldsymbol{r}_m\boldsymbol{r}_m^T$ and $\boldsymbol{J}_F=\sum_{m=1}^n \boldsymbol{r}_m\boldsymbol{r}_m^T$ enables direct application of Theorem \ref{thm:lmi}. The nonlinearities in equations \eqref{eq:TTRD'} and \eqref{eq:TTRD''} are quadratic, so we can directly apply Theorem \ref{thm:lmi}. We also note that for these two-dimensional models, the orthogonal complement satisfying equation (\ref{eq:orthogonal}) is trivial, so we set $\boldsymbol{n}=\boldsymbol{0}$ in applying Theorem \ref{thm:lmi}.  

The results of the application of the optimization procedure described above for solving equation \eqref{eq:optimze_delta} over the given parameter ranges followed by the least-squares fit to $\delta_{\text{p}}=10^ARe^\sigma$ leads to the parameter values $A$ and $\alpha$ shown in Table \ref{tab:shear_flow_com_baggett}. The table indicates good agreement between the simulations and the theory for all three models.

Having obtained good results with the two-dimensional TTRD models, we next consider the three-dimensional BDT shear flow model, 
\begin{widetext}
{\renewcommand{\theequation}{BDT}
\begin{align}
    \frac{d}{dt}\begin{bmatrix} u\\ v\\
    w\end{bmatrix}&=\begin{bmatrix}
    -Re^{-1} & Re^{-1/2} & 0 \\
    0 & -Re^{-1} & Re^{-1/2}\\
    0 & 0 & -Re^{-1}
    \end{bmatrix}\begin{bmatrix} u\\ v\\ w\end{bmatrix}+\norm{\begin{bmatrix} u\\ v\\ w\end{bmatrix}}_2
    \begin{bmatrix}
    0 & -1 & 1 \\
    1 & 0 & 1\\
    -1 & -1 & 0
    \end{bmatrix}
    \begin{bmatrix} u\\ v \\ w\end{bmatrix}.\label{eq:BDT}
\end{align}
}
\addtocounter{equation}{-1}

\end{widetext}The form of the nonlinearity in this model is similar to that in equation \eqref{eq:TTRD}, and therefore we again use Proposition \ref{pro:bound_2norm} and the previously described substitution in order to apply Theorem \ref{thm:lmi}. Since the system is of odd dimension, there is a non-trivial orthogonal complement for the nonlinear term. In particular, we use  $\boldsymbol{n}_{\text{BDT}}^T=\begin{bmatrix}-1 & 1 & 1 \end{bmatrix}$ in the computation of the $\boldsymbol{M}_i$ and $\boldsymbol{T}_j$ in equation \eqref{eq:LMIc}.  Table \ref{tab:shear_flow_com_baggett} shows that the values of $A$ and $\sigma$ obtained through the procedure described above in solving the optimization in equation \eqref{eq:optimze_delta} and fitting the function form for $\delta_{\text{p}}(Re)$ agree well with those obtained through extensive simulations.

The final class of low dimensional models that we analyze in this subsection are the four-dimensional W model 
proposed in \citet{waleffe1995transition} and its three-dimensional variation W' provided in \citep{baggett1997low}.  
We note here that the four-dimensional W model with the coefficients provided in \citep{waleffe1995transition}  is also referred to as the WKH model, e.g. in \citep{kalur2020stability,kalur2020nonlinear} where they perform a related analysis of this particular model. These W and W' models are respectively given by 
\begin{widetext}
{\renewcommand{\theequation}{W}
\begin{align}
    \frac{d}{dt}\begin{bmatrix} u\\ v\\
    w \\ n\end{bmatrix}=\begin{bmatrix}
    - Re^{-1} & 1 & 0 & 0 \\
    0 & - Re^{-1} & 0 & 0\\
    0 & 0 &  -Re^{-1} & 0\\
    0 & 0 & 0 & - Re^{-1}
    \end{bmatrix}\begin{bmatrix} u\\ v\\ w \\ n\end{bmatrix}+
    \begin{bmatrix}
    0 & 0 & - w & -v \\
    0 & 0 &  w & 0\\
     w & - w & 0 & 0\\
    v & 0 & 0 & 0
    \end{bmatrix}
    \begin{bmatrix} u\\ v \\ w \\ n\end{bmatrix},
    \label{eq:W}
\end{align}
}
\end{widetext}
\begin{widetext}
{\renewcommand{\theequation}{W'}
\begin{align}
      \frac{d}{dt}\begin{bmatrix} u\\ v\\
    w\end{bmatrix}=\begin{bmatrix}
    -Re^{-1} & 1 & 0 \\
    0 & -Re^{-1} & 0\\
    0 & 0 & -Re^{-1}
    \end{bmatrix}\begin{bmatrix} u\\ v\\ w\end{bmatrix}+
    \begin{bmatrix}
    0 & 0 & -w \\
    0 & 0 & w\\
    w & -w & 0
    \end{bmatrix}
    \begin{bmatrix} u\\ v \\ w\end{bmatrix}.
    \label{eq:W'}
\end{align}
}
\addtocounter{equation}{-2}
\end{widetext}
Both models allow direct application of Lemma \ref{thm:bound} to bound the nonlinear terms. The analyses for these two models differ in that there exists a non-trivial $\boldsymbol{n}_{\text{W'}}^T=\begin{bmatrix}1 & 1 & 0\end{bmatrix}$ for the nonlinear term in the odd-dimensional model (\ref{eq:W'}) but not for the nonlinear term in the even-dimensional model (\ref{eq:W}). Table \ref{tab:shear_flow_com_baggett}
indicates that the theoretical results and associated optimization problem lead to scalings $\sigma$ for both the W and W' models that are consistent with those obtained through extensive numerical simulations.

\begin{table}[]
    \centering
    \begin{tabular}{c c c c}
    \hline
        Model abbreviation & A & $\sigma$ & \thead{$\sigma$ in \\ Baggett \& Trefethen (1997) \citep{baggett1997low}} \\
        \hline 
        TTRD  & -0.03 & -3.03 & -3\\
        TTRD' & -0.04 & -3.07 & -3\\
        TTRD'' & -0.35 & -1.98 & -2 \\
        BDT & 0.03 & -3.04 & -3\\
        W & -0.45 & -1.98 & -2\\
        W' & -0.38 & -1.94 & -2\\
        \hline
    \end{tabular}
    \caption{$A$ and $\sigma$ fitting to $\delta_{\text{p}}=10^ARe^{\sigma}$ with $\delta_{\text{p}}$ obtained from the current framework for each shear flow model. The obtained $\sigma$ are compared with scaling exponents $\sigma$ reported in Ref. \citep{baggett1997low}. }
    \label{tab:shear_flow_com_baggett}
\end{table}

The results in Table \ref{tab:shear_flow_com_baggett} demonstrate that the scaling exponents $\sigma$ obtained from the current framework are close to the $\sigma$ computed from extensive numerical simulations \citep{baggett1997low}. However, the current framework has the benefit of providing this estimation for the permissible perturbation amplitude without requiring any simulations or experiments. Moreover, the convergence to the origin is guaranteed for \emph{any} perturbation below the obtained permissible perturbation amplitude $\delta_{\text{p}}$, whereas numerical simulations and experiments can only test on a finite set of perturbations and, therefore, do not provide provably definitive results. Given the good agreement with simulation studies for commonly studied low-dimensional shear flow models, we next apply the theory to the more comprehensive nine-dimensional model and discuss the computational complexity of this approach versus SOS based analysis methods. 

\subsection{Application to a 9-D shear flow model and comparison with SOS}
\label{subsec:com_SOS}

In this section, we focus on the nine-dimensional shear flow model~\citep{Moehlis2004}. We first compare the permissible perturbation amplitude $\delta_\text{p}$ obtained through the method proposed in Section \ref{sec:nonlinear_IO} to the values identified using extensive simulations. We then compare our results to the rigorous bounds based on Lyapunov analysis computed through SOS programming. \rev{SOS programming \citep{parrilo2000structured,prajna2002introducing,Papachristodoulou2005} is a widely used tool to search for Lyapunov functions for stability and region of attraction based computations; see e.g., which describe applications in fluid dynamics \citep{goulart2012global,chernyshenko2014polynomial,huang2015sum,lasagna2016sum,Huang2017,fuentes2019global,ahmadi2019framework,lakshmi2020finding}. SOS provides a generalization of the LMI framework as can be used to find higher-order (beyond quadratic) polynomials as the candidate of Lyapunov functions. When the degree of the polynomials in an SOS program is fixed, it is typically solved by converting the SOS constraints to an SDP. Further details of SOS methods and SOS programming can be found in~\citep{parrilo2000structured,Papachristodoulou2005,papachristodoulou2013sostools}.  } The comparison with SOS highlights the computational efficiency of the method and explores the trade-off between the computational efficiency of our LMI based approach and the accuracy that can be obtained through SOS methods, which allow the full representation of the nonlinearity rather than the constraints on its properties detailed in Section \ref{sec:nonlinear_IO}.

The nine-dimensional model is comprised of an eight-dimensional Galerkin model \citep{waleffe1997self} describing the self-sustaining process and an additional mode that enables the full model to capture the change in the mean velocity profile as the flow transitions from the laminar to the turbulent state~\citep{Moehlis2004}. This model has been widely used as a prototype to study stability and transition in shear flows that have no linear instabilities, see e.g. \cite{Moehlis2005, lakshmi2020finding,Kim2008,goulart2012global,chernyshenko2014polynomial,Joglekar2015}.  The dynamics of the nine-mode model are obtained directly from a Galerkin projection of the NS equations \citep{Moehlis2004}. Appendix \ref{sec:appendix_galerkin} provides the details of the derivation of the model, which can be written in the form,
\begin{align}
    \frac{d \boldsymbol{a}}{dt}=-\frac{\boldsymbol{\Xi}}{Re}\boldsymbol{a}+\boldsymbol{J}(\boldsymbol{a})\boldsymbol{\bar{a}}+\boldsymbol{J}(\boldsymbol{\bar{a}})\boldsymbol{a}+\boldsymbol{J}(\boldsymbol{a})\boldsymbol{a},
    \label{eq:galerkin_fluc}
\end{align}
where $\boldsymbol{\bar{a}}$ denotes the laminar flow solution.  We  use the same model coefficients as in \citep{Joglekar2015}, which requires that we use their domain size of $L_x=1.75\pi$ and $L_z=1.2\pi$. Here we describe the role of the various terms, but for the sake of brevity, we refer to equation (\ref{eq:Galerkin_9D_coeff_nonlinear}) in Appendix A for details of each coefficient. The first term on the right-hand side (RHS) of equation (\ref{eq:galerkin_fluc}) is the viscous term, and $\boldsymbol{\Xi}$ is a symmetric positive definite matrix. The second term on the RHS of (\ref{eq:galerkin_fluc}) $\boldsymbol{J}(\boldsymbol{a})\boldsymbol{\bar{a}}$ is an analog to the mean shear term in the linearized NS equations. The resulting shear production mechanism is critical in maintaining turbulence in wall-bounded shear flows \citep{kim2000linear}. The following two terms on the RHS of equation (\ref{eq:galerkin_fluc}), $\boldsymbol{J}(\boldsymbol{\bar{a}})$ and $\boldsymbol{J}(\boldsymbol{a})$, respectively, correspond to the advection by the laminar mean flow and nonlinear advection. The nonlinear advection term is energy-conserving in analogy to the nonlinear advection term in the NS equations, i.e., $\boldsymbol{a}^T\boldsymbol{J}(\boldsymbol{a})\boldsymbol{a}=0$. When the Galerkin model is obtained through data \citep{brunton2016discovering}, this energy-conserving property can be explicitly implemented as a constraint \citep{Loiseau2018}.

In order to apply the theory of Section \ref{sec:nonlinear_IO} we first express the linear terms as
\begin{align}
    \boldsymbol{L}\boldsymbol{a}:=-\frac{\boldsymbol{\Xi}}{Re}\boldsymbol{a}+\boldsymbol{J}(\boldsymbol{a})\boldsymbol{\bar{a}}+\boldsymbol{J}(\boldsymbol{\bar{a}})\boldsymbol{a},
\end{align}
which makes it easy to see that the nonlinear form is exactly that in equation \eqref{eq:nonlinear_af}, i.e. $\boldsymbol{f}:=\boldsymbol{J}(\boldsymbol{a})\boldsymbol{a}$. 
The form of the nonlinearity means that we can directly apply the bounds in Lemma \ref{thm:bound}. The nonlinearity is energy-conserving and of odd dimension, therefore there exists a non-trivial element in the left nullspace of $\boldsymbol{J}(\boldsymbol{a})$. The corresponding element $\boldsymbol{n}^T=\begin{bmatrix}1 & 0 & 0 & 0 & 0 & 0 & 0 & 0 & -1\end{bmatrix}$ is known and can easily be deduced from equations (\ref{eq:9D_coeff_mode1}) and (\ref{eq:9D_coeff_mode9}) in Appendix \ref{sec:appendix_galerkin}.

\begin{figure}
    \centering
   \includegraphics[width=3.4in]{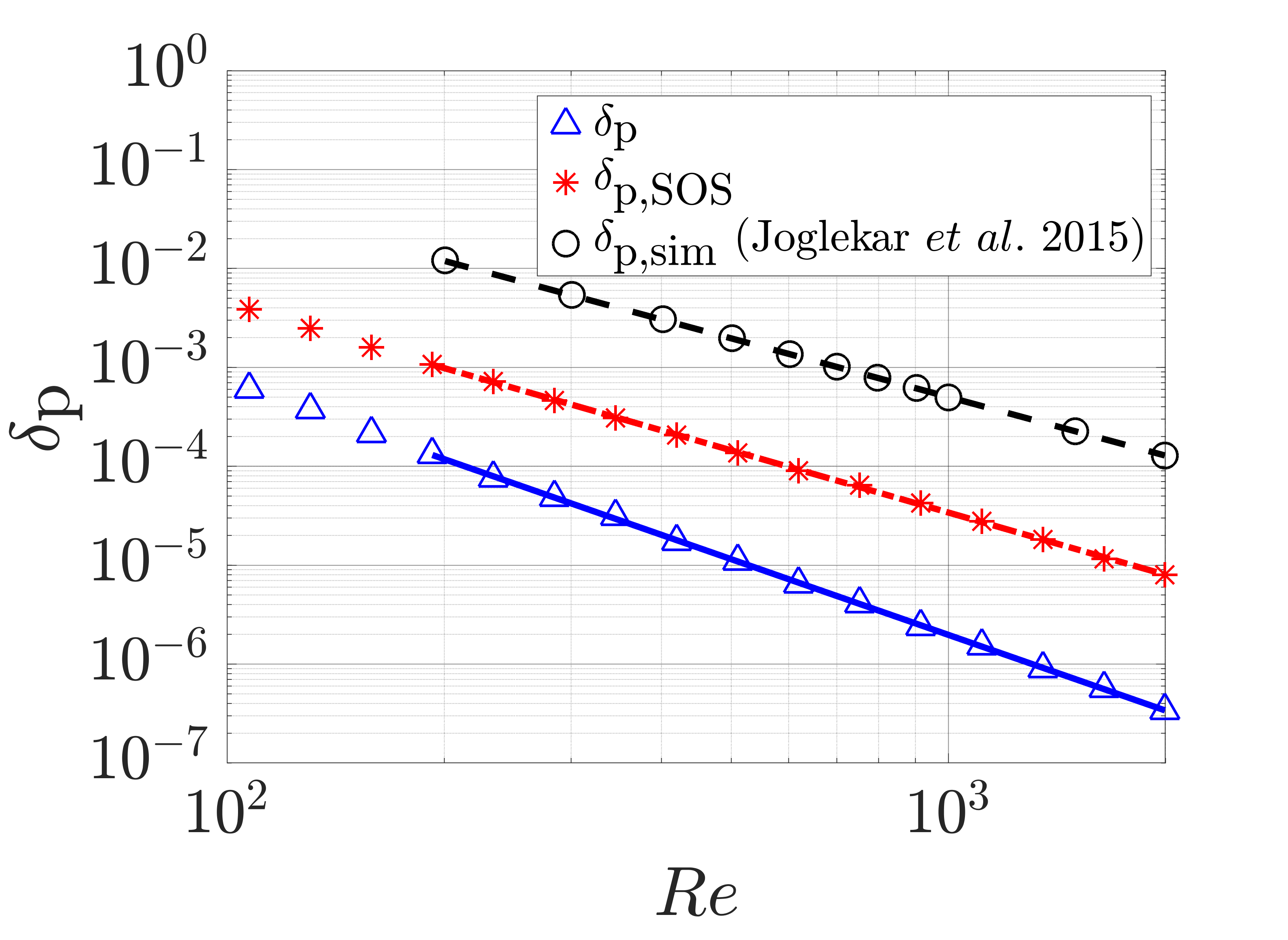}
    \caption{\rev{ Permissible perturbation amplitudes for the nine-dimensional shear flow model \citep{Moehlis2004} in Section \ref{subsec:com_SOS}: $\delta_{\text{p}}$ ({\color{blue}$\hspace{-0.0070in}\vspace{-0.015in}{\triangle}$}) obtained from Theorem \ref{thm:lmi} and equation (\ref{eq:optimze_delta}) displaying $\delta_{\text{p}}=10^{1.92}Re^{-2.54}$ ({\color{blue}$\mline\mline$}); $\delta_{\text{p,SOS}}$ ({\color{red}$\hspace{-0.0070in}\vspace{-0.015in}$\ding{83}}) obtained from the SOS programming in equations (\ref{eq:SOS}) and (\ref{eq:delta_max_SOS}) displaying $\delta_{\text{p,SOS}}=10^{1.80}Re^{-2.09}$ ({\color{red}$\dashdot$}); $\delta_{\text{p,sim}}$ ({\color{black}$\hspace{-0.0070in}\vspace{-0.015in}\bigcirc$}) obtained from simulations displaying $\delta_{\text{p,sim}}=10^{2.61}Re^{-1.97}$ ({\color{black}$\dashed$}) \citep{Joglekar2015}.}}
    \label{fig:delta_re_lmi}
\end{figure}

Having defined the constraint set, we first apply Theorem \ref{thm:lmi} to reproduce results from energy stability analysis using the approach described in section \ref{subsec:permissible_LMI}. The laminar state of this nine-dimensional shear flow model with a larger domain size ($L_x=4\pi$ and $L_z=2\pi$) was shown to be globally asymptotically stable at Reynolds numbers below $7.5$ using classical energy methods. Using the proposed method provides a certification that (\ref{eq:LMI}) is feasible for an arbitrarily large $\delta$ resulting in $\delta_{\text{p}}=\infty$ when $Re<Re_E=7.5$. We note that the energy bound was further improved to $Re_{SOS}=54.1$ through SOS based stability analysis  using fourth-order polynomial Lyapunov functions  \citep{goulart2012global}. However, since the current framework limits the candidate Lyapunov function to a quadratic form (second-order polynomials),  this approach cannot recover the results predicted by the SOS programming with fourth-order polynomials. The LMI based method is, however, far more computationally efficient (as discussed later in this section). Methods that can take advantage of these computational benefits while improving accuracy through higher order Lyaponov functions are a direction of future work.

Figure \ref{fig:delta_re_lmi} next shows the results of the optimization $\delta_{\text{p}}$ at each Reynolds number in the range where there is no proof of global asymptotic stability of the laminar state. In particular, we concentrate on $Re\geq100$ as recent results suggest that the laminar solution of the model is globally asymptotically stable below $Re<80.54$ \citep{lakshmi2020finding}.
We then perform the least-squares fit to the same function $\delta_{\text{p}}(Re)=10^{A}Re^{\sigma}$ and obtain $\delta_{\text{p}}=10^{1.92}Re^{-2.54}$ in the range $Re\in(190,2000)$. These results are plotted alongside the function $\delta_{\text{p,sim}}=10^{2.61}Re^{-1.97}$ reported in Figure 8 of \citep{Joglekar2015}, which are obtained from 10,000 simulations of the same nine-mode model with  randomly chosen initial conditions. The results show that the permissible perturbation amplitude identified using this framework is conservative, however, it has the benefit of providing a rigorous lower bound (Theorem \ref{thm:lmi}) on the results obtained from extensive simulations.

In order to illustrate the effects of constraining rather than fully representing the nonlinearity, we now compare our results to those obtained using a quadratic Lyapunov function obtained through SOS programming. SOS based programs enable the exploration of a larger class of candidate Lyapunov functions; however, these additional degrees of freedom come at the expense of more computational resources; see e.g., \citep{goulart2012global}. The computational complexity increases with the order of the candidate Lyapunov functions. Here, we restrict the candidate Lyapunov functions to quadratic forms $V=\boldsymbol{a}^T\boldsymbol{P}\boldsymbol{a}$ for direct comparison of the accuracy and computational resources associated versus the proposed method based on Theorem \ref{thm:lmi}.  In particular, we employ Theorem 3.7 in Ref. \citep{Anderson2015} to certify local asymptotic stability through checking the conditions,
\begin{subequations}\label{eq:SOS}
\begin{align}
    \boldsymbol{P}-\epsilon\boldsymbol{I}\succeq &0,\\
    \epsilon>&0,\\
    \frac{dV}{dt}+(\delta^2-\boldsymbol{a}^T\boldsymbol{a})\boldsymbol{a}^T\boldsymbol{R}\boldsymbol{a}+\epsilon \boldsymbol{a}^T\boldsymbol{a}\leq& 0,\;\text{and}\label{eq:SOS_dV} \\
    \boldsymbol{R}\succeq& 0.
\end{align}
\end{subequations}
We then define $\delta_{\text{p,SOS}}$ by solving an analogous optimization problem to that in (\ref{eq:optimze_delta}), specifically,
\begin{align}
    &\delta_{\text{p,SOS}}:=\underset{\delta}{\text{max}}\;\delta\sqrt{\frac{\mu_{\text{min}}(\boldsymbol{P})}{\mu_{\text{max}}(\boldsymbol{P})}}\;    \label{eq:delta_max_SOS}\\
    &\text{subject to}\;(\ref{eq:SOS}).\nonumber
\end{align}
Note that the term $(\delta^2-\boldsymbol{a}^T\boldsymbol{a})\boldsymbol{a}^T\boldsymbol{R}\boldsymbol{a}$ in equation (\ref{eq:SOS_dV}) involves a fourth-order polynomial in $\boldsymbol{a}$ and it is this constraint that prevents us from directly formulating the problem as an LMI, which adds to the additional computational complexity. We employ SOSTOOLS version 3.0 \citep{papachristodoulou2013sostools} to implement the inequalities in equation (\ref{eq:SOS}) and test the feasibility of equation \eqref{eq:delta_max_SOS}. SOSTOOLS converts the SOS programming problem into an SDP \citep{prajna2002introducing,papachristodoulou2013sostools}. For comparison purposes, we use the same SDP solver, SeDuMi v1.3, as before.

The resulting $\delta_{\text{p,SOS}}$ values at each Reynolds number and  function $\delta_{\text{p,SOS}}=10^{1.80}Re^{-2.09}$ are provided in Figure \ref{fig:delta_re_lmi} alongside the LMI and simulation results. Clearly, the results obtained from the SOS are closer to the simulation results than those obtained from LMI based method in equation (\ref{eq:optimze_delta}). In particular, the permissible perturbation amplitude $\delta_{\text{p,SOS}}$ shows a scaling exponent $\sigma$ of $-2.09$, which is closer to the $-1.97$ observed in the simulation results in Ref. \citep{Joglekar2015}. However, 
this improved accuracy is achieved at the expense of high computational resources as highlighted in Table \ref{tab:time_compare}. \rev{The results indicate that incorporating more properties of nonlinearity, e.g. those that are captured by the SOS formulation, could improve the performance of the LMI approach. Further analysis of the perturbation structures associated with the lowest permissible perturbations, as discussed in Remark \ref{remark:ROA}, may provide additional insights into the results to provide an understanding of the system stability. This incorporation and analysis require some additional theory and computational tools for efficient implementation, so we leave this as a topic of future work.}

Table \ref{tab:time_compare} compares each of the computational steps contributing to the total computational time of the proposed LMI method to the SOS based solution.  We divide the computation time into the following steps. The `Preprocessing time' describes the time to convert  the problems into an SDP (which is the method of solution in both cases). The computation time used to solve the SDP is reported as the `SDP solver time'. We also report the size of the largest positive semi-definite cone and the number of constraints (for every fixed given $\delta$ and $Re$) to further explain where the differences in the computational times arise.

The values in Table \ref{tab:time_compare} clearly indicate that the LMI based framework in Theorem \ref{thm:lmi} uses substantially less computational time compared with the SOS programming.  Here, we also note that the proposed LMI framework can effectively reduce the size of the largest PSD cone and the number of constraints, resulting in a more efficient estimation for permissible perturbation amplitude. This computational efficiency is achieved through constraining the nonlinearity rather than directly including it, which directly contributes to smaller problem inputs to the  SDP solver. This reduction in the number of inputs to the SDP solver suggests that the LMI framework may also have the benefit of saving the memory, which is another computational bottleneck of SOS \citep{zheng2018fast}. However, as also indicated in Theorem \ref{thm:lmi}, the LMI formulation is currently limited to quadratic Lyapunov functions, which constraints the results that can be obtained.  Further analysis of this trade-off between accuracy and computation along with adapting the method to increase accuracy with less additional computational burden are directions of ongoing work.

\begin{table}[]
    \centering
    \begin{tabular}{c c c}
    \hline
       Method  & LMI & SOS\\
       \hline
      Preprocessing time (s) & 197 & 657837\\
      SDP Solver time (s) & 667 & 17209\\
      Size of the largest PSD cone & 18 & 54\\
      Number of constraints & 74 & 795\\
    \hline
    \end{tabular}
    \caption{Comparison of the proposed LMI framework in Theorem \ref{thm:lmi} and (\ref{eq:optimze_delta}) with SOS programming in equations (\ref{eq:SOS}) and (\ref{eq:delta_max_SOS}) for the same nine-dimensional model of sinusoidal shear flow \citep{Moehlis2004} in Section \ref{subsec:com_SOS}.}
    \label{tab:time_compare}
\end{table}

\section{Conclusions and future work}
\label{sec:conslusion}

\appendix
This work proposes an input--output inspired approach to determining the permissible level of perturbation amplitude to maintain a laminar flow state. The proposed framework partitions the dynamics into a feedback interconnection of the linear and nonlinear dynamics; i.e., a Lur\'e system in which nonlinearity is static feedback. We construct quadratic constraints of the nonlinear term that are restricted by system physics to be energy-conserving (lossless) and to have bounded input--output energy in a local region. These constraints allow us to formulate computation of the region of attraction of the laminar state (a set of safe perturbations) and permissible perturbation amplitude as Linear Matrix Inequalities (LMI), which are solved efficiently through available toolboxes. The proposed framework provides a generalization of both linear analysis and classical energy methods. We apply our approach to a wide class of low dimensional nonlinear shear flow models \citep{trefethen1993hydrodynamic,baggett1995mostly,waleffe1995transition,baggett1997low,Moehlis2004} for a range of Reynolds numbers. The results from our analytically derived bounds on the permissible perturbation amplitude are consistent with the bounds identified through exhaustive simulations  \citep{baggett1997low,Joglekar2015}. However, our results are obtained at a much lower computational cost and have the benefit of providing a provable guarantee that a certain level of perturbation is permissible. 

We perform a more detailed analysis of the nine-mode model of shear flows, which shows that the framework provides more conservative but provably correct results as the model complexity increases. A comparison to SOS based Lyapunov analysis of the full nonlinear system shows that the inherent restriction of the candidate Lyapunov function to a smaller set capturing nonlinearity through constraints on its properties rather than direct description provides improved computational efficiency.  However, this increased efficiency comes at the cost of reduced accuracy, which future work aims to further characterize and mitigate through extensions to the proposed approach.

The accuracy of the approach could potentially be improved through tightening the bounds in Lemma \ref{thm:bound}. One approach that is promising is the direct use of a quadratic form of $\boldsymbol{a}$ to represent $\|\boldsymbol{J}(\boldsymbol{a})\|_{2,2}$, which will render the approach less conservative but require some additional theory and computational tools for efficient implementation. Other forms of nonlinearity are also interesting directions for future work. In particular, the extension to systems with a nonlinearity involving the $l_2$ norm of state variables in Proposition \ref{pro:bound_2norm} here demonstrates its applicability to problems that are not typically straightforward using SOS programming; e.g., a change of variables and additional constraints are required to describe such a nonlinearity as polynomial \citep{papachristodoulou2005analysis}. Generalizing the current framework to a wider class of nonlinear systems \citep{valmorbida2018regional} involving these and other constraints less amenable to polynomial analysis may be a promising direction. 

Other directions for future work involve more detailed analysis of the shape of the region of attraction and extensions to  partial differential equation based models as a step toward analysis of the full NS equations; see e.g., \citep{ahmadi2019framework}.

\section*{Acknowledgements}
The authors gratefully acknowledge support from the US National Science Foundation (NSF) through grant number CBET 1652244 and the Office of Naval Research (ONR) through grant number N00014-18-1-2534. C.L. greatly appreciates support from the Chinese Scholarship Council and would like to acknowledge fruitful discussions with Giovanni Fantuzzi and Chengda Ji on nonlinear system analysis and usage of YALMIP and SDP solvers. In addition, he also greatly appreciates the insightful training on inequalities provided by Zhengqing Tong in preparing for the National High School Mathematics League.

\appendix
\section{Dynamics for the 9D shear flow model in Section \ref{subsec:com_SOS}}
\label{sec:appendix_galerkin}

\begin{figure}
    \centering
    \includegraphics[width=3in]{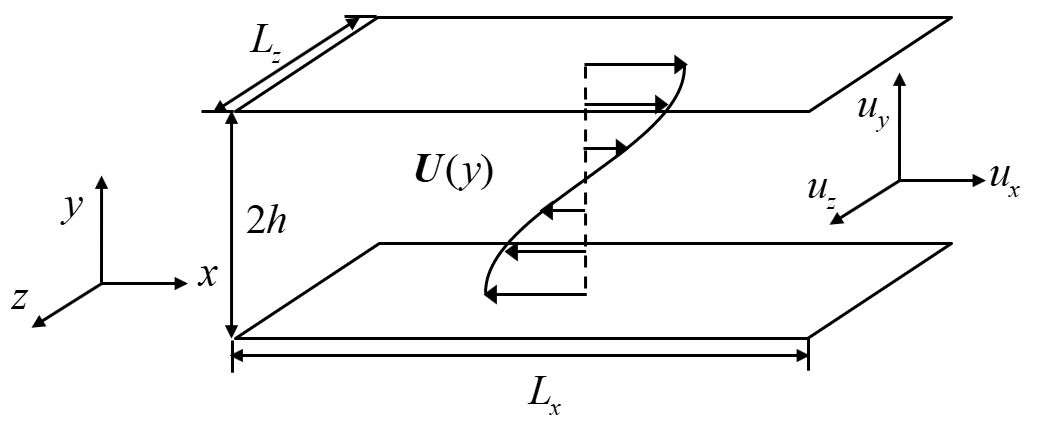}
    \caption{The illustration of sinusoidal shear flow as Refs. \citep{Moehlis2004,Moehlis2005}.}
    \label{fig:illustration}
\end{figure}

The nine-dimensional shear flow model \citep{Moehlis2004} considers the incompressible flow between two parallel flat plates under a sinusoidal body force. Figure \ref{fig:illustration} illustrates this configuration, where $x$, $y$, and $z$ represent the streamwise, wall-normal, and spanwise directions, respectively. The length is non-dimensionalized by $h$, where $h$ is the channel half height. The characteristic velocity $U_0$ is taken to be the laminar velocity resulting from the sinusoidal body force at a distance $h/2$ from the top wall. The time and pressure are, respectively, in units of $h/U_0$ and $U_0^2\rho$, where $\rho$ is the fluid density. The governing equations of the fluid between these two parallel flat plates are described by the incompressible NS equations:
\begin{subequations}
\begin{align}
    \frac{\partial \boldsymbol{u}}{\partial t}=&-(\boldsymbol{u}\cdot \nabla )\boldsymbol{u}-\nabla p+\frac{1}{Re}\nabla^2\boldsymbol{u}+\boldsymbol{F}_S(y),\label{eq:NS}\\
    \nabla \cdot \boldsymbol{u}=&0
\end{align}
\end{subequations}
with the Reynolds number defined as $Re=\frac{U_0 h}{\nu}$, where $\nu$ is the kinematic viscosity.

The boundary conditions are set up as free-slip boundaries at the walls $y=\pm 1$; i.e.,
\begin{subequations}
\begin{align}
    \left. u_y \right\vert_{y=\pm 1}=&0, \\
    \left. \frac{\partial u_x}{\partial y}\right\vert_{y=\pm 1}=&\left.\frac{\partial u_z}{\partial y}\right\vert_{y=\pm 1}=0,
\end{align}
\end{subequations}
where $u_x$, $u_y$, and $u_z$ represent the streamwise, wall-normal, and spanwise velocity, respectively. These free-slip boundary conditions make it easy to construct the Galerkin basis based on physical observations, and the underlying self-sustaining process is demonstrated to be robust no matter whether the boundary is free-slip or no-slip \citep{waleffe1997self}. Following Waleffe \citep{waleffe1997self}, the non-dimensionalized sinusoidal body force $\boldsymbol{F}_S(y)=\frac{\sqrt{2}\pi^2}{4Re}\text{sin}(\pi y/2)\boldsymbol{e}_x$ results in the laminar profile $\boldsymbol{U}(y)=\sqrt{2}\text{sin}(\pi y/2)\boldsymbol{e}_x$ with $\boldsymbol{e}_x$ denoting the unit vector in the streamwise direction. This shear flow with free-slip boundary conditions and sinusoidal body force is also fully resolved to study the large-scale feature of transitional turbulence \citep{chantry2016turbulent,chantry2017universal,tuckerman2020patterns}. In the following, we denote the flow domain $0\leq x\leq L_x$, $-1\leq y\leq 1$, and $0\leq z \leq L_z$ as $\Omega$.

Then, we project the NS equations in (\ref{eq:NS}) to Galerkin modes $\boldsymbol{u}_i, i=1,2,...,9$ that are orthogonal and normalized as:
 \begin{align}
     \int_{\Omega}\boldsymbol{u}_n\cdot\boldsymbol{u}_m d\Omega=2L_xL_z\delta_{mn},
 \end{align}
where $\delta_{mn}$ is Kronecker delta function. These modes satisfy the divergence-free constraint and boundary conditions at the wall. The detail of these modes are reported in the following equation \eqref{eq:mode_9D}, which can be also seen in equations (7)-(17) in Ref. \citep{Moehlis2004} and Appendix C in Ref. \citep{goulart2012global}:
\begin{subequations}
\begingroup
\allowdisplaybreaks
\begin{align}
    \boldsymbol{u}_1&:=\begin{bmatrix}
    \sqrt{2}\text{sin}(\pi y/2)\\
    0\\
    0
    \end{bmatrix},\\
    \boldsymbol{u}_2&:=\begin{bmatrix}
    \text{cos}^2(\pi y/2)\text{cos}(\gamma z)\\
    0\\
    0\\
    \end{bmatrix}\cdot \frac{4}{\sqrt{3}},\\
    \boldsymbol{u}_3&:=\begin{bmatrix}
    0\\
    2\gamma \text{cos}(\pi y/2)\text{cos}(\gamma z)\\
    \pi \text{sin}(\pi y/2) \text{sin}(\gamma z)
    \end{bmatrix}\cdot \frac{2}{\sqrt{4\gamma^2+\pi^2}},\\
    \boldsymbol{u}_4&:=\begin{bmatrix}
    0\\
    0\\
    \text{cos}(\alpha x)\text{cos}^2(\pi y/2)
    \end{bmatrix}\cdot\frac{4}{\sqrt{3}},\\
    \boldsymbol{u}_5&:=\begin{bmatrix}
    0\\0\\
    2\text{sin}(\alpha x) \text{sin}(\pi y/2)
    \end{bmatrix},\\
    \boldsymbol{u}_6&:=\begin{bmatrix}
    -\gamma \text{cos}(\alpha x)\text{cos}^2(\frac{\pi y}{2})\text{sin}(\gamma z)\\
    0\\
    \alpha \text{sin}(\alpha x)\text{cos}^2(\frac{\pi y}{2}) \text{cos}(\gamma z)
    \end{bmatrix}\cdot \frac{4\sqrt{2}}{\sqrt{3}\;\kappa_{\alpha\gamma}},\\
    \boldsymbol{u}_7&:=\begin{bmatrix}
    \gamma \text{sin}(\alpha x)\text{sin}(\pi y/2) \text{sin}(\gamma z)\\
    0\\
    \alpha \text{cos}(\alpha x) \text{sin}(\pi y/2) \text{cos}(\gamma z)
    \end{bmatrix}\cdot \frac{2\sqrt{2}}{\kappa_{\alpha\gamma}},\\
    \boldsymbol{u}_8&:=\begin{bmatrix}
    \pi \alpha \text{sin}(\alpha x) \text{sin}(\frac{\pi y}{2}) \text{sin}(\gamma z)\\
    2(\alpha^2+\gamma^2) \text{cos}(\alpha x) \text{cos}(\frac{\pi y}{2}) \text{sin}(\gamma z)\\
    -\pi \gamma \text{cos}(\alpha x) \text{sin}(\frac{\pi y}{2}) \text{cos}(\gamma z)
    \end{bmatrix}\cdot N_8,\\
    \boldsymbol{u}_9&:=\begin{bmatrix}
    \sqrt{2}\text{sin}(3\pi y/2)\\
    0\\
    0
    \end{bmatrix},
\end{align}
\label{eq:mode_9D}
\endgroup
\end{subequations}\noindent where $\alpha:=2\pi/L_x$, $\beta:=\pi/2$, $\gamma :=2\pi/L_z$, $\kappa_{\alpha \gamma}:=\sqrt{\alpha^2+\gamma^2}$, and
\begin{align}
    N_8:=\frac{2\sqrt{2}}{\sqrt{(\alpha^2+\gamma^2)(4\alpha^2+4\gamma^2+\pi^2)}}.
\end{align}

 Through expanding the velocity under these Galerkin modes $\boldsymbol{u}=\sum_{i=1}^9\tilde{a}_i\boldsymbol{u}_i$, substituting this expansion into the momentum equation (\ref{eq:NS}) and enforcing the residue to be orthogonal to each Galerkin mode, we obtain the Galerkin projection of the original governing equations as a nine-dimensional dynamical system:
 \begin{align}
     \frac{d \tilde{a}_i}{d t}=-\frac{\xi_{ij}}{Re}\tilde{a}_j+N_{ijk}\tilde{a}_j\tilde{a}_k+F_{i},
     \label{eq:Galerkin}
 \end{align}
 where each coefficient is obtained through:
 \begin{subequations}
  \begin{align}
     \xi_{ij}:=&\frac{\int_{\Omega}(-\nabla^2\boldsymbol{u}_j)\cdot \boldsymbol{u}_i d\Omega}{\int_{\Omega}\boldsymbol{u}_i\cdot \boldsymbol{u}_i d\Omega},\\
     N_{ijk}:=&\frac{-\int_{\Omega}[\boldsymbol{u}_j\cdot \nabla \boldsymbol{u}_k]\cdot \boldsymbol{u}_i d\Omega}{\int_{\Omega}\boldsymbol{u}_i\cdot \boldsymbol{u}_i d\Omega},\;\;\text{and}\\
     F_i:=&\frac{\int_{\Omega}\boldsymbol{F}_S\cdot \boldsymbol{u}_i d\Omega}{\int_{\Omega}\boldsymbol{u}_i\cdot \boldsymbol{u}_i d\Omega}.
 \end{align}
  \end{subequations}
 The pressure term in equation (\ref{eq:NS}) has no contribution to the Galerkin projection results as these modes are divergence-free, vanish at the wall, and satisfy periodic boundary conditions in wall parallel directions. Here, we rewrite equation (\ref{eq:Galerkin}) as:
\begin{align}
     \frac{d\boldsymbol{\tilde{a}}}{dt}=-\frac{\boldsymbol{\Xi}}{Re}\boldsymbol{\tilde{a}}+\boldsymbol{J}(\boldsymbol{\tilde{a}})\boldsymbol{\tilde{a}}+\boldsymbol{F},
\end{align}
 where we define entries of a positive definite matrix as $[\boldsymbol{\Xi}]_{i,j}:=\xi_{ij}$, entries of the state-dependent matrix as $[\boldsymbol{J}(\boldsymbol{\tilde{a}})]_{i,j}:=N_{ijk}\tilde{a}_k$, and entries of the forcing vectors as $[\boldsymbol{F}]_i:=F_i$. 
 
 For completeness of this paper, we also document the details of $\boldsymbol{\Xi}$ and $\boldsymbol{J}(\boldsymbol{\tilde{a}})$ of this Galerkin model in the following equations (\ref{eq:Galerkin_9D_coeff_viscous}) and (\ref{eq:Galerkin_9D_coeff_nonlinear}), which were also reported in (21)-(32) of Ref. \citep{Moehlis2004} and Appendix C in Ref. \citep{goulart2012global}: 
 \begin{align}
    \boldsymbol{\Xi}=&\text{diag}(\beta^2,\frac{4\beta^2}{3}+\gamma^2,\kappa_{\beta \gamma}^2,\frac{3\alpha^2+4\beta^2}{3},\kappa_{\alpha\beta}^2,\nonumber\\
    &\frac{3\alpha^2+4\beta^2+3\gamma^2}{3},\kappa_{\alpha \beta\gamma}^2,\kappa_{\alpha \beta\gamma}^2,9\beta^2),
    \label{eq:Galerkin_9D_coeff_viscous}
 \end{align} and
\begingroup
\allowdisplaybreaks
 \begin{subequations}
 \begin{align}
    [\boldsymbol{J}(\boldsymbol{\tilde{a}})\boldsymbol{\tilde{a}}]_1=&\sqrt{\frac{3}{2}}\frac{\beta \gamma}{\kappa_{\beta \gamma}}\tilde{a}_2\tilde{a}_3-\sqrt{\frac{3}{2}}\frac{\beta \gamma }{\kappa_{\alpha \beta \gamma }}\tilde{a}_6\tilde{a}_8,\label{eq:9D_coeff_mode1}\\
    [\boldsymbol{J}(\boldsymbol{\tilde{a}})\boldsymbol{\tilde{a}}]_2=&\frac{10}{3\sqrt{6}}\frac{\gamma^2}{\kappa_{\alpha \gamma}}\tilde{a}_4\tilde{a}_6-\frac{\gamma^2}{\sqrt{6}\kappa_{\alpha \gamma}}\tilde{a}_5\tilde{a}_7\nonumber\\
    &-\frac{\alpha \beta \gamma}{\sqrt{6}\kappa_{\alpha \gamma}\kappa_{\alpha \beta \gamma }}\tilde{a}_5\tilde{a}_8\nonumber\\
    &-\sqrt{\frac{3}{2}}\frac{\beta \gamma }{\kappa_{\beta \gamma}}(\tilde{a}_1\tilde{a}_3+\tilde{a}_3\tilde{a}_9),\\
    [\boldsymbol{J}(\boldsymbol{\tilde{a}})\boldsymbol{\tilde{a}}]_3=&\sqrt{\frac{2}{3}}\frac{\alpha \beta \gamma }{\kappa_{\alpha \gamma}\kappa_{\beta \gamma }}(\tilde{a}_5\tilde{a}_6+\tilde{a}_4\tilde{a}_7)\nonumber\\
    &+\frac{\beta^2(3\alpha^2+\gamma^2)-3\gamma^2\kappa_{\alpha\gamma}^2}{\sqrt{6}\kappa_{\alpha \gamma}\kappa_{\beta \gamma }\kappa_{\alpha \beta \gamma}}\tilde{a}_4\tilde{a}_8,\\
    [\boldsymbol{J}(\boldsymbol{\tilde{a}})\boldsymbol{\tilde{a}}]_4=&-\frac{\alpha}{\sqrt{6}}(\tilde{a}_1\tilde{a}_5+\tilde{a}_5\tilde{a}_9)-\frac{10}{3\sqrt{6}}\frac{\alpha^2}{\kappa_{\alpha \gamma}}\tilde{a}_2\tilde{a}_6\nonumber\\
    &-\sqrt{\frac{3}{2}}\frac{\alpha \beta \gamma}{\kappa_{\alpha \gamma}\kappa_{\beta \gamma}}\tilde{a}_3\tilde{a}_7\nonumber\\
    &-\sqrt{\frac{3}{2}}\frac{\alpha^2\beta^2}{\kappa_{\alpha \gamma}\kappa_{\beta \gamma }\kappa_{\alpha \beta \gamma}}\tilde{a}_3\tilde{a}_8,\\
    [\boldsymbol{J}(\boldsymbol{\tilde{a}})\boldsymbol{\tilde{a}}]_5=&\frac{\alpha}{\sqrt{6}}(\tilde{a}_1\tilde{a}_4+\tilde{a}_4\tilde{a}_9)+\sqrt{\frac{2}{3}}\frac{\alpha \beta \gamma }{\kappa_{\alpha \gamma }\kappa_{\beta \gamma}}\tilde{a}_3\tilde{a}_6\nonumber\\
    &+\frac{\alpha^2}{\sqrt{6}\kappa_{\alpha \gamma}}\tilde{a}_2\tilde{a}_7-\frac{\alpha \beta\gamma}{\sqrt{6}\kappa_{\alpha \gamma}\kappa_{\alpha \beta \gamma}}\tilde{a}_2\tilde{a}_8,\\
    [\boldsymbol{J}(\boldsymbol{\tilde{a}})\boldsymbol{\tilde{a}}]_6=&\frac{10}{3\sqrt{6}}\frac{\alpha^2-\gamma^2}{\kappa_{\alpha \gamma}}\tilde{a}_2\tilde{a}_4-\sqrt{\frac{2}{3}}\frac{2\alpha \beta \gamma}{\kappa_{\alpha \gamma}\kappa_{\beta \gamma}}\tilde{a}_3\tilde{a}_5\nonumber\\
    &+\frac{\alpha}{\sqrt{6}}(\tilde{a}_1\tilde{a}_7+\tilde{a}_7\tilde{a}_9)\nonumber\\
    &+\sqrt{\frac{3}{2}}\frac{\beta \gamma}{\kappa_{\alpha \beta \gamma}}(\tilde{a}_1\tilde{a}_8+\tilde{a}_8\tilde{a}_9),\\
    [\boldsymbol{J}(\boldsymbol{\tilde{a}})\boldsymbol{\tilde{a}}]_7=&\frac{\alpha \beta \gamma}{\sqrt{6}\kappa_{\alpha \gamma}\kappa_{\beta \gamma}}\tilde{a}_3\tilde{a}_4+\frac{-\alpha^2+\gamma^2}{\sqrt{6}\kappa_{\alpha \gamma}}\tilde{a}_2\tilde{a}_5\nonumber\\
    &-\frac{\alpha}{\sqrt{6}}(\tilde{a}_1\tilde{a}_6+\tilde{a}_6\tilde{a}_9),\\
    [\boldsymbol{J}(\boldsymbol{\tilde{a}})\boldsymbol{\tilde{a}}]_8=&\frac{\gamma^2(3\alpha^2-\beta^2+3\gamma^2)}{\sqrt{6}\kappa_{\alpha \gamma}\kappa_{\beta \gamma}\kappa_{\alpha \beta \gamma}}\tilde{a}_3\tilde{a}_4\nonumber\\
    &+\sqrt{\frac{2}{3}}\frac{\alpha \beta \gamma}{\kappa_{\alpha \gamma }\kappa_{\alpha \beta \gamma}}\tilde{a}_2\tilde{a}_5,\\
    [\boldsymbol{J}(\boldsymbol{\tilde{a}})\boldsymbol{\tilde{a}}]_9=&\sqrt{\frac{3}{2}}\frac{\beta \gamma }{\kappa_{\beta \gamma}}\tilde{a}_2\tilde{a}_3-\sqrt{\frac{3}{2}}\frac{\beta \gamma }{\kappa_{\alpha \beta \gamma}}\tilde{a}_6\tilde{a}_8,
    \label{eq:9D_coeff_mode9}
\end{align}
\label{eq:Galerkin_9D_coeff_nonlinear}
\end{subequations}
\endgroup
where $[\boldsymbol{J}(\boldsymbol{\tilde{a}})\boldsymbol{\tilde{a}}]_m:=\boldsymbol{e}_m^T\boldsymbol{J}(\boldsymbol{\tilde{a}})\boldsymbol{\tilde{a}},\;m=1,2,...,9$ is the $m^{\text{th}}$ component of $\boldsymbol{J}(\boldsymbol{\tilde{a}})\boldsymbol{\tilde{a}}$, and  $\kappa_{\alpha \beta}:=\sqrt{\alpha^2+\beta^2}$, $\kappa_{\beta \gamma}:=\sqrt{\beta^2+\gamma^2}$ and $\kappa_{\alpha \beta\gamma}:=\sqrt{\alpha^2+\beta^2+\gamma^2}$.

The laminar profile $\boldsymbol{U}(y)$ in this model corresponds to a fixed point $\boldsymbol{\bar{a}}=\begin{bmatrix}1 & 0 & 0 & 0 & 0 & 0 & 0& 0 & 0\end{bmatrix}^T$, and it satisfies:
\begin{align}
    -\frac{\boldsymbol{\Xi}}{Re}\boldsymbol{\bar{a}}+\boldsymbol{\boldsymbol{J}(\bar{a})}\boldsymbol{\bar{a}}+\boldsymbol{F}=\boldsymbol{0}.
    \label{eq:laminar_balance}
\end{align}
We can perform a decomposition of Galerkin coefficients similar to Reynolds decomposition: \begin{align}
    \boldsymbol{\tilde{a}}=\boldsymbol{\bar{a}}+\boldsymbol{a},
\end{align}
so as to shift the laminar state to the origin of fluctuating coefficients $\boldsymbol{a}$. The resulting dynamical system for these fluctuating coefficients is
\begin{align}
    \frac{d \boldsymbol{a}}{dt}=-\frac{\boldsymbol{\Xi}}{Re}\boldsymbol{a}+\boldsymbol{J}(\boldsymbol{a})\boldsymbol{\bar{a}}+\boldsymbol{J}(\boldsymbol{\bar{a}})\boldsymbol{a}+\boldsymbol{J}(\boldsymbol{a})\boldsymbol{a},
    \label{eq:galerkin_fluc_append}
\end{align}
which gives equation \eqref{eq:galerkin_fluc} in section \ref{subsec:com_SOS}.


\bibliography{main}

\begin{thebibliography}{101}%
\makeatletter
\providecommand \@ifxundefined [1]{%
 \@ifx{#1\undefined}
}%
\providecommand \@ifnum [1]{%
 \ifnum #1\expandafter \@firstoftwo
 \else \expandafter \@secondoftwo
 \fi
}%
\providecommand \@ifx [1]{%
 \ifx #1\expandafter \@firstoftwo
 \else \expandafter \@secondoftwo
 \fi
}%
\providecommand \natexlab [1]{#1}%
\providecommand \enquote  [1]{``#1''}%
\providecommand \bibnamefont  [1]{#1}%
\providecommand \bibfnamefont [1]{#1}%
\providecommand \citenamefont [1]{#1}%
\providecommand \href@noop [0]{\@secondoftwo}%
\providecommand \href [0]{\begingroup \@sanitize@url \@href}%
\providecommand \@href[1]{\@@startlink{#1}\@@href}%
\providecommand \@@href[1]{\endgroup#1\@@endlink}%
\providecommand \@sanitize@url [0]{\catcode `\\12\catcode `\$12\catcode
  `\&12\catcode `\#12\catcode `\^12\catcode `\_12\catcode `\%12\relax}%
\providecommand \@@startlink[1]{}%
\providecommand \@@endlink[0]{}%
\providecommand \url  [0]{\begingroup\@sanitize@url \@url }%
\providecommand \@url [1]{\endgroup\@href {#1}{\urlprefix }}%
\providecommand \urlprefix  [0]{URL }%
\providecommand \Eprint [0]{\href }%
\providecommand \doibase [0]{http://dx.doi.org/}%
\providecommand \selectlanguage [0]{\@gobble}%
\providecommand \bibinfo  [0]{\@secondoftwo}%
\providecommand \bibfield  [0]{\@secondoftwo}%
\providecommand \translation [1]{[#1]}%
\providecommand \BibitemOpen [0]{}%
\providecommand \bibitemStop [0]{}%
\providecommand \bibitemNoStop [0]{.\EOS\space}%
\providecommand \EOS [0]{\spacefactor3000\relax}%
\providecommand \BibitemShut  [1]{\csname bibitem#1\endcsname}%
\let\auto@bib@innerbib\@empty
\bibitem [{\citenamefont {Drazin}\ and\ \citenamefont
  {Reid}(2004)}]{drazin2004hydrodynamic}%
  \BibitemOpen
  \bibfield  {author} {\bibinfo {author} {\bibfnamefont {P.~G.}\ \bibnamefont
  {Drazin}}\ and\ \bibinfo {author} {\bibfnamefont {W.~H.}\ \bibnamefont
  {Reid}},\ }\href@noop {} {\emph {\bibinfo {title} {Hydrodynamic stability}}}\
  (\bibinfo  {publisher} {Cambridge university press},\ \bibinfo {year}
  {2004})\BibitemShut {NoStop}%
\bibitem [{\citenamefont {Schmid}\ and\ \citenamefont
  {Henningson}(2012)}]{schmid2012stability}%
  \BibitemOpen
  \bibfield  {author} {\bibinfo {author} {\bibfnamefont {P.~J.}\ \bibnamefont
  {Schmid}}\ and\ \bibinfo {author} {\bibfnamefont {D.~S.}\ \bibnamefont
  {Henningson}},\ }\href@noop {} {\emph {\bibinfo {title} {Stability and
  transition in shear flows}}},\ Vol.\ \bibinfo {volume} {142}\ (\bibinfo
  {publisher} {Springer Science \& Business Media},\ \bibinfo {year}
  {2012})\BibitemShut {NoStop}%
\bibitem [{\citenamefont {Romanov}(1973)}]{romanov1973stability}%
  \BibitemOpen
  \bibfield  {author} {\bibinfo {author} {\bibfnamefont {V.~A.}\ \bibnamefont
  {Romanov}},\ }\bibfield  {title} {\enquote {\bibinfo {title} {Stability of
  plane-parallel {C}ouette flow},}\ }\href@noop {} {\bibfield  {journal}
  {\bibinfo  {journal} {Funct. Anal. Appl.}\ }\textbf {\bibinfo {volume} {7}},\
  \bibinfo {pages} {137--146} (\bibinfo {year} {1973})}\BibitemShut {NoStop}%
\bibitem [{\citenamefont {Tillmark}\ and\ \citenamefont
  {Alfredsson}(1992)}]{tillmark1992experiments}%
  \BibitemOpen
  \bibfield  {author} {\bibinfo {author} {\bibfnamefont {N.}~\bibnamefont
  {Tillmark}}\ and\ \bibinfo {author} {\bibfnamefont {P.~H.}\ \bibnamefont
  {Alfredsson}},\ }\bibfield  {title} {\enquote {\bibinfo {title} {Experiments
  on transition in plane {C}ouette flow},}\ }\href@noop {} {\bibfield
  {journal} {\bibinfo  {journal} {J. Fluid Mech.}\ }\textbf {\bibinfo {volume}
  {235}},\ \bibinfo {pages} {89--102} (\bibinfo {year} {1992})}\BibitemShut
  {NoStop}%
\bibitem [{\citenamefont {Waleffe}(1995)}]{waleffe1995transition}%
  \BibitemOpen
  \bibfield  {author} {\bibinfo {author} {\bibfnamefont {F.}~\bibnamefont
  {Waleffe}},\ }\bibfield  {title} {\enquote {\bibinfo {title} {Transition in
  shear flows. nonlinear normality versus non-normal linearity},}\ }\href@noop
  {} {\bibfield  {journal} {\bibinfo  {journal} {Phys. Fluids}\ }\textbf
  {\bibinfo {volume} {7}},\ \bibinfo {pages} {3060--3066} (\bibinfo {year}
  {1995})}\BibitemShut {NoStop}%
\bibitem [{\citenamefont {Reddy}\ and\ \citenamefont
  {Henningson}(1993)}]{reddy1993energy}%
  \BibitemOpen
  \bibfield  {author} {\bibinfo {author} {\bibfnamefont {S.~C.}\ \bibnamefont
  {Reddy}}\ and\ \bibinfo {author} {\bibfnamefont {D.~S.}\ \bibnamefont
  {Henningson}},\ }\bibfield  {title} {\enquote {\bibinfo {title} {Energy
  growth in viscous channel flows},}\ }\href@noop {} {\bibfield  {journal}
  {\bibinfo  {journal} {J. Fluid Mech.}\ }\textbf {\bibinfo {volume} {252}},\
  \bibinfo {pages} {209--238} (\bibinfo {year} {1993})}\BibitemShut {NoStop}%
\bibitem [{\citenamefont {Trefethen}\ \emph {et~al.}(1993)\citenamefont
  {Trefethen}, \citenamefont {Trefethen}, \citenamefont {Reddy},\ and\
  \citenamefont {Driscoll}}]{trefethen1993hydrodynamic}%
  \BibitemOpen
  \bibfield  {author} {\bibinfo {author} {\bibfnamefont {L.~N.}\ \bibnamefont
  {Trefethen}}, \bibinfo {author} {\bibfnamefont {A.~E.}\ \bibnamefont
  {Trefethen}}, \bibinfo {author} {\bibfnamefont {S.~C.}\ \bibnamefont
  {Reddy}}, \ and\ \bibinfo {author} {\bibfnamefont {T.~A.}\ \bibnamefont
  {Driscoll}},\ }\bibfield  {title} {\enquote {\bibinfo {title} {Hydrodynamic
  stability without eigenvalues},}\ }\href@noop {} {\bibfield  {journal}
  {\bibinfo  {journal} {Science}\ }\textbf {\bibinfo {volume} {261}},\ \bibinfo
  {pages} {578--584} (\bibinfo {year} {1993})}\BibitemShut {NoStop}%
\bibitem [{\citenamefont {Henningson}\ and\ \citenamefont
  {Reddy}(1994)}]{henningson1994role}%
  \BibitemOpen
  \bibfield  {author} {\bibinfo {author} {\bibfnamefont {D.~S.}\ \bibnamefont
  {Henningson}}\ and\ \bibinfo {author} {\bibfnamefont {S.~C.}\ \bibnamefont
  {Reddy}},\ }\bibfield  {title} {\enquote {\bibinfo {title} {On the role of
  linear mechanisms in transition to turbulence},}\ }\href@noop {} {\bibfield
  {journal} {\bibinfo  {journal} {Phys. Fluids}\ }\textbf {\bibinfo {volume}
  {6}},\ \bibinfo {pages} {1396--1398} (\bibinfo {year} {1994})}\BibitemShut
  {NoStop}%
\bibitem [{\citenamefont {Trefethen}\ and\ \citenamefont
  {Embree}(2005)}]{trefethen2005spectra}%
  \BibitemOpen
  \bibfield  {author} {\bibinfo {author} {\bibfnamefont {L.~N.}\ \bibnamefont
  {Trefethen}}\ and\ \bibinfo {author} {\bibfnamefont {M.}~\bibnamefont
  {Embree}},\ }\href@noop {} {\emph {\bibinfo {title} {Spectra and
  pseudospectra: the behavior of nonnormal matrices and operators}}}\ (\bibinfo
   {publisher} {Princeton University Press},\ \bibinfo {year}
  {2005})\BibitemShut {NoStop}%
\bibitem [{\citenamefont {Joseph}(2013)}]{joseph2013stability}%
  \BibitemOpen
  \bibfield  {author} {\bibinfo {author} {\bibfnamefont {D.~D.}\ \bibnamefont
  {Joseph}},\ }\href@noop {} {\emph {\bibinfo {title} {Stability of fluid
  motions I}}},\ Vol.~\bibinfo {volume} {27}\ (\bibinfo  {publisher} {Springer
  Science \& Business Media},\ \bibinfo {year} {2013})\BibitemShut {NoStop}%
\bibitem [{\citenamefont {Straughan}(2013)}]{straughan2013energy}%
  \BibitemOpen
  \bibfield  {author} {\bibinfo {author} {\bibfnamefont {B.}~\bibnamefont
  {Straughan}},\ }\href@noop {} {\emph {\bibinfo {title} {The energy method,
  stability, and nonlinear convection}}},\ Vol.~\bibinfo {volume} {91}\
  (\bibinfo  {publisher} {Springer Science \& Business Media},\ \bibinfo {year}
  {2013})\BibitemShut {NoStop}%
\bibitem [{\citenamefont {Goulart}\ and\ \citenamefont
  {Chernyshenko}(2012)}]{goulart2012global}%
  \BibitemOpen
  \bibfield  {author} {\bibinfo {author} {\bibfnamefont {P.~J.}\ \bibnamefont
  {Goulart}}\ and\ \bibinfo {author} {\bibfnamefont {S.}~\bibnamefont
  {Chernyshenko}},\ }\bibfield  {title} {\enquote {\bibinfo {title} {Global
  stability analysis of fluid flows using sum-of-squares},}\ }\href@noop {}
  {\bibfield  {journal} {\bibinfo  {journal} {Physica D}\ }\textbf {\bibinfo
  {volume} {241}},\ \bibinfo {pages} {692--704} (\bibinfo {year}
  {2012})}\BibitemShut {NoStop}%
\bibitem [{\citenamefont {Chernyshenko}\ \emph {et~al.}(2014)\citenamefont
  {Chernyshenko}, \citenamefont {Goulart}, \citenamefont {Huang},\ and\
  \citenamefont {Papachristodoulou}}]{chernyshenko2014polynomial}%
  \BibitemOpen
  \bibfield  {author} {\bibinfo {author} {\bibfnamefont {S.~I.}\ \bibnamefont
  {Chernyshenko}}, \bibinfo {author} {\bibfnamefont {P.}~\bibnamefont
  {Goulart}}, \bibinfo {author} {\bibfnamefont {D.}~\bibnamefont {Huang}}, \
  and\ \bibinfo {author} {\bibfnamefont {A.}~\bibnamefont
  {Papachristodoulou}},\ }\bibfield  {title} {\enquote {\bibinfo {title}
  {Polynomial sum of squares in fluid dynamics: a review with a look ahead},}\
  }\href@noop {} {\bibfield  {journal} {\bibinfo  {journal} {Phil. Trans. R.
  Soc. A}\ }\textbf {\bibinfo {volume} {372}},\ \bibinfo {pages} {20130350}
  (\bibinfo {year} {2014})}\BibitemShut {NoStop}%
\bibitem [{\citenamefont {Huang}\ \emph {et~al.}(2015)\citenamefont {Huang},
  \citenamefont {Chernyshenko}, \citenamefont {Goulart}, \citenamefont
  {Lasagna}, \citenamefont {Tutty},\ and\ \citenamefont
  {Fuentes}}]{huang2015sum}%
  \BibitemOpen
  \bibfield  {author} {\bibinfo {author} {\bibfnamefont {D.}~\bibnamefont
  {Huang}}, \bibinfo {author} {\bibfnamefont {S.}~\bibnamefont {Chernyshenko}},
  \bibinfo {author} {\bibfnamefont {P.}~\bibnamefont {Goulart}}, \bibinfo
  {author} {\bibfnamefont {D.}~\bibnamefont {Lasagna}}, \bibinfo {author}
  {\bibfnamefont {O.}~\bibnamefont {Tutty}}, \ and\ \bibinfo {author}
  {\bibfnamefont {F.}~\bibnamefont {Fuentes}},\ }\bibfield  {title} {\enquote
  {\bibinfo {title} {Sum-of-squares of polynomials approach to nonlinear
  stability of fluid flows: an example of application},}\ }\href@noop {}
  {\bibfield  {journal} {\bibinfo  {journal} {Proc. R. Soc. A}\ }\textbf
  {\bibinfo {volume} {471}},\ \bibinfo {pages} {20150622} (\bibinfo {year}
  {2015})}\BibitemShut {NoStop}%
\bibitem [{\citenamefont {Fuentes}\ \emph {et~al.}(2019)\citenamefont
  {Fuentes}, \citenamefont {Goluskin},\ and\ \citenamefont
  {Chernyshenko}}]{fuentes2019global}%
  \BibitemOpen
  \bibfield  {author} {\bibinfo {author} {\bibfnamefont {F.}~\bibnamefont
  {Fuentes}}, \bibinfo {author} {\bibfnamefont {D.}~\bibnamefont {Goluskin}}, \
  and\ \bibinfo {author} {\bibfnamefont {S.}~\bibnamefont {Chernyshenko}},\
  }\bibfield  {title} {\enquote {\bibinfo {title} {Global stability of fluid
  flows despite transient growth of energy},}\ }\href@noop {} {\bibfield
  {journal} {\bibinfo  {journal} {arXiv preprint arXiv:1911.09079}\ } (\bibinfo
  {year} {2019})}\BibitemShut {NoStop}%
\bibitem [{\citenamefont {Prajna}\ \emph {et~al.}(2002)\citenamefont {Prajna},
  \citenamefont {Papachristodoulou},\ and\ \citenamefont
  {Parrilo}}]{prajna2002introducing}%
  \BibitemOpen
  \bibfield  {author} {\bibinfo {author} {\bibfnamefont {S.}~\bibnamefont
  {Prajna}}, \bibinfo {author} {\bibfnamefont {A.}~\bibnamefont
  {Papachristodoulou}}, \ and\ \bibinfo {author} {\bibfnamefont {P.~A.}\
  \bibnamefont {Parrilo}},\ }\bibfield  {title} {\enquote {\bibinfo {title}
  {Introducing {SOSTOOLS}: {A} general purpose sum of squares programming
  solver},}\ }in\ \href@noop {} {\emph {\bibinfo {booktitle} {Proceedings of
  the 41st IEEE Conference on Decision and Control, 2002.}}},\ Vol.~\bibinfo
  {volume} {1}\ (\bibinfo {organization} {IEEE},\ \bibinfo {year} {2002})\ pp.\
  \bibinfo {pages} {741--746}\BibitemShut {NoStop}%
\bibitem [{\citenamefont {Papachristodoulou}\ \emph {et~al.}(2013)\citenamefont
  {Papachristodoulou}, \citenamefont {Anderson}, \citenamefont {Valmorbida},
  \citenamefont {Prajna}, \citenamefont {Seiler},\ and\ \citenamefont
  {Parrilo}}]{papachristodoulou2013sostools}%
  \BibitemOpen
  \bibfield  {author} {\bibinfo {author} {\bibfnamefont {A.}~\bibnamefont
  {Papachristodoulou}}, \bibinfo {author} {\bibfnamefont {J.}~\bibnamefont
  {Anderson}}, \bibinfo {author} {\bibfnamefont {G.}~\bibnamefont
  {Valmorbida}}, \bibinfo {author} {\bibfnamefont {S.}~\bibnamefont {Prajna}},
  \bibinfo {author} {\bibfnamefont {P.}~\bibnamefont {Seiler}}, \ and\ \bibinfo
  {author} {\bibfnamefont {P.}~\bibnamefont {Parrilo}},\ }\bibfield  {title}
  {\enquote {\bibinfo {title} {{SOSTOOLS} version 3.00 sum of squares
  optimization toolbox for {MATLAB}},}\ }\href@noop {} {\bibfield  {journal}
  {\bibinfo  {journal} {arXiv preprint arXiv:1310.4716}\ } (\bibinfo {year}
  {2013})}\BibitemShut {NoStop}%
\bibitem [{\citenamefont {Baggett}\ and\ \citenamefont
  {Trefethen}(1997)}]{baggett1997low}%
  \BibitemOpen
  \bibfield  {author} {\bibinfo {author} {\bibfnamefont {J.~S.}\ \bibnamefont
  {Baggett}}\ and\ \bibinfo {author} {\bibfnamefont {L.~N.}\ \bibnamefont
  {Trefethen}},\ }\bibfield  {title} {\enquote {\bibinfo {title}
  {Low-dimensional models of subcritical transition to turbulence},}\
  }\href@noop {} {\bibfield  {journal} {\bibinfo  {journal} {Phys. Fluids}\
  }\textbf {\bibinfo {volume} {9}},\ \bibinfo {pages} {1043--1053} (\bibinfo
  {year} {1997})}\BibitemShut {NoStop}%
\bibitem [{\citenamefont {Kreiss}\ \emph {et~al.}(1994)\citenamefont {Kreiss},
  \citenamefont {Lundbladh},\ and\ \citenamefont
  {Henningson}}]{kreiss1994bounds}%
  \BibitemOpen
  \bibfield  {author} {\bibinfo {author} {\bibfnamefont {G.}~\bibnamefont
  {Kreiss}}, \bibinfo {author} {\bibfnamefont {A.}~\bibnamefont {Lundbladh}}, \
  and\ \bibinfo {author} {\bibfnamefont {D.~S.}\ \bibnamefont {Henningson}},\
  }\bibfield  {title} {\enquote {\bibinfo {title} {Bounds for threshold
  amplitudes in subcritical shear flows},}\ }\href@noop {} {\bibfield
  {journal} {\bibinfo  {journal} {J. Fluid Mech.}\ }\textbf {\bibinfo {volume}
  {270}},\ \bibinfo {pages} {175--198} (\bibinfo {year} {1994})}\BibitemShut
  {NoStop}%
\bibitem [{\citenamefont {Reddy}\ \emph {et~al.}(1998)\citenamefont {Reddy},
  \citenamefont {Schmid}, \citenamefont {Baggett},\ and\ \citenamefont
  {Henningson}}]{reddy1998stability}%
  \BibitemOpen
  \bibfield  {author} {\bibinfo {author} {\bibfnamefont {S.~C.}\ \bibnamefont
  {Reddy}}, \bibinfo {author} {\bibfnamefont {P.~J.}\ \bibnamefont {Schmid}},
  \bibinfo {author} {\bibfnamefont {J.~S.}\ \bibnamefont {Baggett}}, \ and\
  \bibinfo {author} {\bibfnamefont {D.~S.}\ \bibnamefont {Henningson}},\
  }\bibfield  {title} {\enquote {\bibinfo {title} {On stability of streamwise
  streaks and transition thresholds in plane channel flows},}\ }\href@noop {}
  {\bibfield  {journal} {\bibinfo  {journal} {J. Fluid Mech.}\ }\textbf
  {\bibinfo {volume} {365}},\ \bibinfo {pages} {269--303} (\bibinfo {year}
  {1998})}\BibitemShut {NoStop}%
\bibitem [{\citenamefont {Schneider}\ \emph {et~al.}(2007)\citenamefont
  {Schneider}, \citenamefont {Eckhardt},\ and\ \citenamefont
  {Yorke}}]{Schneider2007}%
  \BibitemOpen
  \bibfield  {author} {\bibinfo {author} {\bibfnamefont {T.~M.}\ \bibnamefont
  {Schneider}}, \bibinfo {author} {\bibfnamefont {B.}~\bibnamefont {Eckhardt}},
  \ and\ \bibinfo {author} {\bibfnamefont {J.~A.}\ \bibnamefont {Yorke}},\
  }\bibfield  {title} {\enquote {\bibinfo {title} {{Turbulence transition and
  the edge of chaos in pipe flow}},}\ }\href {\doibase
  10.1103/PhysRevLett.99.034502} {\bibfield  {journal} {\bibinfo  {journal}
  {Phys. Rev. Lett.}\ }\textbf {\bibinfo {volume} {99}},\ \bibinfo {pages}
  {1--4} (\bibinfo {year} {2007})}\BibitemShut {NoStop}%
\bibitem [{\citenamefont {Eckhardt}\ \emph {et~al.}(2007)\citenamefont
  {Eckhardt}, \citenamefont {Schneider}, \citenamefont {Hof},\ and\
  \citenamefont {Westerweel}}]{eckhardt2007turbulence}%
  \BibitemOpen
  \bibfield  {author} {\bibinfo {author} {\bibfnamefont {B.}~\bibnamefont
  {Eckhardt}}, \bibinfo {author} {\bibfnamefont {T.~M.}\ \bibnamefont
  {Schneider}}, \bibinfo {author} {\bibfnamefont {B.}~\bibnamefont {Hof}}, \
  and\ \bibinfo {author} {\bibfnamefont {J.}~\bibnamefont {Westerweel}},\
  }\bibfield  {title} {\enquote {\bibinfo {title} {Turbulence transition in
  pipe flow},}\ }\href@noop {} {\bibfield  {journal} {\bibinfo  {journal}
  {Annu. Rev. Fluid Mech.}\ }\textbf {\bibinfo {volume} {39}},\ \bibinfo
  {pages} {447--468} (\bibinfo {year} {2007})}\BibitemShut {NoStop}%
\bibitem [{\citenamefont {Schneider}\ \emph {et~al.}(2010)\citenamefont
  {Schneider}, \citenamefont {Marinc},\ and\ \citenamefont
  {Eckhardt}}]{Schneider2010}%
  \BibitemOpen
  \bibfield  {author} {\bibinfo {author} {\bibfnamefont {T.~M.}\ \bibnamefont
  {Schneider}}, \bibinfo {author} {\bibfnamefont {D.}~\bibnamefont {Marinc}}, \
  and\ \bibinfo {author} {\bibfnamefont {B.}~\bibnamefont {Eckhardt}},\
  }\bibfield  {title} {\enquote {\bibinfo {title} {{Localized edge states
  nucleate turbulence in extended plane Couette cells}},}\ }\href {\doibase
  10.1017/S0022112009993144} {\bibfield  {journal} {\bibinfo  {journal} {J.
  Fluid Mech.}\ }\textbf {\bibinfo {volume} {646}},\ \bibinfo {pages}
  {441--451} (\bibinfo {year} {2010})}\BibitemShut {NoStop}%
\bibitem [{\citenamefont {Chantry}\ and\ \citenamefont
  {Schneider}(2014)}]{Chantry2014}%
  \BibitemOpen
  \bibfield  {author} {\bibinfo {author} {\bibfnamefont {M.}~\bibnamefont
  {Chantry}}\ and\ \bibinfo {author} {\bibfnamefont {T.~M.}\ \bibnamefont
  {Schneider}},\ }\bibfield  {title} {\enquote {\bibinfo {title} {{Studying
  edge geometry in transiently turbulent shear flows}},}\ }\href {\doibase
  10.1017/jfm.2014.150} {\bibfield  {journal} {\bibinfo  {journal} {J. Fluid
  Mech.}\ }\textbf {\bibinfo {volume} {747}},\ \bibinfo {pages} {506--517}
  (\bibinfo {year} {2014})}\BibitemShut {NoStop}%
\bibitem [{\citenamefont {Grossmann}(2000)}]{grossmann2000onset}%
  \BibitemOpen
  \bibfield  {author} {\bibinfo {author} {\bibfnamefont {S.}~\bibnamefont
  {Grossmann}},\ }\bibfield  {title} {\enquote {\bibinfo {title} {The onset of
  shear flow turbulence},}\ }\href@noop {} {\bibfield  {journal} {\bibinfo
  {journal} {Rev. Mod. Phys.}\ }\textbf {\bibinfo {volume} {72}},\ \bibinfo
  {pages} {603} (\bibinfo {year} {2000})}\BibitemShut {NoStop}%
\bibitem [{\citenamefont {Hof}\ \emph {et~al.}(2003)\citenamefont {Hof},
  \citenamefont {Juel},\ and\ \citenamefont {Mullin}}]{hof2003scaling}%
  \BibitemOpen
  \bibfield  {author} {\bibinfo {author} {\bibfnamefont {B.}~\bibnamefont
  {Hof}}, \bibinfo {author} {\bibfnamefont {A.}~\bibnamefont {Juel}}, \ and\
  \bibinfo {author} {\bibfnamefont {T.}~\bibnamefont {Mullin}},\ }\bibfield
  {title} {\enquote {\bibinfo {title} {Scaling of the turbulence transition
  threshold in a pipe},}\ }\href@noop {} {\bibfield  {journal} {\bibinfo
  {journal} {Phys. Rev. Lett.}\ }\textbf {\bibinfo {volume} {91}},\ \bibinfo
  {pages} {244502} (\bibinfo {year} {2003})}\BibitemShut {NoStop}%
\bibitem [{\citenamefont {Peixinho}\ and\ \citenamefont
  {Mullin}(2007)}]{peixinho2007finite}%
  \BibitemOpen
  \bibfield  {author} {\bibinfo {author} {\bibfnamefont {J.}~\bibnamefont
  {Peixinho}}\ and\ \bibinfo {author} {\bibfnamefont {T.}~\bibnamefont
  {Mullin}},\ }\bibfield  {title} {\enquote {\bibinfo {title} {Finite-amplitude
  thresholds for transition in pipe flow},}\ }\href@noop {} {\bibfield
  {journal} {\bibinfo  {journal} {J. Fluid Mech.}\ }\textbf {\bibinfo {volume}
  {582}},\ \bibinfo {pages} {169--178} (\bibinfo {year} {2007})}\BibitemShut
  {NoStop}%
\bibitem [{\citenamefont {Mullin}(2011)}]{mullin2011experimental}%
  \BibitemOpen
  \bibfield  {author} {\bibinfo {author} {\bibfnamefont {T.}~\bibnamefont
  {Mullin}},\ }\bibfield  {title} {\enquote {\bibinfo {title} {Experimental
  studies of transition to turbulence in a pipe},}\ }\href@noop {} {\bibfield
  {journal} {\bibinfo  {journal} {Annu. Rev. Fluid Mech.}\ }\textbf {\bibinfo
  {volume} {43}},\ \bibinfo {pages} {1--24} (\bibinfo {year}
  {2011})}\BibitemShut {NoStop}%
\bibitem [{\citenamefont {Khalil}(2002)}]{khalil2002nonlinear}%
  \BibitemOpen
  \bibfield  {author} {\bibinfo {author} {\bibfnamefont {H.~K.}\ \bibnamefont
  {Khalil}},\ }\href@noop {} {\emph {\bibinfo {title} {Nonlinear systems}}}\
  (\bibinfo  {publisher} {Upper Saddle River},\ \bibinfo {year}
  {2002})\BibitemShut {NoStop}%
\bibitem [{\citenamefont {Fantuzzi}\ \emph {et~al.}(2016)\citenamefont
  {Fantuzzi}, \citenamefont {Goluskin}, \citenamefont {Huang},\ and\
  \citenamefont {Chernyshenko}}]{fantuzzi2016bounds}%
  \BibitemOpen
  \bibfield  {author} {\bibinfo {author} {\bibfnamefont {G.}~\bibnamefont
  {Fantuzzi}}, \bibinfo {author} {\bibfnamefont {D.}~\bibnamefont {Goluskin}},
  \bibinfo {author} {\bibfnamefont {D.}~\bibnamefont {Huang}}, \ and\ \bibinfo
  {author} {\bibfnamefont {S.~I.}\ \bibnamefont {Chernyshenko}},\ }\bibfield
  {title} {\enquote {\bibinfo {title} {Bounds for deterministic and stochastic
  dynamical systems using sum-of-squares optimization},}\ }\href@noop {}
  {\bibfield  {journal} {\bibinfo  {journal} {SIAM J. Appl. Dyn. Syst.}\
  }\textbf {\bibinfo {volume} {15}},\ \bibinfo {pages} {1962--1988} (\bibinfo
  {year} {2016})}\BibitemShut {NoStop}%
\bibitem [{\citenamefont {Lasagna}\ \emph {et~al.}(2016)\citenamefont
  {Lasagna}, \citenamefont {Huang}, \citenamefont {Tutty},\ and\ \citenamefont
  {Chernyshenko}}]{lasagna2016sum}%
  \BibitemOpen
  \bibfield  {author} {\bibinfo {author} {\bibfnamefont {D.}~\bibnamefont
  {Lasagna}}, \bibinfo {author} {\bibfnamefont {D.}~\bibnamefont {Huang}},
  \bibinfo {author} {\bibfnamefont {O.~R.}\ \bibnamefont {Tutty}}, \ and\
  \bibinfo {author} {\bibfnamefont {S.}~\bibnamefont {Chernyshenko}},\
  }\bibfield  {title} {\enquote {\bibinfo {title} {Sum-of-squares approach to
  feedback control of laminar wake flows},}\ }\href@noop {} {\bibfield
  {journal} {\bibinfo  {journal} {J. Fluid Mech.}\ }\textbf {\bibinfo {volume}
  {809}},\ \bibinfo {pages} {628--663} (\bibinfo {year} {2016})}\BibitemShut
  {NoStop}%
\bibitem [{\citenamefont {Huang}\ \emph {et~al.}(2017)\citenamefont {Huang},
  \citenamefont {Jin}, \citenamefont {Lasagna}, \citenamefont {Chernyshenko},\
  and\ \citenamefont {Tutty}}]{Huang2017}%
  \BibitemOpen
  \bibfield  {author} {\bibinfo {author} {\bibfnamefont {D.}~\bibnamefont
  {Huang}}, \bibinfo {author} {\bibfnamefont {B.}~\bibnamefont {Jin}}, \bibinfo
  {author} {\bibfnamefont {D.}~\bibnamefont {Lasagna}}, \bibinfo {author}
  {\bibfnamefont {S.}~\bibnamefont {Chernyshenko}}, \ and\ \bibinfo {author}
  {\bibfnamefont {O.}~\bibnamefont {Tutty}},\ }\bibfield  {title} {\enquote
  {\bibinfo {title} {Expensive control of long-time averages using sum of
  squares and its application to a laminar wake flow},}\ }\href {\doibase
  10.1109/TCST.2016.2638881} {\bibfield  {journal} {\bibinfo  {journal} {IEEE
  Trans. Control Syst. Technol.}\ }\textbf {\bibinfo {volume} {25}},\ \bibinfo
  {pages} {2073--2086} (\bibinfo {year} {2017})}\BibitemShut {NoStop}%
\bibitem [{\citenamefont {Lakshmi}\ \emph {et~al.}(2020)\citenamefont
  {Lakshmi}, \citenamefont {Fantuzzi}, \citenamefont {Fern{\'a}ndez-Caballero},
  \citenamefont {Hwang},\ and\ \citenamefont
  {Chernyshenko}}]{lakshmi2020finding}%
  \BibitemOpen
  \bibfield  {author} {\bibinfo {author} {\bibfnamefont {M.~V.}\ \bibnamefont
  {Lakshmi}}, \bibinfo {author} {\bibfnamefont {G.}~\bibnamefont {Fantuzzi}},
  \bibinfo {author} {\bibfnamefont {J.~D.}\ \bibnamefont
  {Fern{\'a}ndez-Caballero}}, \bibinfo {author} {\bibfnamefont
  {Y.}~\bibnamefont {Hwang}}, \ and\ \bibinfo {author} {\bibfnamefont {S.~I.}\
  \bibnamefont {Chernyshenko}},\ }\bibfield  {title} {\enquote {\bibinfo
  {title} {Finding extremal periodic orbits with polynomial optimization, with
  application to a nine-mode model of shear flow},}\ }\href@noop {} {\bibfield
  {journal} {\bibinfo  {journal} {SIAM J. Appl. Dyn. Syst.}\ }\textbf {\bibinfo
  {volume} {19}},\ \bibinfo {pages} {763--787} (\bibinfo {year}
  {2020})}\BibitemShut {NoStop}%
\bibitem [{\citenamefont {Zheng}\ \emph {et~al.}(2018)\citenamefont {Zheng},
  \citenamefont {Fantuzzi},\ and\ \citenamefont
  {Papachristodoulou}}]{zheng2018fast}%
  \BibitemOpen
  \bibfield  {author} {\bibinfo {author} {\bibfnamefont {Y.}~\bibnamefont
  {Zheng}}, \bibinfo {author} {\bibfnamefont {G.}~\bibnamefont {Fantuzzi}}, \
  and\ \bibinfo {author} {\bibfnamefont {A.}~\bibnamefont
  {Papachristodoulou}},\ }\bibfield  {title} {\enquote {\bibinfo {title} {Fast
  {ADMM} for sum-of-squares programs using partial orthogonality},}\
  }\href@noop {} {\bibfield  {journal} {\bibinfo  {journal} {IEEE Trans. Autom.
  Control}\ }\textbf {\bibinfo {volume} {64}},\ \bibinfo {pages} {3869--3876}
  (\bibinfo {year} {2018})}\BibitemShut {NoStop}%
\bibitem [{\citenamefont {Kerswell}\ \emph {et~al.}(2014)\citenamefont
  {Kerswell}, \citenamefont {Pringle},\ and\ \citenamefont
  {Willis}}]{Kerswell2014}%
  \BibitemOpen
  \bibfield  {author} {\bibinfo {author} {\bibfnamefont {R.~R.}\ \bibnamefont
  {Kerswell}}, \bibinfo {author} {\bibfnamefont {C.~C.}\ \bibnamefont
  {Pringle}}, \ and\ \bibinfo {author} {\bibfnamefont {A.~P.}\ \bibnamefont
  {Willis}},\ }\bibfield  {title} {\enquote {\bibinfo {title} {{An optimization
  approach for analysing nonlinear stability with transition to turbulence in
  fluids as an exemplar}},}\ }\href@noop {} {\bibfield  {journal} {\bibinfo
  {journal} {Rep. Prog. Phys.}\ }\textbf {\bibinfo {volume} {77}} (\bibinfo
  {year} {2014})}\BibitemShut {NoStop}%
\bibitem [{\citenamefont {Kerswell}(2018)}]{Kerswell2018}%
  \BibitemOpen
  \bibfield  {author} {\bibinfo {author} {\bibfnamefont {R.~R.}\ \bibnamefont
  {Kerswell}},\ }\bibfield  {title} {\enquote {\bibinfo {title} {Nonlinear
  nonmodal stability theory},}\ }\href {\doibase
  10.1146/annurev-fluid-122316-045042} {\bibfield  {journal} {\bibinfo
  {journal} {Annu. Rev. Fluid Mech.}\ }\textbf {\bibinfo {volume} {50}},\
  \bibinfo {pages} {319--345} (\bibinfo {year} {2018})}\BibitemShut {NoStop}%
\bibitem [{\citenamefont {Pringle}\ and\ \citenamefont
  {Kerswell}(2010)}]{Pringle2010}%
  \BibitemOpen
  \bibfield  {author} {\bibinfo {author} {\bibfnamefont {C.~C.~T.}\
  \bibnamefont {Pringle}}\ and\ \bibinfo {author} {\bibfnamefont {R.~R.}\
  \bibnamefont {Kerswell}},\ }\bibfield  {title} {\enquote {\bibinfo {title}
  {{Using nonlinear transient growth to construct the minimal seed for shear
  flow turbulence}},}\ }\href {\doibase 10.1103/PhysRevLett.105.154502}
  {\bibfield  {journal} {\bibinfo  {journal} {Phys. Rev. Lett.}\ }\textbf
  {\bibinfo {volume} {105}},\ \bibinfo {pages} {1--4} (\bibinfo {year}
  {2010})}\BibitemShut {NoStop}%
\bibitem [{\citenamefont {Duguet}\ \emph {et~al.}(2010)\citenamefont {Duguet},
  \citenamefont {Brandt},\ and\ \citenamefont {Larsson}}]{duguet2010towards}%
  \BibitemOpen
  \bibfield  {author} {\bibinfo {author} {\bibfnamefont {Y.}~\bibnamefont
  {Duguet}}, \bibinfo {author} {\bibfnamefont {L.}~\bibnamefont {Brandt}}, \
  and\ \bibinfo {author} {\bibfnamefont {B.~R.~J.}\ \bibnamefont {Larsson}},\
  }\bibfield  {title} {\enquote {\bibinfo {title} {Towards minimal
  perturbations in transitional plane {C}ouette flow},}\ }\href@noop {}
  {\bibfield  {journal} {\bibinfo  {journal} {Phys. Rev. E}\ }\textbf {\bibinfo
  {volume} {82}},\ \bibinfo {pages} {026316} (\bibinfo {year}
  {2010})}\BibitemShut {NoStop}%
\bibitem [{\citenamefont {Pringle}\ \emph {et~al.}(2012)\citenamefont
  {Pringle}, \citenamefont {Willis},\ and\ \citenamefont
  {Kerswell}}]{Pringle2012}%
  \BibitemOpen
  \bibfield  {author} {\bibinfo {author} {\bibfnamefont {C.~C.~T.}\
  \bibnamefont {Pringle}}, \bibinfo {author} {\bibfnamefont {A.~P.}\
  \bibnamefont {Willis}}, \ and\ \bibinfo {author} {\bibfnamefont {R.~R.}\
  \bibnamefont {Kerswell}},\ }\bibfield  {title} {\enquote {\bibinfo {title}
  {Minimal seeds for shear flow turbulence: using nonlinear transient growth to
  touch the edge of chaos},}\ }\href {\doibase 10.1017/jfm.2012.192} {\bibfield
   {journal} {\bibinfo  {journal} {J. Fluid Mech.}\ }\textbf {\bibinfo {volume}
  {702}},\ \bibinfo {pages} {415--443} (\bibinfo {year} {2012})}\BibitemShut
  {NoStop}%
\bibitem [{\citenamefont {Rabin}\ \emph {et~al.}(2012)\citenamefont {Rabin},
  \citenamefont {Caulfield},\ and\ \citenamefont
  {Kerswell}}]{rabin2012triggering}%
  \BibitemOpen
  \bibfield  {author} {\bibinfo {author} {\bibfnamefont {S.~M.~E.}\
  \bibnamefont {Rabin}}, \bibinfo {author} {\bibfnamefont {C.~P.}\ \bibnamefont
  {Caulfield}}, \ and\ \bibinfo {author} {\bibfnamefont {R.~R.}\ \bibnamefont
  {Kerswell}},\ }\bibfield  {title} {\enquote {\bibinfo {title} {Triggering
  turbulence efficiently in plane {C}ouette flow},}\ }\href@noop {} {\bibfield
  {journal} {\bibinfo  {journal} {J. Fluid Mech.}\ }\textbf {\bibinfo {volume}
  {712}},\ \bibinfo {pages} {244--272} (\bibinfo {year} {2012})}\BibitemShut
  {NoStop}%
\bibitem [{\citenamefont {Duguet}\ \emph {et~al.}(2013)\citenamefont {Duguet},
  \citenamefont {Monokrousos}, \citenamefont {Brandt},\ and\ \citenamefont
  {Henningson}}]{Duguet2013}%
  \BibitemOpen
  \bibfield  {author} {\bibinfo {author} {\bibfnamefont {Y.}~\bibnamefont
  {Duguet}}, \bibinfo {author} {\bibfnamefont {A.}~\bibnamefont {Monokrousos}},
  \bibinfo {author} {\bibfnamefont {L.}~\bibnamefont {Brandt}}, \ and\ \bibinfo
  {author} {\bibfnamefont {D.~S.}\ \bibnamefont {Henningson}},\ }\bibfield
  {title} {\enquote {\bibinfo {title} {Minimal transition thresholds in plane
  {C}ouette flow},}\ }\href@noop {} {\bibfield  {journal} {\bibinfo  {journal}
  {Phys. Fluids}\ }\textbf {\bibinfo {volume} {25}},\ \bibinfo {pages} {084103}
  (\bibinfo {year} {2013})}\BibitemShut {NoStop}%
\bibitem [{\citenamefont {Gebhardt}\ and\ \citenamefont
  {Grossmann}(1994)}]{gebhardt1994chaos}%
  \BibitemOpen
  \bibfield  {author} {\bibinfo {author} {\bibfnamefont {T.}~\bibnamefont
  {Gebhardt}}\ and\ \bibinfo {author} {\bibfnamefont {S.}~\bibnamefont
  {Grossmann}},\ }\bibfield  {title} {\enquote {\bibinfo {title} {Chaos
  transition despite linear stability},}\ }\href@noop {} {\bibfield  {journal}
  {\bibinfo  {journal} {Phys. Rev. E}\ }\textbf {\bibinfo {volume} {50}},\
  \bibinfo {pages} {3705} (\bibinfo {year} {1994})}\BibitemShut {NoStop}%
\bibitem [{\citenamefont {Baggett}\ \emph {et~al.}(1995)\citenamefont
  {Baggett}, \citenamefont {Driscoll},\ and\ \citenamefont
  {Trefethen}}]{baggett1995mostly}%
  \BibitemOpen
  \bibfield  {author} {\bibinfo {author} {\bibfnamefont {J.~S.}\ \bibnamefont
  {Baggett}}, \bibinfo {author} {\bibfnamefont {T.~A.}\ \bibnamefont
  {Driscoll}}, \ and\ \bibinfo {author} {\bibfnamefont {L.~N.}\ \bibnamefont
  {Trefethen}},\ }\bibfield  {title} {\enquote {\bibinfo {title} {A mostly
  linear model of transition to turbulence},}\ }\href@noop {} {\bibfield
  {journal} {\bibinfo  {journal} {Phys. Fluids}\ }\textbf {\bibinfo {volume}
  {7}},\ \bibinfo {pages} {833--838} (\bibinfo {year} {1995})}\BibitemShut
  {NoStop}%
\bibitem [{\citenamefont {Moehlis}\ \emph {et~al.}(2004)\citenamefont
  {Moehlis}, \citenamefont {Faisst},\ and\ \citenamefont
  {Eckhardt}}]{Moehlis2004}%
  \BibitemOpen
  \bibfield  {author} {\bibinfo {author} {\bibfnamefont {J.}~\bibnamefont
  {Moehlis}}, \bibinfo {author} {\bibfnamefont {H.}~\bibnamefont {Faisst}}, \
  and\ \bibinfo {author} {\bibfnamefont {B.}~\bibnamefont {Eckhardt}},\
  }\bibfield  {title} {\enquote {\bibinfo {title} {{A low-dimensional model for
  turbulent shear flows}},}\ }\href {\doibase 10.1088/1367-2630/6/1/056}
  {\bibfield  {journal} {\bibinfo  {journal} {New J. Phys.}\ }\textbf {\bibinfo
  {volume} {6}},\ \bibinfo {pages} {1--17} (\bibinfo {year}
  {2004})}\BibitemShut {NoStop}%
\bibitem [{\citenamefont {Moehlis}\ \emph {et~al.}(2005)\citenamefont
  {Moehlis}, \citenamefont {Faisst},\ and\ \citenamefont
  {Eckhardt}}]{Moehlis2005}%
  \BibitemOpen
  \bibfield  {author} {\bibinfo {author} {\bibfnamefont {J.}~\bibnamefont
  {Moehlis}}, \bibinfo {author} {\bibfnamefont {H.}~\bibnamefont {Faisst}}, \
  and\ \bibinfo {author} {\bibfnamefont {B.}~\bibnamefont {Eckhardt}},\
  }\bibfield  {title} {\enquote {\bibinfo {title} {Periodic orbits and chaotic
  sets in a low-dimensional model for shear flows},}\ }\href {\doibase
  10.1137/040606144} {\bibfield  {journal} {\bibinfo  {journal} {SIAM J. Appl.
  Dyn. Syst.}\ }\textbf {\bibinfo {volume} {4}},\ \bibinfo {pages} {352--376}
  (\bibinfo {year} {2005})}\BibitemShut {NoStop}%
\bibitem [{\citenamefont {Lebovitz}\ and\ \citenamefont
  {Mariotti}(2013)}]{lebovitz2013edges}%
  \BibitemOpen
  \bibfield  {author} {\bibinfo {author} {\bibfnamefont {N.}~\bibnamefont
  {Lebovitz}}\ and\ \bibinfo {author} {\bibfnamefont {G.}~\bibnamefont
  {Mariotti}},\ }\bibfield  {title} {\enquote {\bibinfo {title} {Edges in
  models of shear flow},}\ }\href@noop {} {\bibfield  {journal} {\bibinfo
  {journal} {J. Fluid Mech.}\ }\textbf {\bibinfo {volume} {721}},\ \bibinfo
  {pages} {386--402} (\bibinfo {year} {2013})}\BibitemShut {NoStop}%
\bibitem [{\citenamefont {Joglekar}\ \emph {et~al.}(2015)\citenamefont
  {Joglekar}, \citenamefont {Feudel},\ and\ \citenamefont
  {Yorke}}]{Joglekar2015}%
  \BibitemOpen
  \bibfield  {author} {\bibinfo {author} {\bibfnamefont {M.}~\bibnamefont
  {Joglekar}}, \bibinfo {author} {\bibfnamefont {U.}~\bibnamefont {Feudel}}, \
  and\ \bibinfo {author} {\bibfnamefont {J.~A.}\ \bibnamefont {Yorke}},\
  }\bibfield  {title} {\enquote {\bibinfo {title} {Geometry of the edge of
  chaos in a low-dimensional turbulent shear flow model},}\ }\href@noop {}
  {\bibfield  {journal} {\bibinfo  {journal} {Phys. Rev. E}\ }\textbf {\bibinfo
  {volume} {91}},\ \bibinfo {pages} {052903} (\bibinfo {year}
  {2015})}\BibitemShut {NoStop}%
\bibitem [{\citenamefont {Kim}\ and\ \citenamefont {Moehlis}(2008)}]{Kim2008}%
  \BibitemOpen
  \bibfield  {author} {\bibinfo {author} {\bibfnamefont {L.}~\bibnamefont
  {Kim}}\ and\ \bibinfo {author} {\bibfnamefont {J.}~\bibnamefont {Moehlis}},\
  }\bibfield  {title} {\enquote {\bibinfo {title} {{Characterizing the edge of
  chaos for a shear flow model}},}\ }\href {\doibase
  10.1103/PhysRevE.78.036315} {\bibfield  {journal} {\bibinfo  {journal} {Phys.
  Rev. E}\ }\textbf {\bibinfo {volume} {78}},\ \bibinfo {pages} {1--9}
  (\bibinfo {year} {2008})}\BibitemShut {NoStop}%
\bibitem [{\citenamefont {Bamieh}\ and\ \citenamefont
  {Dahleh}(2001)}]{Bamieh2001}%
  \BibitemOpen
  \bibfield  {author} {\bibinfo {author} {\bibfnamefont {B.}~\bibnamefont
  {Bamieh}}\ and\ \bibinfo {author} {\bibfnamefont {M.}~\bibnamefont
  {Dahleh}},\ }\bibfield  {title} {\enquote {\bibinfo {title} {{Energy
  amplification in channel flows with stochastic excitation}},}\ }\href
  {\doibase 10.1063/1.1398044} {\bibfield  {journal} {\bibinfo  {journal}
  {Phys. Fluids}\ }\textbf {\bibinfo {volume} {13}},\ \bibinfo {pages}
  {3258--3269} (\bibinfo {year} {2001})}\BibitemShut {NoStop}%
\bibitem [{\citenamefont {Ahmadi}\ \emph {et~al.}(2019)\citenamefont {Ahmadi},
  \citenamefont {Valmorbida}, \citenamefont {Gayme},\ and\ \citenamefont
  {Papachristodoulou}}]{ahmadi2019framework}%
  \BibitemOpen
  \bibfield  {author} {\bibinfo {author} {\bibfnamefont {M.}~\bibnamefont
  {Ahmadi}}, \bibinfo {author} {\bibfnamefont {G.}~\bibnamefont {Valmorbida}},
  \bibinfo {author} {\bibfnamefont {D.}~\bibnamefont {Gayme}}, \ and\ \bibinfo
  {author} {\bibfnamefont {A.}~\bibnamefont {Papachristodoulou}},\ }\bibfield
  {title} {\enquote {\bibinfo {title} {A framework for input--output analysis
  of wall-bounded shear flows},}\ }\href@noop {} {\bibfield  {journal}
  {\bibinfo  {journal} {J. Fluid Mech.}\ }\textbf {\bibinfo {volume} {873}},\
  \bibinfo {pages} {742--785} (\bibinfo {year} {2019})}\BibitemShut {NoStop}%
\bibitem [{\citenamefont {Jovanovi{\'{c}}}\ and\ \citenamefont
  {Bamieh}(2005)}]{Jovanovic2005}%
  \BibitemOpen
  \bibfield  {author} {\bibinfo {author} {\bibfnamefont {M.~R.}\ \bibnamefont
  {Jovanovi{\'{c}}}}\ and\ \bibinfo {author} {\bibfnamefont {B.}~\bibnamefont
  {Bamieh}},\ }\bibfield  {title} {\enquote {\bibinfo {title} {{Componentwise
  energy amplification in channel flows}},}\ }\href {\doibase
  10.1017/S0022112005004295} {\bibfield  {journal} {\bibinfo  {journal} {J.
  Fluid Mech.}\ }\textbf {\bibinfo {volume} {534}},\ \bibinfo {pages}
  {145--183} (\bibinfo {year} {2005})}\BibitemShut {NoStop}%
\bibitem [{\citenamefont {Mc{K}eon}\ and\ \citenamefont
  {Sharma}(2010)}]{McKeon2010}%
  \BibitemOpen
  \bibfield  {author} {\bibinfo {author} {\bibfnamefont {B.~J.}\ \bibnamefont
  {Mc{K}eon}}\ and\ \bibinfo {author} {\bibfnamefont {A.~S.}\ \bibnamefont
  {Sharma}},\ }\bibfield  {title} {\enquote {\bibinfo {title} {{A
  critical-layer framework for turbulent pipe flow}},}\ }\href {\doibase
  10.1017/S002211201000176X} {\bibfield  {journal} {\bibinfo  {journal} {J.
  Fluid Mech.}\ }\textbf {\bibinfo {volume} {658}},\ \bibinfo {pages}
  {336--382} (\bibinfo {year} {2010})}\BibitemShut {NoStop}%
\bibitem [{\citenamefont {Jovanovi{\'c}}(2021)}]{jovanovic2020bypass}%
  \BibitemOpen
  \bibfield  {author} {\bibinfo {author} {\bibfnamefont {M.~R.}\ \bibnamefont
  {Jovanovi{\'c}}},\ }\bibfield  {title} {\enquote {\bibinfo {title} {From
  bypass transition to flow control and data-driven turbulence modeling: {A}n
  input-output viewpoint},}\ }\href {\doibase
  10.1146/annurev-fluid-010719-060244} {\bibfield  {journal} {\bibinfo
  {journal} {Annu. Rev. Fluid Mech.}\ }\textbf {\bibinfo {volume} {53}}
  (\bibinfo {year} {2021}),\ 10.1146/annurev-fluid-010719-060244}\BibitemShut
  {NoStop}%
\bibitem [{\citenamefont {Sharma}\ and\ \citenamefont
  {Mc{K}eon}(2013)}]{Sharma2013}%
  \BibitemOpen
  \bibfield  {author} {\bibinfo {author} {\bibfnamefont {A.~S.}\ \bibnamefont
  {Sharma}}\ and\ \bibinfo {author} {\bibfnamefont {B.~J.}\ \bibnamefont
  {Mc{K}eon}},\ }\bibfield  {title} {\enquote {\bibinfo {title} {{On coherent
  structure in wall turbulence}},}\ }\href {\doibase 10.1017/jfm.2013.286}
  {\bibfield  {journal} {\bibinfo  {journal} {J. Fluid Mech.}\ }\textbf
  {\bibinfo {volume} {728}},\ \bibinfo {pages} {196--238} (\bibinfo {year}
  {2013})}\BibitemShut {NoStop}%
\bibitem [{\citenamefont {McKeon}\ \emph {et~al.}(2013)\citenamefont {McKeon},
  \citenamefont {Sharma},\ and\ \citenamefont
  {Jacobi}}]{mckeon2013experimental}%
  \BibitemOpen
  \bibfield  {author} {\bibinfo {author} {\bibfnamefont {B.~J.}\ \bibnamefont
  {McKeon}}, \bibinfo {author} {\bibfnamefont {A.~S.}\ \bibnamefont {Sharma}},
  \ and\ \bibinfo {author} {\bibfnamefont {I.}~\bibnamefont {Jacobi}},\
  }\bibfield  {title} {\enquote {\bibinfo {title} {Experimental manipulation of
  wall turbulence: a systems approach},}\ }\href@noop {} {\bibfield  {journal}
  {\bibinfo  {journal} {Phys. Fluids}\ }\textbf {\bibinfo {volume} {25}},\
  \bibinfo {pages} {031301} (\bibinfo {year} {2013})}\BibitemShut {NoStop}%
\bibitem [{\citenamefont {Mc{K}eon}(2017)}]{mckeon2017engine}%
  \BibitemOpen
  \bibfield  {author} {\bibinfo {author} {\bibfnamefont {B.}~\bibnamefont
  {Mc{K}eon}},\ }\bibfield  {title} {\enquote {\bibinfo {title} {The engine
  behind (wall) turbulence: perspectives on scale interactions},}\ }\href@noop
  {} {\bibfield  {journal} {\bibinfo  {journal} {J. Fluid Mech.}\ }\textbf
  {\bibinfo {volume} {817}},\ \bibinfo {pages} {P1} (\bibinfo {year}
  {2017})}\BibitemShut {NoStop}%
\bibitem [{\citenamefont {Liu}\ and\ \citenamefont
  {Gayme}(2019)}]{liu2019vorticity}%
  \BibitemOpen
  \bibfield  {author} {\bibinfo {author} {\bibfnamefont {C.}~\bibnamefont
  {Liu}}\ and\ \bibinfo {author} {\bibfnamefont {D.~F.}\ \bibnamefont
  {Gayme}},\ }\bibfield  {title} {\enquote {\bibinfo {title} {Convective
  velocities of vorticity fluctuations in turbulent channel flows: an
  input-output approach},}\ }in\ \href@noop {} {\emph {\bibinfo {booktitle}
  {Proceedings of the Eleventh International Symposium on Turbulence and Shear
  Flow Phenomenon}}}\ (\bibinfo  {publisher} {Southampton, UK},\ \bibinfo
  {year} {2019})\BibitemShut {NoStop}%
\bibitem [{\citenamefont {Liu}\ and\ \citenamefont
  {Gayme}(2020)}]{liu2019input}%
  \BibitemOpen
  \bibfield  {author} {\bibinfo {author} {\bibfnamefont {C.}~\bibnamefont
  {Liu}}\ and\ \bibinfo {author} {\bibfnamefont {D.~F.}\ \bibnamefont
  {Gayme}},\ }\bibfield  {title} {\enquote {\bibinfo {title} {An input-output
  based analysis of convective velocity in turbulent channels},}\ }\href@noop
  {} {\bibfield  {journal} {\bibinfo  {journal} {J. Fluid Mech.}\ }\textbf
  {\bibinfo {volume} {888}},\ \bibinfo {pages} {A32} (\bibinfo {year}
  {2020})}\BibitemShut {NoStop}%
\bibitem [{\citenamefont {Kalman}(1963)}]{kalman1963lyapunov}%
  \BibitemOpen
  \bibfield  {author} {\bibinfo {author} {\bibfnamefont {R.~E.}\ \bibnamefont
  {Kalman}},\ }\bibfield  {title} {\enquote {\bibinfo {title} {Lyapunov
  functions for the problem of {L}ur'e in automatic control},}\ }\href@noop {}
  {\bibfield  {journal} {\bibinfo  {journal} {Proc. Natl. Acad. Sci.}\ }\textbf
  {\bibinfo {volume} {49}},\ \bibinfo {pages} {201} (\bibinfo {year}
  {1963})}\BibitemShut {NoStop}%
\bibitem [{\citenamefont {Boyd}\ \emph {et~al.}(1994)\citenamefont {Boyd},
  \citenamefont {El~Ghaoui}, \citenamefont {Feron},\ and\ \citenamefont
  {Balakrishnan}}]{boyd1994linear}%
  \BibitemOpen
  \bibfield  {author} {\bibinfo {author} {\bibfnamefont {S.}~\bibnamefont
  {Boyd}}, \bibinfo {author} {\bibfnamefont {L.}~\bibnamefont {El~Ghaoui}},
  \bibinfo {author} {\bibfnamefont {E.}~\bibnamefont {Feron}}, \ and\ \bibinfo
  {author} {\bibfnamefont {V.}~\bibnamefont {Balakrishnan}},\ }\href@noop {}
  {\emph {\bibinfo {title} {Linear matrix inequalities in system and control
  theory}}},\ Vol.~\bibinfo {volume} {15}\ (\bibinfo  {publisher} {{SIAM}},\
  \bibinfo {year} {1994})\BibitemShut {NoStop}%
\bibitem [{\citenamefont {Li}\ \emph {et~al.}(2007)\citenamefont {Li},
  \citenamefont {Heath},\ and\ \citenamefont {Lennox}}]{li2007improved}%
  \BibitemOpen
  \bibfield  {author} {\bibinfo {author} {\bibfnamefont {G.}~\bibnamefont
  {Li}}, \bibinfo {author} {\bibfnamefont {W.~P.}\ \bibnamefont {Heath}}, \
  and\ \bibinfo {author} {\bibfnamefont {B.}~\bibnamefont {Lennox}},\
  }\bibfield  {title} {\enquote {\bibinfo {title} {An improved stability
  criterion for a class of {L}ur’e systems},}\ }in\ \href@noop {} {\emph
  {\bibinfo {booktitle} {2007 46th IEEE Conference on Decision and Control}}}\
  (\bibinfo {organization} {IEEE},\ \bibinfo {year} {2007})\ pp.\ \bibinfo
  {pages} {4483--4488}\BibitemShut {NoStop}%
\bibitem [{\citenamefont {Li}\ \emph {et~al.}(2008)\citenamefont {Li},
  \citenamefont {Heath},\ and\ \citenamefont {Lennox}}]{li2008concise}%
  \BibitemOpen
  \bibfield  {author} {\bibinfo {author} {\bibfnamefont {G.}~\bibnamefont
  {Li}}, \bibinfo {author} {\bibfnamefont {W.~P.}\ \bibnamefont {Heath}}, \
  and\ \bibinfo {author} {\bibfnamefont {B.}~\bibnamefont {Lennox}},\
  }\bibfield  {title} {\enquote {\bibinfo {title} {Concise stability conditions
  for systems with static nonlinear feedback expressed by a quadratic
  program},}\ }\href@noop {} {\bibfield  {journal} {\bibinfo  {journal} {IET
  Control Theory Appl.}\ }\textbf {\bibinfo {volume} {2}},\ \bibinfo {pages}
  {554--563} (\bibinfo {year} {2008})}\BibitemShut {NoStop}%
\bibitem [{\citenamefont {Popov}(1961)}]{popov1961absolute}%
  \BibitemOpen
  \bibfield  {author} {\bibinfo {author} {\bibfnamefont {V.~M.}\ \bibnamefont
  {Popov}},\ }\bibfield  {title} {\enquote {\bibinfo {title} {Absolute
  stability of nonlinear systems of automatic control},}\ }\href@noop {}
  {\bibfield  {journal} {\bibinfo  {journal} {Autom. Remote Control}\ }\textbf
  {\bibinfo {volume} {22}},\ \bibinfo {pages} {857--875} (\bibinfo {year}
  {1961})}\BibitemShut {NoStop}%
\bibitem [{\citenamefont {Zames}(1966)}]{zames1966input}%
  \BibitemOpen
  \bibfield  {author} {\bibinfo {author} {\bibfnamefont {G.}~\bibnamefont
  {Zames}},\ }\bibfield  {title} {\enquote {\bibinfo {title} {On the
  input-output stability of time-varying nonlinear feedback systems--part {II}:
  Conditions involving circles in the frequency plane and sector
  nonlinearities},}\ }\href@noop {} {\bibfield  {journal} {\bibinfo  {journal}
  {IEEE Trans. Autom. Control}\ }\textbf {\bibinfo {volume} {11}},\ \bibinfo
  {pages} {465--476} (\bibinfo {year} {1966})}\BibitemShut {NoStop}%
\bibitem [{\citenamefont {van~der Schaft}(2000)}]{van2000l2}%
  \BibitemOpen
  \bibfield  {author} {\bibinfo {author} {\bibfnamefont {A.~J.}\ \bibnamefont
  {van~der Schaft}},\ }\href@noop {} {\emph {\bibinfo {title} {$L_2$-gain and
  passivity techniques in nonlinear control}}},\ Vol.~\bibinfo {volume} {2}\
  (\bibinfo  {publisher} {Springer},\ \bibinfo {year} {2000})\BibitemShut
  {NoStop}%
\bibitem [{\citenamefont {Ortega}\ \emph {et~al.}(2013)\citenamefont {Ortega},
  \citenamefont {Perez}, \citenamefont {Nicklasson},\ and\ \citenamefont
  {Sira-Ramirez}}]{ortega2013passivity}%
  \BibitemOpen
  \bibfield  {author} {\bibinfo {author} {\bibfnamefont {R.}~\bibnamefont
  {Ortega}}, \bibinfo {author} {\bibfnamefont {J.~A.~L.}\ \bibnamefont
  {Perez}}, \bibinfo {author} {\bibfnamefont {P.~J.}\ \bibnamefont
  {Nicklasson}}, \ and\ \bibinfo {author} {\bibfnamefont {H.~J.}\ \bibnamefont
  {Sira-Ramirez}},\ }\href@noop {} {\emph {\bibinfo {title} {Passivity-based
  control of {E}uler-{L}agrange systems: mechanical, electrical and
  electromechanical applications}}}\ (\bibinfo  {publisher} {Springer},\
  \bibinfo {year} {2013})\BibitemShut {NoStop}%
\bibitem [{\citenamefont {Sharma}\ \emph {et~al.}(2011)\citenamefont {Sharma},
  \citenamefont {Morrison}, \citenamefont {McKeon}, \citenamefont {Limebeer},
  \citenamefont {Koberg},\ and\ \citenamefont
  {Sherwin}}]{sharma2011relaminarisation}%
  \BibitemOpen
  \bibfield  {author} {\bibinfo {author} {\bibfnamefont {A.}~\bibnamefont
  {Sharma}}, \bibinfo {author} {\bibfnamefont {J.}~\bibnamefont {Morrison}},
  \bibinfo {author} {\bibfnamefont {B.}~\bibnamefont {McKeon}}, \bibinfo
  {author} {\bibfnamefont {D.}~\bibnamefont {Limebeer}}, \bibinfo {author}
  {\bibfnamefont {W.}~\bibnamefont {Koberg}}, \ and\ \bibinfo {author}
  {\bibfnamefont {S.}~\bibnamefont {Sherwin}},\ }\bibfield  {title} {\enquote
  {\bibinfo {title} {Relaminarisation of {R}e$_\tau$= 100 channel flow with
  globally stabilising linear feedback control},}\ }\href@noop {} {\bibfield
  {journal} {\bibinfo  {journal} {Phys. Fluids}\ }\textbf {\bibinfo {volume}
  {23}},\ \bibinfo {pages} {125105} (\bibinfo {year} {2011})}\BibitemShut
  {NoStop}%
\bibitem [{\citenamefont {Damaren}(2016)}]{Damaren2016}%
  \BibitemOpen
  \bibfield  {author} {\bibinfo {author} {\bibfnamefont {C.~J.}\ \bibnamefont
  {Damaren}},\ }\bibfield  {title} {\enquote {\bibinfo {title}
  {Laminar–turbulent transition control using passivity analysis of the
  {Orr–Sommerfeld} equation},}\ }\href {\doibase 10.2514/1.g001763}
  {\bibfield  {journal} {\bibinfo  {journal} {J. Guid. Control Dyn.}\ }\textbf
  {\bibinfo {volume} {39}},\ \bibinfo {pages} {1602--1613} (\bibinfo {year}
  {2016})}\BibitemShut {NoStop}%
\bibitem [{\citenamefont {Damaren}(2018)}]{Damaren2018}%
  \BibitemOpen
  \bibfield  {author} {\bibinfo {author} {\bibfnamefont {C.~J.}\ \bibnamefont
  {Damaren}},\ }\bibfield  {title} {\enquote {\bibinfo {title} {Transition
  control of the {B}lasius boundary layer using passivity},}\ }\href@noop {}
  {\bibfield  {journal} {\bibinfo  {journal} {Aerosp. Syst.}\ }\textbf
  {\bibinfo {volume} {2}},\ \bibinfo {pages} {21--31} (\bibinfo {year}
  {2018})}\BibitemShut {NoStop}%
\bibitem [{\citenamefont {Heins}\ \emph {et~al.}(2016)\citenamefont {Heins},
  \citenamefont {Jones},\ and\ \citenamefont {Sharma}}]{heins2016passivity}%
  \BibitemOpen
  \bibfield  {author} {\bibinfo {author} {\bibfnamefont {P.~H.}\ \bibnamefont
  {Heins}}, \bibinfo {author} {\bibfnamefont {B.~L.}\ \bibnamefont {Jones}}, \
  and\ \bibinfo {author} {\bibfnamefont {A.~S.}\ \bibnamefont {Sharma}},\
  }\bibfield  {title} {\enquote {\bibinfo {title} {Passivity-based
  output-feedback control of turbulent channel flow},}\ }\href@noop {}
  {\bibfield  {journal} {\bibinfo  {journal} {Automatica}\ }\textbf {\bibinfo
  {volume} {69}},\ \bibinfo {pages} {348--355} (\bibinfo {year}
  {2016})}\BibitemShut {NoStop}%
\bibitem [{\citenamefont {Park}(1997)}]{park1997revisited}%
  \BibitemOpen
  \bibfield  {author} {\bibinfo {author} {\bibfnamefont {P.}~\bibnamefont
  {Park}},\ }\bibfield  {title} {\enquote {\bibinfo {title} {A revisited
  {P}opov criterion for nonlinear {L}ur'e systems with sector-restrictions},}\
  }\href@noop {} {\bibfield  {journal} {\bibinfo  {journal} {Int. J. Control}\
  }\textbf {\bibinfo {volume} {68}},\ \bibinfo {pages} {461--470} (\bibinfo
  {year} {1997})}\BibitemShut {NoStop}%
\bibitem [{\citenamefont {Park}\ \emph {et~al.}(2019)\citenamefont {Park},
  \citenamefont {Lee},\ and\ \citenamefont {Park}}]{park2019less}%
  \BibitemOpen
  \bibfield  {author} {\bibinfo {author} {\bibfnamefont {J.}~\bibnamefont
  {Park}}, \bibinfo {author} {\bibfnamefont {S.~Y.}\ \bibnamefont {Lee}}, \
  and\ \bibinfo {author} {\bibfnamefont {P.}~\bibnamefont {Park}},\ }\bibfield
  {title} {\enquote {\bibinfo {title} {A less conservative stability criterion
  for discrete-time {L}ur'e systems with sector and slope restrictions},}\
  }\href@noop {} {\bibfield  {journal} {\bibinfo  {journal} {IEEE Trans. Autom.
  Control}\ }\textbf {\bibinfo {volume} {64}},\ \bibinfo {pages} {4391--4395}
  (\bibinfo {year} {2019})}\BibitemShut {NoStop}%
\bibitem [{\citenamefont {Weissenberger}(1968)}]{weissenberger1968application}%
  \BibitemOpen
  \bibfield  {author} {\bibinfo {author} {\bibfnamefont {S.}~\bibnamefont
  {Weissenberger}},\ }\bibfield  {title} {\enquote {\bibinfo {title}
  {Application of results from the absolute stability problem to the
  computation of finite stability domains},}\ }\href@noop {} {\bibfield
  {journal} {\bibinfo  {journal} {IEEE Trans. Autom. Control}\ }\textbf
  {\bibinfo {volume} {13}},\ \bibinfo {pages} {124--125} (\bibinfo {year}
  {1968})}\BibitemShut {NoStop}%
\bibitem [{\citenamefont {Hindi}\ and\ \citenamefont
  {Boyd}(1998)}]{hindi1998analysis}%
  \BibitemOpen
  \bibfield  {author} {\bibinfo {author} {\bibfnamefont {H.}~\bibnamefont
  {Hindi}}\ and\ \bibinfo {author} {\bibfnamefont {S.}~\bibnamefont {Boyd}},\
  }\bibfield  {title} {\enquote {\bibinfo {title} {Analysis of linear systems
  with saturation using convex optimization},}\ }in\ \href@noop {} {\emph
  {\bibinfo {booktitle} {Proceedings of the 37th IEEE Conference on Decision
  and Control}}},\ Vol.~\bibinfo {volume} {1}\ (\bibinfo {organization}
  {IEEE},\ \bibinfo {year} {1998})\ pp.\ \bibinfo {pages}
  {903--908}\BibitemShut {NoStop}%
\bibitem [{\citenamefont {Valmorbida}\ \emph {et~al.}(2018)\citenamefont
  {Valmorbida}, \citenamefont {Drummond},\ and\ \citenamefont
  {Duncan}}]{valmorbida2018regional}%
  \BibitemOpen
  \bibfield  {author} {\bibinfo {author} {\bibfnamefont {G.}~\bibnamefont
  {Valmorbida}}, \bibinfo {author} {\bibfnamefont {R.}~\bibnamefont
  {Drummond}}, \ and\ \bibinfo {author} {\bibfnamefont {S.~R.}\ \bibnamefont
  {Duncan}},\ }\bibfield  {title} {\enquote {\bibinfo {title} {Regional
  analysis of slope-restricted {L}urie systems},}\ }\href@noop {} {\bibfield
  {journal} {\bibinfo  {journal} {IEEE Trans. Autom. Control}\ }\textbf
  {\bibinfo {volume} {64}},\ \bibinfo {pages} {1201--1208} (\bibinfo {year}
  {2018})}\BibitemShut {NoStop}%
\bibitem [{\citenamefont {Kalur}\ \emph
  {et~al.}(2020{\natexlab{a}})\citenamefont {Kalur}, \citenamefont {Seiler},\
  and\ \citenamefont {Hemati}}]{kalur2020stability}%
  \BibitemOpen
  \bibfield  {author} {\bibinfo {author} {\bibfnamefont {A.}~\bibnamefont
  {Kalur}}, \bibinfo {author} {\bibfnamefont {P.}~\bibnamefont {Seiler}}, \
  and\ \bibinfo {author} {\bibfnamefont {M.~S.}\ \bibnamefont {Hemati}},\
  }\bibfield  {title} {\enquote {\bibinfo {title} {Stability and performance
  analysis of nonlinear and non-normal systems using quadratic constraints},}\
  }in\ \href@noop {} {\emph {\bibinfo {booktitle} {AIAA Scitech 2020 Forum}}}\
  (\bibinfo {year} {2020})\ p.\ \bibinfo {pages} {0833}\BibitemShut {NoStop}%
\bibitem [{\citenamefont {Kalur}\ \emph
  {et~al.}(2020{\natexlab{b}})\citenamefont {Kalur}, \citenamefont {Seiler},\
  and\ \citenamefont {Hemati}}]{kalur2020nonlinear}%
  \BibitemOpen
  \bibfield  {author} {\bibinfo {author} {\bibfnamefont {A.}~\bibnamefont
  {Kalur}}, \bibinfo {author} {\bibfnamefont {P.}~\bibnamefont {Seiler}}, \
  and\ \bibinfo {author} {\bibfnamefont {M.~S.}\ \bibnamefont {Hemati}},\
  }\bibfield  {title} {\enquote {\bibinfo {title} {Nonlinear stability analysis
  of transitional flows using quadratic constraints},}\ }\href@noop {}
  {\bibfield  {journal} {\bibinfo  {journal} {arXiv preprint arXiv:2004.05440}\
  } (\bibinfo {year} {2020}{\natexlab{b}})}\BibitemShut {NoStop}%
\bibitem [{\citenamefont {Sharma}(2009)}]{sharma2009model}%
  \BibitemOpen
  \bibfield  {author} {\bibinfo {author} {\bibfnamefont {A.~S.}\ \bibnamefont
  {Sharma}},\ }\bibfield  {title} {\enquote {\bibinfo {title} {Model reduction
  of turbulent fluid flows using the supply rate},}\ }\href@noop {} {\bibfield
  {journal} {\bibinfo  {journal} {Int. J. Bifurc. Chaos}\ }\textbf {\bibinfo
  {volume} {19}},\ \bibinfo {pages} {1267--1278} (\bibinfo {year}
  {2009})}\BibitemShut {NoStop}%
\bibitem [{\citenamefont {Constantin}\ and\ \citenamefont
  {Doering}(1995)}]{Constantin1995}%
  \BibitemOpen
  \bibfield  {author} {\bibinfo {author} {\bibfnamefont {P.}~\bibnamefont
  {Constantin}}\ and\ \bibinfo {author} {\bibfnamefont {C.~R.}\ \bibnamefont
  {Doering}},\ }\bibfield  {title} {\enquote {\bibinfo {title} {{Variational
  bounds on energy dissipation in incompressible flows. II. Channel flow}},}\
  }\href {\doibase 10.1103/PhysRevE.51.3192} {\bibfield  {journal} {\bibinfo
  {journal} {Phys. Rev. E}\ }\textbf {\bibinfo {volume} {51}},\ \bibinfo
  {pages} {3192--3198} (\bibinfo {year} {1995})}\BibitemShut {NoStop}%
\bibitem [{\citenamefont {Eves}(1980)}]{eves1980elementary}%
  \BibitemOpen
  \bibfield  {author} {\bibinfo {author} {\bibfnamefont {H.~W.}\ \bibnamefont
  {Eves}},\ }\href@noop {} {\emph {\bibinfo {title} {Elementary matrix
  theory}}}\ (\bibinfo  {publisher} {Courier Corporation},\ \bibinfo {year}
  {1980})\BibitemShut {NoStop}%
\bibitem [{\citenamefont {Holm}\ \emph {et~al.}(1985)\citenamefont {Holm},
  \citenamefont {Marsden}, \citenamefont {Ratiu},\ and\ \citenamefont
  {Weinstein}}]{holm1985nonlinear}%
  \BibitemOpen
  \bibfield  {author} {\bibinfo {author} {\bibfnamefont {D.~D.}\ \bibnamefont
  {Holm}}, \bibinfo {author} {\bibfnamefont {J.~E.}\ \bibnamefont {Marsden}},
  \bibinfo {author} {\bibfnamefont {T.}~\bibnamefont {Ratiu}}, \ and\ \bibinfo
  {author} {\bibfnamefont {A.}~\bibnamefont {Weinstein}},\ }\bibfield  {title}
  {\enquote {\bibinfo {title} {Nonlinear stability of fluid and plasma
  equilibria},}\ }\href@noop {} {\bibfield  {journal} {\bibinfo  {journal}
  {Phys. Rep.}\ }\textbf {\bibinfo {volume} {123}},\ \bibinfo {pages} {1--116}
  (\bibinfo {year} {1985})}\BibitemShut {NoStop}%
\bibitem [{\citenamefont {Salmon}(1988)}]{salmon1988hamiltonian}%
  \BibitemOpen
  \bibfield  {author} {\bibinfo {author} {\bibfnamefont {R.}~\bibnamefont
  {Salmon}},\ }\bibfield  {title} {\enquote {\bibinfo {title} {Hamiltonian
  fluid mechanics},}\ }\href@noop {} {\bibfield  {journal} {\bibinfo  {journal}
  {Annu. Rev. Fluid Mech.}\ }\textbf {\bibinfo {volume} {20}},\ \bibinfo
  {pages} {225--256} (\bibinfo {year} {1988})}\BibitemShut {NoStop}%
\bibitem [{\citenamefont {Morrison}(1998)}]{morrison1998hamiltonian}%
  \BibitemOpen
  \bibfield  {author} {\bibinfo {author} {\bibfnamefont {P.~J.}\ \bibnamefont
  {Morrison}},\ }\bibfield  {title} {\enquote {\bibinfo {title} {Hamiltonian
  description of the ideal fluid},}\ }\href@noop {} {\bibfield  {journal}
  {\bibinfo  {journal} {Rev. Mod. Phys.}\ }\textbf {\bibinfo {volume} {70}},\
  \bibinfo {pages} {467} (\bibinfo {year} {1998})}\BibitemShut {NoStop}%
\bibitem [{\citenamefont {Mu}\ and\ \citenamefont {Wu}(2001)}]{mu2001arnol}%
  \BibitemOpen
  \bibfield  {author} {\bibinfo {author} {\bibfnamefont {M.}~\bibnamefont
  {Mu}}\ and\ \bibinfo {author} {\bibfnamefont {Y.~H.}\ \bibnamefont {Wu}},\
  }\bibfield  {title} {\enquote {\bibinfo {title} {Arnol'd nonlinear stability
  theorems and their application to the atmosphere and oceans},}\ }\href@noop
  {} {\bibfield  {journal} {\bibinfo  {journal} {Surv. Geophys.}\ }\textbf
  {\bibinfo {volume} {22}},\ \bibinfo {pages} {383--426} (\bibinfo {year}
  {2001})}\BibitemShut {NoStop}%
\bibitem [{\citenamefont {Horn}\ and\ \citenamefont
  {Johnson}(2012)}]{horn2012matrix}%
  \BibitemOpen
  \bibfield  {author} {\bibinfo {author} {\bibfnamefont {R.~A.}\ \bibnamefont
  {Horn}}\ and\ \bibinfo {author} {\bibfnamefont {C.~R.}\ \bibnamefont
  {Johnson}},\ }\href@noop {} {\emph {\bibinfo {title} {Matrix analysis}}}\
  (\bibinfo  {publisher} {Cambridge university press},\ \bibinfo {year}
  {2012})\BibitemShut {NoStop}%
\bibitem [{\citenamefont {Gallier}(2010)}]{gallier2010schur}%
  \BibitemOpen
  \bibfield  {author} {\bibinfo {author} {\bibfnamefont {J.}~\bibnamefont
  {Gallier}},\ }\bibfield  {title} {\enquote {\bibinfo {title} {The {S}chur
  complement and symmetric positive semidefinite (and definite) matrices},}\
  }\href@noop {} {\bibfield  {journal} {\bibinfo  {journal} {Penn Engineering}\
  } (\bibinfo {year} {2010})}\BibitemShut {NoStop}%
\bibitem [{\citenamefont {L{\"o}fberg}(2004)}]{lofberg2004yalmip}%
  \BibitemOpen
  \bibfield  {author} {\bibinfo {author} {\bibfnamefont {J.}~\bibnamefont
  {L{\"o}fberg}},\ }\bibfield  {title} {\enquote {\bibinfo {title} {{YALMIP}: A
  toolbox for modeling and optimization in {MATLAB}},}\ }in\ \href@noop {}
  {\emph {\bibinfo {booktitle} {Proceedings of the CACSD Conference}}},\
  Vol.~\bibinfo {volume} {3}\ (\bibinfo {organization} {Taipei, Taiwan},\
  \bibinfo {year} {2004})\BibitemShut {NoStop}%
\bibitem [{\citenamefont {Sturm}(1999)}]{sturm1999using}%
  \BibitemOpen
  \bibfield  {author} {\bibinfo {author} {\bibfnamefont {J.~F.}\ \bibnamefont
  {Sturm}},\ }\bibfield  {title} {\enquote {\bibinfo {title} {Using {SeDuMi}
  1.02, a {MATLAB} toolbox for optimization over symmetric cones},}\
  }\href@noop {} {\bibfield  {journal} {\bibinfo  {journal} {Optim. Methods
  Softw.}\ }\textbf {\bibinfo {volume} {11}},\ \bibinfo {pages} {625--653}
  (\bibinfo {year} {1999})}\BibitemShut {NoStop}%
\bibitem [{\citenamefont {Vandenberghe}\ and\ \citenamefont
  {Boyd}(1996)}]{vandenberghe1996semidefinite}%
  \BibitemOpen
  \bibfield  {author} {\bibinfo {author} {\bibfnamefont {L.}~\bibnamefont
  {Vandenberghe}}\ and\ \bibinfo {author} {\bibfnamefont {S.}~\bibnamefont
  {Boyd}},\ }\bibfield  {title} {\enquote {\bibinfo {title} {Semidefinite
  programming},}\ }\href@noop {} {\bibfield  {journal} {\bibinfo  {journal}
  {SIAM Rev.}\ }\textbf {\bibinfo {volume} {38}},\ \bibinfo {pages} {49--95}
  (\bibinfo {year} {1996})}\BibitemShut {NoStop}%
\bibitem [{\citenamefont {Boyd}\ and\ \citenamefont
  {Vandenberghe}(2004)}]{boyd2004convex}%
  \BibitemOpen
  \bibfield  {author} {\bibinfo {author} {\bibfnamefont {S.}~\bibnamefont
  {Boyd}}\ and\ \bibinfo {author} {\bibfnamefont {L.}~\bibnamefont
  {Vandenberghe}},\ }\href@noop {} {\emph {\bibinfo {title} {Convex
  optimization}}}\ (\bibinfo  {publisher} {Cambridge university press},\
  \bibinfo {year} {2004})\BibitemShut {NoStop}%
\bibitem [{\citenamefont {Parrilo}(2000)}]{parrilo2000structured}%
  \BibitemOpen
  \bibfield  {author} {\bibinfo {author} {\bibfnamefont {P.~A.}\ \bibnamefont
  {Parrilo}},\ }\emph {\bibinfo {title} {Structured semidefinite programs and
  semialgebraic geometry methods in robustness and optimization}},\ \href@noop
  {} {Ph.D. thesis},\ \bibinfo  {school} {California Institute of Technology}
  (\bibinfo {year} {2000})\BibitemShut {NoStop}%
\bibitem [{\citenamefont {Papachristodoulou}\ and\ \citenamefont
  {Prajna}(2005{\natexlab{a}})}]{Papachristodoulou2005}%
  \BibitemOpen
  \bibfield  {author} {\bibinfo {author} {\bibfnamefont {A.}~\bibnamefont
  {Papachristodoulou}}\ and\ \bibinfo {author} {\bibfnamefont {S.}~\bibnamefont
  {Prajna}},\ }\bibfield  {title} {\enquote {\bibinfo {title} {A tutorial on
  sum of squares techniques for systems analysis},}\ }in\ \href@noop {} {\emph
  {\bibinfo {booktitle} {Proceedings of the 2005, American Control Conference,
  2005.}}}\ (\bibinfo {organization} {IEEE},\ \bibinfo {year} {2005})\ pp.\
  \bibinfo {pages} {2686--2700}\BibitemShut {NoStop}%
\bibitem [{\citenamefont {Waleffe}(1997)}]{waleffe1997self}%
  \BibitemOpen
  \bibfield  {author} {\bibinfo {author} {\bibfnamefont {F.}~\bibnamefont
  {Waleffe}},\ }\bibfield  {title} {\enquote {\bibinfo {title} {On a
  self-sustaining process in shear flows},}\ }\href@noop {} {\bibfield
  {journal} {\bibinfo  {journal} {Phys. Fluids}\ }\textbf {\bibinfo {volume}
  {9}},\ \bibinfo {pages} {883--900} (\bibinfo {year} {1997})}\BibitemShut
  {NoStop}%
\bibitem [{\citenamefont {Kim}\ and\ \citenamefont
  {Lim}(2000)}]{kim2000linear}%
  \BibitemOpen
  \bibfield  {author} {\bibinfo {author} {\bibfnamefont {J.}~\bibnamefont
  {Kim}}\ and\ \bibinfo {author} {\bibfnamefont {J.}~\bibnamefont {Lim}},\
  }\bibfield  {title} {\enquote {\bibinfo {title} {A linear process in
  wall-bounded turbulent shear flows},}\ }\href@noop {} {\bibfield  {journal}
  {\bibinfo  {journal} {Phys. Fluids}\ }\textbf {\bibinfo {volume} {12}},\
  \bibinfo {pages} {1885--1888} (\bibinfo {year} {2000})}\BibitemShut {NoStop}%
\bibitem [{\citenamefont {Brunton}\ \emph {et~al.}(2016)\citenamefont
  {Brunton}, \citenamefont {Proctor},\ and\ \citenamefont
  {Kutz}}]{brunton2016discovering}%
  \BibitemOpen
  \bibfield  {author} {\bibinfo {author} {\bibfnamefont {S.~L.}\ \bibnamefont
  {Brunton}}, \bibinfo {author} {\bibfnamefont {J.~L.}\ \bibnamefont
  {Proctor}}, \ and\ \bibinfo {author} {\bibfnamefont {J.~N.}\ \bibnamefont
  {Kutz}},\ }\bibfield  {title} {\enquote {\bibinfo {title} {Discovering
  governing equations from data by sparse identification of nonlinear dynamical
  systems},}\ }\href@noop {} {\bibfield  {journal} {\bibinfo  {journal} {Proc.
  Natl. Acad. Sci.}\ }\textbf {\bibinfo {volume} {113}},\ \bibinfo {pages}
  {3932--3937} (\bibinfo {year} {2016})}\BibitemShut {NoStop}%
\bibitem [{\citenamefont {Loiseau}\ and\ \citenamefont
  {Brunton}(2018)}]{Loiseau2018}%
  \BibitemOpen
  \bibfield  {author} {\bibinfo {author} {\bibfnamefont {J.~C.}\ \bibnamefont
  {Loiseau}}\ and\ \bibinfo {author} {\bibfnamefont {S.~L.}\ \bibnamefont
  {Brunton}},\ }\bibfield  {title} {\enquote {\bibinfo {title} {{Constrained
  sparse Galerkin regression}},}\ }\href {\doibase 10.1017/jfm.2017.823}
  {\bibfield  {journal} {\bibinfo  {journal} {J. Fluid Mech.}\ }\textbf
  {\bibinfo {volume} {838}},\ \bibinfo {pages} {42--67} (\bibinfo {year}
  {2018})}\BibitemShut {NoStop}%
\bibitem [{\citenamefont {Anderson}\ and\ \citenamefont
  {Papachristodoulou}(2015)}]{Anderson2015}%
  \BibitemOpen
  \bibfield  {author} {\bibinfo {author} {\bibfnamefont {J.}~\bibnamefont
  {Anderson}}\ and\ \bibinfo {author} {\bibfnamefont {A.}~\bibnamefont
  {Papachristodoulou}},\ }\bibfield  {title} {\enquote {\bibinfo {title}
  {{Advances in computational Lyapunov analysis using sum-of-squares
  programming}},}\ }\href {\doibase 10.3934/dcdsb.2015.20.2361} {\bibfield
  {journal} {\bibinfo  {journal} {Discrete Cont. Dyn.-B}\ }\textbf {\bibinfo
  {volume} {20}},\ \bibinfo {pages} {2361--2381} (\bibinfo {year}
  {2015})}\BibitemShut {NoStop}%
\bibitem [{\citenamefont {Papachristodoulou}\ and\ \citenamefont
  {Prajna}(2005{\natexlab{b}})}]{papachristodoulou2005analysis}%
  \BibitemOpen
  \bibfield  {author} {\bibinfo {author} {\bibfnamefont {A.}~\bibnamefont
  {Papachristodoulou}}\ and\ \bibinfo {author} {\bibfnamefont {S.}~\bibnamefont
  {Prajna}},\ }\bibfield  {title} {\enquote {\bibinfo {title} {Analysis of
  non-polynomial systems using the sum of squares decomposition},}\ }in\
  \href@noop {} {\emph {\bibinfo {booktitle} {Positive polynomials in
  control}}}\ (\bibinfo  {publisher} {Springer},\ \bibinfo {year} {2005})\ pp.\
  \bibinfo {pages} {23--43}\BibitemShut {NoStop}%
\bibitem [{\citenamefont {Chantry}\ \emph {et~al.}(2016)\citenamefont
  {Chantry}, \citenamefont {Tuckerman},\ and\ \citenamefont
  {Barkley}}]{chantry2016turbulent}%
  \BibitemOpen
  \bibfield  {author} {\bibinfo {author} {\bibfnamefont {M.}~\bibnamefont
  {Chantry}}, \bibinfo {author} {\bibfnamefont {L.~S.}\ \bibnamefont
  {Tuckerman}}, \ and\ \bibinfo {author} {\bibfnamefont {D.}~\bibnamefont
  {Barkley}},\ }\bibfield  {title} {\enquote {\bibinfo {title}
  {Turbulent--laminar patterns in shear flows without walls},}\ }\href@noop {}
  {\bibfield  {journal} {\bibinfo  {journal} {J. Fluid Mech.}\ }\textbf
  {\bibinfo {volume} {791}},\ \bibinfo {pages} {R8} (\bibinfo {year}
  {2016})}\BibitemShut {NoStop}%
\bibitem [{\citenamefont {Chantry}\ \emph {et~al.}(2017)\citenamefont
  {Chantry}, \citenamefont {Tuckerman},\ and\ \citenamefont
  {Barkley}}]{chantry2017universal}%
  \BibitemOpen
  \bibfield  {author} {\bibinfo {author} {\bibfnamefont {M.}~\bibnamefont
  {Chantry}}, \bibinfo {author} {\bibfnamefont {L.~S.}\ \bibnamefont
  {Tuckerman}}, \ and\ \bibinfo {author} {\bibfnamefont {D.}~\bibnamefont
  {Barkley}},\ }\bibfield  {title} {\enquote {\bibinfo {title} {Universal
  continuous transition to turbulence in a planar shear flow},}\ }\href@noop {}
  {\bibfield  {journal} {\bibinfo  {journal} {J. Fluid Mech.}\ }\textbf
  {\bibinfo {volume} {824}},\ \bibinfo {pages} {R1} (\bibinfo {year}
  {2017})}\BibitemShut {NoStop}%
\bibitem [{\citenamefont {Tuckerman}\ \emph {et~al.}(2020)\citenamefont
  {Tuckerman}, \citenamefont {Chantry},\ and\ \citenamefont
  {Barkley}}]{tuckerman2020patterns}%
  \BibitemOpen
  \bibfield  {author} {\bibinfo {author} {\bibfnamefont {L.~S.}\ \bibnamefont
  {Tuckerman}}, \bibinfo {author} {\bibfnamefont {M.}~\bibnamefont {Chantry}},
  \ and\ \bibinfo {author} {\bibfnamefont {D.}~\bibnamefont {Barkley}},\
  }\bibfield  {title} {\enquote {\bibinfo {title} {Patterns in wall-bounded
  shear flows},}\ }\href@noop {} {\bibfield  {journal} {\bibinfo  {journal}
  {Annu. Rev. Fluid Mech.}\ }\textbf {\bibinfo {volume} {52}},\ \bibinfo
  {pages} {343--367} (\bibinfo {year} {2020})}\BibitemShut {NoStop}%
\end{thebibliography}%


%

\end{document}